\documentclass[aps,twocolumn,pra,superscriptaddress,footinbib]{revtex4-1}
\usepackage{amsmath,amssymb,bm}
\usepackage{graphicx}
\usepackage{epstopdf}
\usepackage{latexsym}
\usepackage[normalem]{ulem} 
\usepackage{appendix}
\usepackage{dsfont}
\usepackage[usenames,dvipsnames]{color}
\usepackage{hyperref}
\usepackage{natbib}
\usepackage{braket}
\usepackage{placeins}
\usepackage{soul}
\begin{document}

\newcommand{\Rev }[1]{{\color{blue}{#1}\normalcolor}} 
\newcommand{\Com}[1]{{\color{red}{#1}\normalcolor}} 

\newcommand{\ketbra}[2]{|#1\rangle\langle #2|}
\newcommand{\normord}[1]{\mathopen{:}\,#1\,\mathopen{:}}

\author{D. Barberena}
\affiliation{JILA, NIST, Department of Physics, University of Colorado,  Boulder, CO 80309, USA}
\affiliation{Center for Theory of Quantum Matter, University of Colorado, Boulder, CO 80309, USA}
\author{R.~J. Lewis-Swan}
\affiliation{JILA, NIST, Department of Physics, University of Colorado,  Boulder, CO 80309, USA}
\affiliation{Center for Theory of Quantum Matter, University of Colorado, Boulder, CO 80309, USA}
\author{J.~K. Thompson}
\affiliation{JILA, NIST, Department of Physics, University of Colorado,  Boulder, CO 80309, USA}
\affiliation{Center for Theory of Quantum Matter, University of Colorado, Boulder, CO 80309, USA}
\author{A.~M. Rey}
\affiliation{JILA, NIST, Department of Physics, University of Colorado,  Boulder, CO 80309, USA}
\affiliation{Center for Theory of Quantum Matter, University of Colorado, Boulder, CO 80309, USA}

\title{Atom-light entanglement for precise field sensing in the optical domain}
\date{\today}

\begin{abstract}
Macroscopic arrays of cold atoms trapped in optical  cavities can reach the strong atom-light  collective coupling regime thanks to the simultaneous interactions of the cavity mode with  the atomic ensemble. In a recent work \cite{PRL} we reported a protocol that takes advantage of the strong and collective atom-light interactions in cavity QED systems for precise electric field sensing in the optical domain. We showed that it can provide  between $10$-$20$~dB of metrological gain over the standard quantum limit  in current cavity QED experiments operating with long-lived alkaline-earth atoms. Here, we give a more in depth discussion of the protocol using both exact analytical calculations and numerical simulations, and describe the precise conditions under  which the predicted enhancement holds after thoroughly  accounting for both photon loss and spontaneous emission,  natural decoherence mechanisms in current experiments.  The analysis presented here not only serves to benchmark the protocol and its utility in cavity QED arrays but also sets the conditions required for its applicability in other experimental platforms such as arrays of trapped ions.
\end{abstract}


\maketitle  

\section{Introduction}
Quantum sensing is emerging as an area with great promise, particularly in the context of leveraging quantum effects for real-world technological advances. 
Towards this end, much of the effort in this field has been directed to demonstrations of sensing beyond the Standard Quantum Limit (SQL), which bounds the sensitivity of classical devices with respect to measuring or inferring small perturbations. Efforts to surpass the SQL by harnessing quantum effects such as entanglement and non-classical correlations are taking place in a diverse range of platforms and are allowing for improved phase-estimation in state-of-the art interferometers. A pioneering example is the use of squeezed states of light \cite{walls_quantum_2008} for gravitational wave detection in the Advanced LIGO experiment \cite{Caves1981,AdvancedLigo2016,Aasi2013}, for axion-like dark matter searches in microwave cavities \cite{Malnou2019}, and also a proof-of principle experiment to sense small mechanical displacements in a trapped ion system \cite{Burd2019} (in this case using phonons instead of photons). Similarly,  demonstrations of non-classical atom-light states, including Schr\"{o}dinger cat states, have been also been demonstrated in microwave cavities  using Rydberg atoms \cite{Penasa_2016}, superconducting qubits \cite{Vlastakis2013} and phonon-like trapped ion analogs \cite{Didi_2018}.


A crucial limit to any quantum-enhanced technology is decoherence due to undesirable coupling to an environment. Specifically in the context of quantum metrology, states which posses sub-SQL sensitivity are intrinsically fragile to decoherence. In fact, there is a delicate tradeoff between enhanced metrological utility and increased susceptibility to decoherence, which can quickly degrade any quantum advantage \cite{Huelga1997}. To ensure that coherent or even dissipative processes responsible for the creation of the non-classical state are much faster than any undesirable decoherence rates, experiments using light-matter interactions with single qubits 
have been forced to  operate in the strong coupling regime \cite{Schuster_2007,Girvin_2014,Hacker_2019}. Here, the atom-light interaction strength $2g$, is larger than the decay rates of the qubit, $\gamma$, and resonator $\kappa$. 

Optical cavities traditionally do not fall into this paradigm, and much of the recent focus in this platform has been on the generation of entangled atomic states \cite{Leroux2010c,Schleier-Smith2010b,Hosten_2016,Hosten_2016b,Cox_2016} with efforts directed towards their use for enhanced optical frequency standards  using long-lived clock states  \cite{Norcia_2018,Hu_2017,RLS_2018,Ludlow2015}.
However, one of the key benefits of optical cavity platforms is that they can potentially host a very large number of atoms that coherently interact with a single electromagnetic mode. This effectively causes a collective enhancement of the interaction strength and can lead to the realization of strong \emph{collective} coupling $g\sqrt{N} \gg \kappa, \gamma$.  

In Ref.\cite{PRL} we predicted that strong collective coupling in an optical cavity can be used to prepare entangled atom-light cat-states for quantum-enhanced sensing of weak fields in the optical domain. We showed that collective  atom-light interactions  can provide  between $10$-$20$~dB of metrological gain over the standard quantum limit  in current cavity QED experiments operating with long-lived alkaline-earth atoms. Moreover, we demonstrated that by generating the entanglement via an interaction between two different subsystems, combined with a readout protocol based on time-reversal of the entangling dynamics, one can  extract nearly optimal sensitivity using only readily accessible observables such as atomic inversion. 

In this companion article we elaborate on these results in detail. We also present a more generic treatment of the dynamics that is applicable in a broader parameter regime and which extends beyond the scope of the simpler perturbative approaches used in Ref.\cite{PRL}. In Sec.~\ref{sec:IdealProtocol} we introduce the  dispersive atom-light interaction Hamiltonian  that is the basis of our protocol and describe how to use it to generate metrologically useful states. Further, we outline the basic time-reversal protocol that underpins our proposed experimental realization. Then, in Sec.~\ref{sec:InteractionEngineering} we outline how to engineer the dispersive interaction from the fundamental Tavis-Cummings Hamiltonian that describes the natural atom-light coupling in an optical cavity. Finally, in Sec.~\ref{sec:Dissipation} we present a detailed analysis of the experimentally achievable sensitivity in the presence of cavity decay, atomic spontaneous emission and relevant technical noise. We also discuse a possible generalization of the time-reversal protocol to counteract these effects.

\section{Atom-light quantum sensor \label{sec:IdealProtocol}}

\begin{figure}
    \centering
    \includegraphics[width=8.5cm]{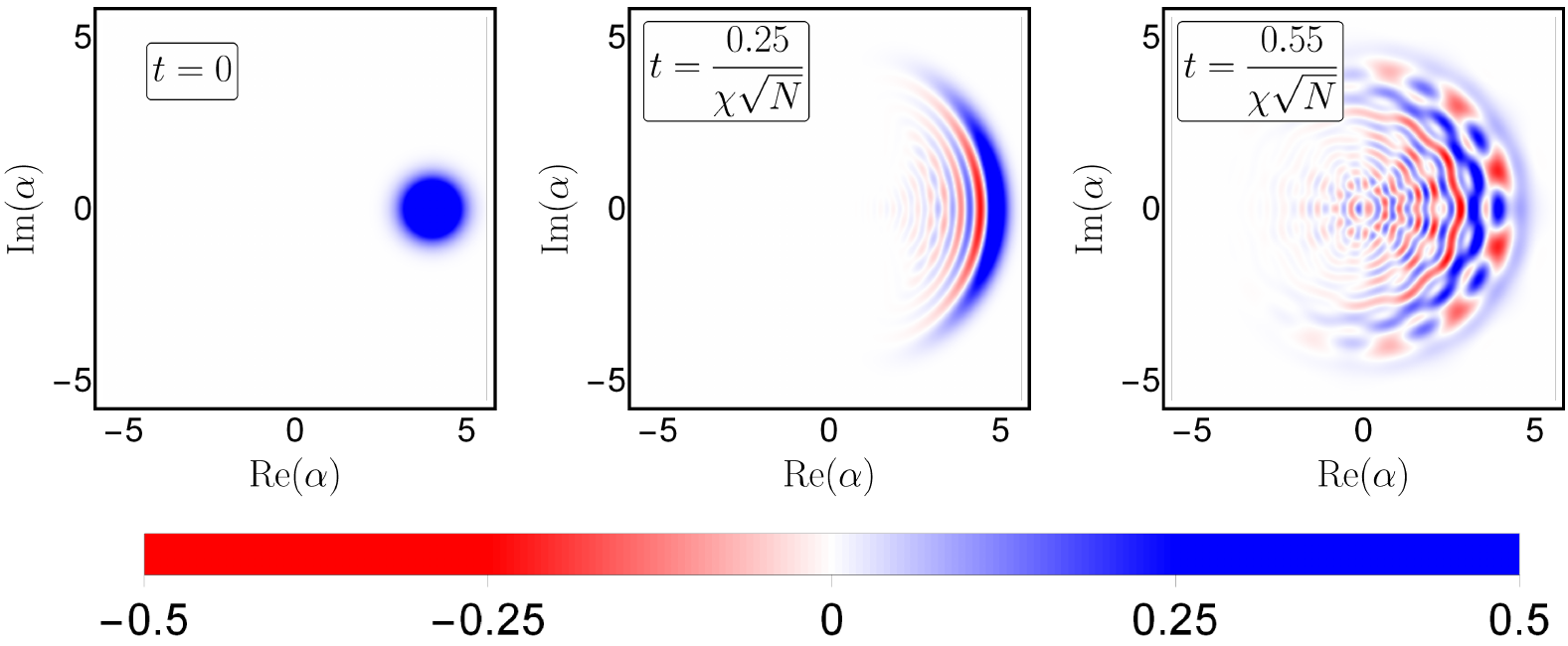}
    \caption{ Wigner function of analogous bosonic generalized cat state $\ket{\psi_B} \propto \sum_{{m_z}=-N/2}^{N/2} c_{m_z} \vert \alpha e^{-i\omega_{m_z} t} \rangle$ with $\omega_{m_z} = \chi {m_z}$ at various evolution times $t$. Sensitivity to displacements increases left to right as it is relates to the smallest scale structure observed in the Wigner function.}
    \label{fig:WignerFuncs}
\end{figure}

Our aim is to use atom-light interactions in an optical cavity to generate states that are useful for quantum-enhanced sensing of small coherent displacements of the cavity field. Specifically, we want to dynamically generate resource states which are capable of sensing small displacements beyond the standard quantum limit (SQL) \cite{Braunstein1994,Jaekel_1990}. For clarity, we reiterate that we are referring to the SQL of displacements of a bosonic system, in contrast to the SQL of phase-shifts, which is the more typical use in the literature. In this context, the SQL is defined with respect to the sensitivity attainable using quasi-classical states, in particular a bosonic coherent state. Intuitively, the sensitivity achievable with a coherent state is bounded by its rms width $\sigma$ in phase-space (see Fig.~\ref{fig:WignerFuncs}), corresponding to the vacuum noise \cite{walls_quantum_2008}. Hence, the SQL is given by $(\delta\beta)^2 \geq 1/\sigma^2 = 1/4$, where $\delta\beta$ is the precision to which a displacement $\beta$ can be estimated. Numerous investigations have demonstrated that by introducing correlations and 
non-classicality, the quantum projection noise of a bosonic state can be manipulated to achieve precision beyond the SQL to the so-called Heisenberg limit \cite{Zurek2001,Pezze_2018,Giovannetti_2006,Yurke_1986,Holland_1993}, which typically scales as $(\delta\beta)^2 \sim 1/\bar{n}$ where $\bar{n}$ is the average particle number of the state. It is also useful to compare these bounds on bosonic displacements with the analogous metrological bounds on phase shifts $\phi$. For the latter, the SQL scales like $(\delta\phi)^2\sim 1/\bar{n}$ while the Heisenberg limit possesses the improved scaling $(\delta\phi)^2\sim 1/\bar{n}^2$. In fact, under certain conditions the bounds on phase shifts and displacements can be related using the relation $(\delta\beta)^2\sim(\sqrt{\bar{n}}\delta\phi)^2$~\cite{Toscano_2006}.

It has previously been shown within the context of, e.g., microwave cavities and circuit-QED, that dispersively coupling a single qubit to a bosonic field can be used to dynamically prepare superposition states which are highly sensitive to small coherent displacements \cite{Toscano_2006,Penasa_2016}. Here, we extrapolate to the case of many qubits and consider a dispersive atom-light coupling of the form:
\begin{equation}
    \hat{H} = \chi \hat{a}^{\dagger}\hat{a}\hat{S}_z , \label{eqn:Hal}
\end{equation}
where $\hat{a}$ ($\hat{a}^{\dagger}$) is the destruction (creation) operator of a single cavity mode, $\hat{S}_{x,y,z} = (1/2)\sum_{j=1}^N \hat{\sigma}^j_{x,y,z}$ are collective spin operators defined as the sum over individual Pauli operators $\hat{\sigma}^j_{x,y,z}$ acting on atom $j$, and $\chi$ characterizes the strength of the atom-light interaction. Note we set $\hbar = 1$ throughout this paper. Furthermore, we remark that this Hamiltonian is not exclusive of atom-light systems and can arise in, e.g., trapped ion setups, where the center of mass of the ions takes the place of the cavity mode.

Entangled atom-light states can readily be generated by evolution under Eq.~(\ref{eqn:Hal}) starting from the initial product state: 
\begin{equation}\label{eqn:IniState}
    \ket{\psi_0}=\ket{N/2_x}\otimes\ket{\alpha},
\end{equation} where $\ket{N/2_x}=\sum_{m=-N/2}^{N/2} c_m\ket{m_z}$ is a spin coherent state of $N$ spin-$1/2$s fully polarized along $x$, $\ket{m_z}$ is a spin basis state such that $\hat{S}_z\ket{m_z}=m_z\ket{m_z}$ and $\ket{\alpha}$ is a bosonic coherent state with amplitude $\alpha\in\mathds{R}$. Since $\hat{H}$ is invariant under $\hat{a}\to\hat{a} e^{i\phi}$, the choice $\alpha\in\mathds{R}$ entails no loss of generality and in fact defines the phase reference from which all other phases are measured. 

The dispersive interaction generates rotations of the initial bosonic state at a rate set by   the  $z$ spin projection of the initial atomic state \cite{PRL}: 
\begin{equation}
    \ket{\psi_t}=\sum_{m_z} c_{m_z}\ket{m_z}\otimes\ket{\alpha e^{-i\omega_{m_z}t}} , \label{eqn:ALcat}
\end{equation}
where $\omega_{m_z }= \chi {m_z}$.

The generated superposition state, Eq.~(\ref{eqn:ALcat}), can be identified as a generalized cat-state \cite{Schneider_1998,Zurek2001,Toscano_2006}. Such states are appreciated to have great metrological potential \cite{Zurek2001,Munro_2002}, because they exhibit fine structure in phase-space which makes them quickly distinguishable upon perturbation. In particular, while the spin degree of freedom is essential to the measurement protocol outlined below, much of the metrological sensitivity of the state $\ket{\psi_t}$ can be understood by considering an analogous purely bosonic state $\ket{\psi_B} \propto \sum_{{m_z}=-N/2}^{N/2} c_{m_z} \vert \alpha e^{-i\omega_{m_z} t} \rangle$. The state $\ket{\psi_B}$ is meant to serve as a toy model of the full state $\ket{\psi_t}$, with the added benefit that visualization is much simpler for $\ket{\psi_B}$, as will be discussed in the next paragraph. For clarity, we remark that $\ket{\psi_B}$ is \emph{not} obtained from $\ket{\psi_t}$ by tracing over the spin degrees of freedom. Here, we choose to weight the superposition of bosonic coherent states by the same coefficients $c_{m_z}$ simply to make the analogy closer. 

The structure of the state $\ket{\psi_B}$ and corresponding metrological utility can be best visualized using the associated Wigner function \cite{HILLERY_1984,Agarwal_2004}:
\begin{equation}
    W_B(\zeta) = \frac{2}{\pi} e^{2\vert\zeta\vert^2} \int d^2\beta ~ \langle -\beta \vert \psi_B \rangle \langle \psi_B \vert \beta \rangle e^{-2(\beta\zeta^* - \beta^*\zeta)} ,
\end{equation}
defined in a bosonic phase-space with respect to the coherent state basis. In Fig.~\ref{fig:WignerFuncs} we plot $W_B(\zeta)$ at three different times $t$ for a simple example with $\alpha = 4$ and $N = 10$. At $t=0$ the Wigner function is that of the initial coherent state, which is a Gaussian centered at $\zeta = \alpha$ and with rms width $\sigma = 1/2$ corresponding to the vacuum noise. For relatively short-times, $t\lesssim 1/\chi\sqrt{N}$, we expect many of the coherent states in the superposition of $\ket{\psi_B}$ to have significant overlap. This is illustrated for an example $t = 0.25(\chi\sqrt{N})^{-1}$, for which we see the Wigner function appears to be composed of concentric crescents. These crescents highlight that the Wigner function has oscillatory structure at length scales smaller than the original vacuum noise. At longer times, the coherent states begin to distinctly disperse and even more complex fine structure emerges in phase-space as shown for the example at $t = 0.55(\chi\sqrt{N})^{-1}$. The sub-SQL sensitivity of these latter two states is a direct result of the fine structure appearing on scales smaller than typical vacuum noise. Specifically, the fine structure means that the application of a small displacement, at a scale below the SQL, can still make the final perturbed state rapidly orthogonal to the original \cite{Zurek2001}. 

A quantitative assessment of the metrological utility of the spin boson cat state $\ket{\psi_t}$ for any perturbation is given by the Cr\`amer-Rao bound \cite{Braunstein1994,Helstrom_1976}. In particular, the sensitivity $\delta\beta$ to a small coherent displacement $\beta$ is bounded by the quantum Fisher information $\mathcal{F}_Q$ (QFI) as $(\delta\beta)^2\geq(\mathcal{F}_Q)^{-1}$. For a pure state, $\mathcal{F}_Q$ is proportional to the quantum variance of the operator that generates the perturbation \cite{Jarzyna_2012}, evaluated with respect to the aforementioned state.  For the specific case of  displacements along the real axis in phase-space generated by the displacement operator $\mathcal{D} = e^{i\beta\hat{Y}}$  with  $\hat{Y} = i(\hat{a}^{\dagger} - \hat{a})$,  then  $\mathcal{F}_{\mathrm{Q}} = 4\langle (\Delta \hat{Y})^2 \rangle$ where $\langle (\Delta \hat{O} )^2 \rangle \equiv \langle \hat{O}^2 \rangle - \langle \hat{O} \rangle^2$. The choice of $\hat{Y}$ is motivated because Fig.~\ref{fig:WignerFuncs}(b) shows that the largest variance (and hence largest QFI) is at $90^\circ$ with respect to the initial displacement, at least for short times. 

The QFI of the state $\ket{\psi_t}$, Eq.~(\ref{eqn:ALcat}), can be evaluated exactly. For simplicity, this is most easily accomplished by considering the equivalent evolution generating $\ket{\psi_t}$ in the Heisenberg picture. Specifically, we compute the time-evolved annihilation operator:
\begin{equation}
    \hat{a}(t) \equiv e^{i\hat{H}t}\hat{a}(0)e^{-i\hat{H}t}=\hat{a}(0) e^{i\chi\hat{S}_z t} . 
\end{equation}
The relevant moments for the QFI are:
\begin{align}
    \begin{split}
        \braket{\hat{a}(t)}&=\alpha \Big[\cos\Big(\frac{\chi t}{2}\Big)\Big]^{N}\approx \alpha e^{-\frac{N\chi^2 t^2}{8}}, \\[5pt]
        \braket{\hat{a}(t)^2}&=\alpha^2 \Big[\cos(\chi t)\Big]^{N}\approx \alpha^2 e^{-\frac{N\chi^2t^2}{2}} , \\[5pt]
        \braket{\hat{a}^{\dagger}(t)\hat{a}(t)}&=\alpha^2 ,
    \end{split}
\end{align}
which leads to the final result,
\begin{equation}\label{eqn:IdealFisherInformation}
    \mathcal{F}_{\mathrm{Q}}=4+8\alpha^2\Big(1-\cos(\chi t)^{N}\Big)\approx4+8\alpha^2\Big(1-e^{-\frac{N\chi^2t^2}{2}}\Big).
\end{equation}

For short times ($\chi\sqrt{N}t\ll 1$), the QFI simplifies to
\begin{equation}
     \mathcal{F}_{\mathrm{Q}}\approx 4+4N\chi^2\alpha^2 t^2,
\end{equation}
When $\chi\sqrt{N}t\gg 1$, the QFI generically saturates  to
\begin{equation}
    \mathrm{sat}(\mathcal{F}_{\mathrm{Q}}) \approx 4+8\alpha^2,
\end{equation}
apart from rare revivals at $\chi t\approx \pi n$ for $n\in\mathbb{Z}$ where it returns to the SQL $\mathcal{F}_{Q}=4$. A special exception to this is when both $n$ and $N$ are odd, where at $\chi t\approx n\pi$ for $n\in\mathbb{Z}$ the QFI $\mathcal{F}_{Q}$ further increases to $4+16\alpha^2$. 


While the QFI sets a lower bound  on $\delta\beta$, in practice the attainable sensitivity is dictated by the available measurements which can be implemented to infer $\beta$. In particular, a tradeoff in the use of powerful entangled states such as  cat-states is that they typically require sophisticated measurements to saturate the Cramer-Rao bound. This can include parity or fidelity measurements \cite{Bollinger_1996,Macri_2016}, and construction of full distribution functions of observables \cite{Strobel2014}, all of which require single-particle resolution. 

Reflective of this, we find that in our case, measurement of simple observables with respect to the perturbed state $\ket{\psi_{\beta}} = \mathcal{D}(\beta)\ket{\psi_t}$, where $\beta \in \mathbb{R}$,  do not capture the effects of the displacement. To be concrete: Measurement of the cavity quadratures $\hat{X} = \hat{a} + \hat{a}^{\dagger}$ or $\hat{Y} = i(\hat{a}^{\dagger} - \hat{a})$ do not provide sub-SQL sensitivity (see Appendix \ref{app:IdealExpectationValues}), whilst spin observables such as $\hat{S}_{x,y,z}$ are completely insensitive as the perturbation commutes with them. 


Recently, it has been recognized that a powerful approach to overcome this technical obstacle is to use time-reversal of the entangling dynamics after application of the perturbation \cite{Yurke_1986,Hudelist_2014,Linnemann_2016,Hosten_2016,Penasa_2016,Macri_2016,Davis_2016,Szigeti_2017,Wrubel_2018} (also known as an interaction-based readout scheme \cite{Huang_2018,Haine_2018,Nolan_2017,Mirkhalaf_2018}). Typically, if the initial prepared state is Gaussian then reversal of the nonlinear dynamics may allow the perturbation to be inferred efficiently in simple observables, such as the cavity quadratures or spin projections $\hat{S}_{x,y,z}$. 

\begin{figure}
    \centering
    \includegraphics[width=8cm]{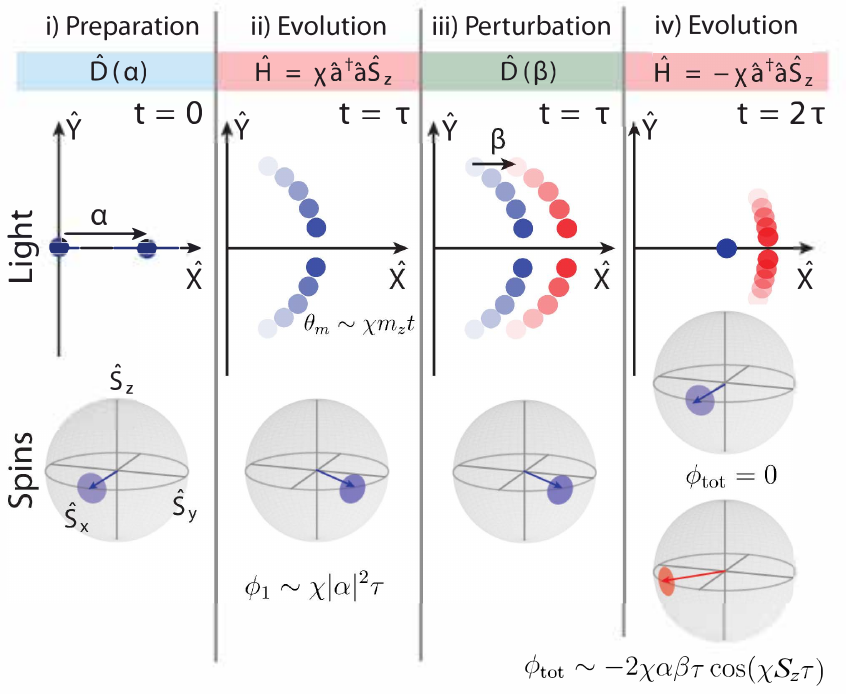}
    \caption{Preparation of generalized cat-state $\vert\psi^{\mathrm{SB}}_{\mathrm{cat}}\rangle$ and interferometric protocol.
    (i) The cavity is injected with a coherent field $\alpha$ and the collective spin is fully polarized along $\hat{x}$ (blue circles). (ii) Fluctuations in the spin projection combined with the dispersive interaction drive a rotation of the initial bosonic coherent state into a superposition at angles $\theta_m \sim \chi m_z \tau$. Conversely, the large cavity occupation rotates the collective Bloch vector by $\phi_1 \sim \chi \vert\alpha\vert^2 \tau$ about $\hat{z}$. (iii) The cavity field is coherently displaced by $\beta$ (red circles). The spin degree of freedom is unaffected. (iv) By reversing the sign of the dispersive interaction the initial rotations are undone. If $\beta \neq 0$ the final cavity state (red circles) does not return to the original coherent state. Similarly, the collective spin rotates back under the evolution by $\phi_2 \sim -\chi\vert\alpha  e^{-i\chi S_z \tau} + \beta\vert \tau$ about $\hat{z}$, leading to an overall rotation $\phi_{\mathrm{tot}} = \phi_1 + \phi_2 \sim -2\chi\alpha\beta \tau\mathrm{cos}(\chi S_z \tau)$ relative to the initial state along $\hat{x}$.}
    \label{fig:Protocol}
\end{figure}

In light of this, we propose the following sensing protocol: (i) prepare the cavity in a coherent state of real amplitude $\alpha$ and all pseudospins polarized along $\hat{x}$, (ii) evolve with $\hat{H}$ [Eq.~(\ref{eqn:Hal})] for time $\tau$, (iii) coherently displace the cavity by small $\beta$, (iv) evolve with $-\hat{H}$ for time $\tau$, and (v) measure an observable $\hat{M}$ (at final time $2\tau$). Note that we use $\tau$ instead of $t$ whenever we refer to the time reversal protocol. The state at the end of all three steps is then described by
\begin{equation}
    \ket{\psi_0}\rightarrow \ket{\psi_f}=e^{i\hat{H}\tau}\ket{\psi_{\beta}}=e^{i\hat{H}\tau}\mathcal{D}(\beta)e^{-i\hat{H}\tau}\ket{\psi_0}.
\end{equation}

We will choose to measure either spin projection $\hat{M}=\hat{S}_{x}$ or $\hat{S}_y$. We can motivate this choice, in particular compared to, e.g., the optical quadratures $\hat{X}$ or $\hat{Y}$, by considering the dynamics of the protocol within a semi-classical approximation. The first evolution describes a rotation of the collective pseudospin Bloch vector about $\hat{z}$ by an angle $\phi_1 \sim \chi \vert \alpha \vert^2 \tau$, induced by the large coherent bosonic amplitude. Reversal of the dynamics, after the small displacement of the cavity field, rotates the Bloch vector in the opposing direction by $\phi_2 \sim -\chi \vert \alpha e^{-i\chi S_z \tau} + \beta \vert^2 \tau$ about $\hat{z}$. Here, the additional phase added to the $\alpha$ term accounts for dynamics of the cavity field during the first evolution period, while $S_z$ is a semi-classical fluctuation $\sim \sqrt{N}$ of the (conserved) inversion due to quantum projection noise. Collectively, these two rotations add up to yield a residual rotation of the Bloch vector from its initial configuration, by $\phi_{\mathrm{tot}} = \phi_1 + \phi_2 \sim -2\chi\alpha\beta \tau \mathrm{cos}(\chi S_z \tau)$ about $\hat{z}$. The rotation angle scales with the coherent amplitude $\alpha$, which thus amplifies the effect of the perturbation $\beta$. The resulting collective spin precession can be tracked by measuring the mean spin projections $\hat{S}_x$ or $\hat{S}_y$. The correction $\propto \mathrm{cos}(\chi S_z \tau)$ will lead to a slow decay in the observable signal as the interaction period $\tau$ increases, and is a result of residual atom-light entanglement at the end of the protocol.

\begin{figure}
    \centering
    \includegraphics[width=0.48\textwidth]{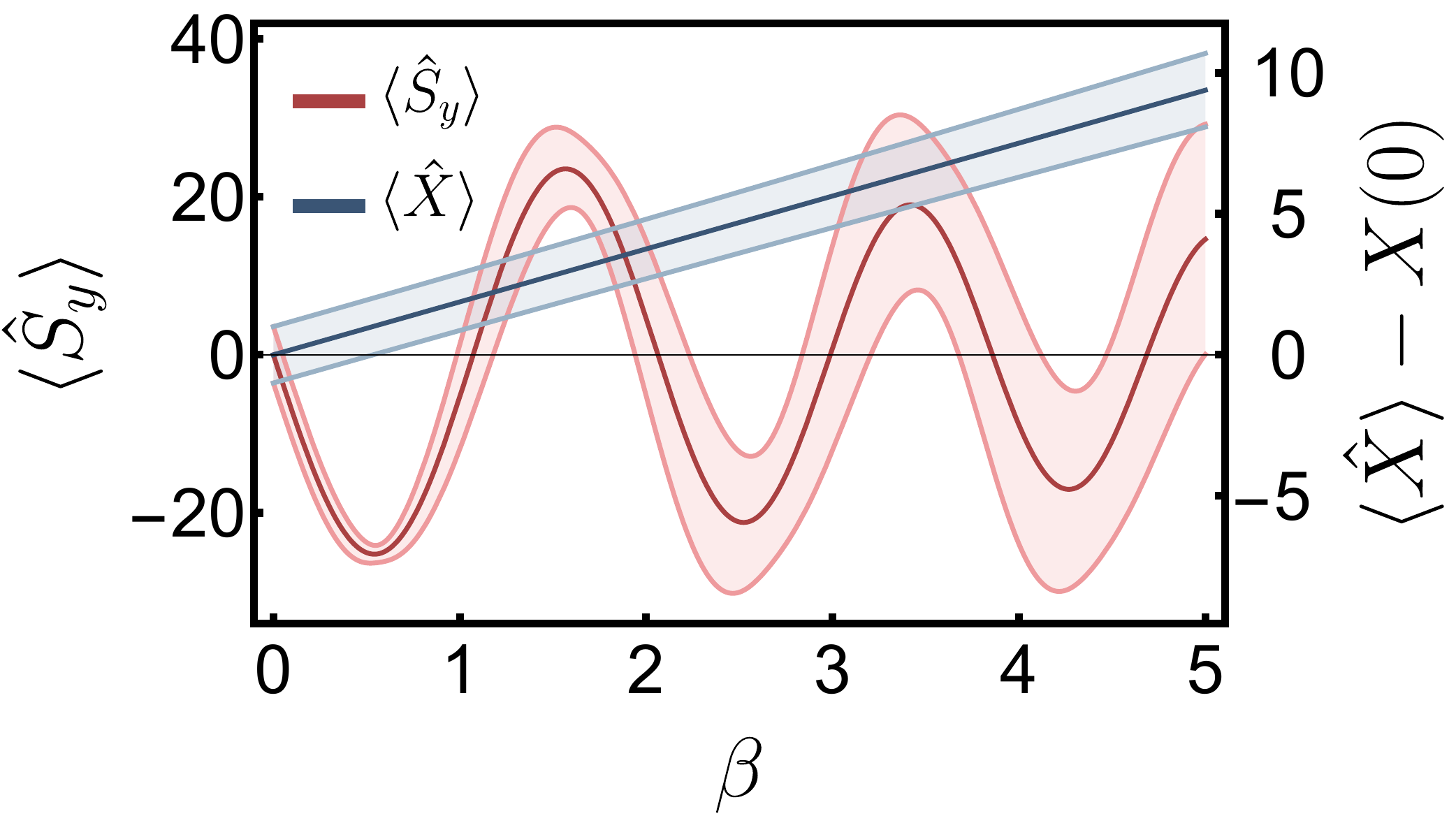}
    \includegraphics[width=0.48\textwidth]{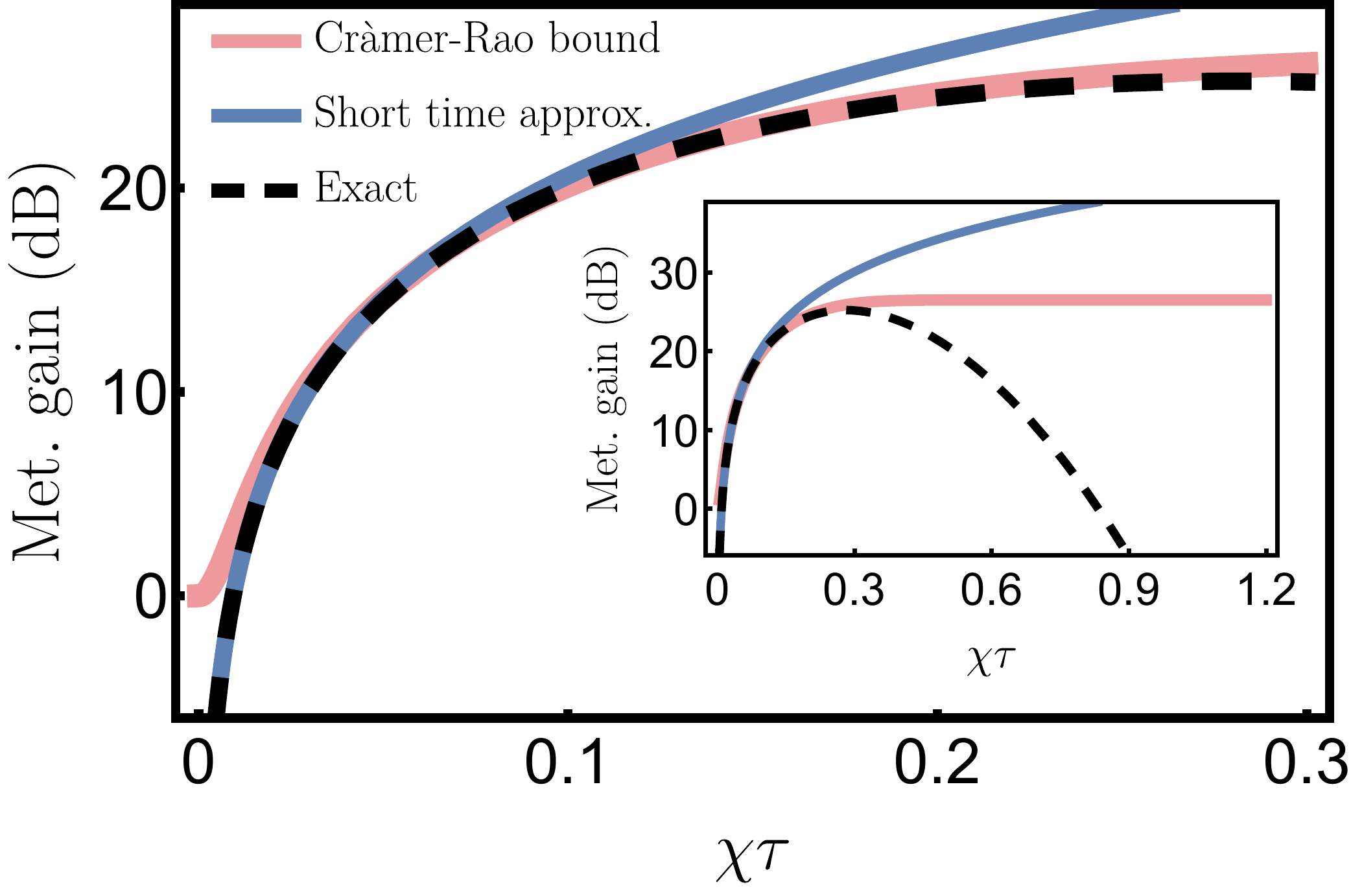}
    \caption{(a) Dependence of measurement observable on $\beta$ for fixed $\chi\tau=0.1$: $\braket{\hat{S}_y}(2\tau)$ (red, oscillatory) and $\braket{\hat{X}}(2\tau)$ (blue, linear). Shaded regions indicate rms fluctuations due to quantum noise, i.e. $\sqrt{\langle [\Delta\hat{S}_y(2\tau)]^2 \rangle}$ and $\sqrt{\langle [\Delta\hat{X}(2\tau)]^2 \rangle}$. The period of oscillations in $\braket{\hat{S}_y}(2\tau)$ is enhanced by the amplitude $\alpha$, allowing a more precise inference of $\beta$. (b) Comparison of attainable metrological gain relative to the SQL as a function of interaction time $\tau$ for $N=51$ and $\alpha=15$. The short time approximation (upper line, solid blue) is given by Eq.~(\ref{AppSensitivity}), while the exact result (lower solid red line) is given by Eq.(\ref{eqn:SensIdealExact}).}
    \label{fig:IdealSensitivity}
\end{figure}

For comparison, following the same protocol the optical quadratures evolve as $X(2\tau) \sim X(0) + 2\beta\mathrm{cos}(\chi S_z \tau)$ and $Y(2\tau) = Y(0)$ respectively. Clearly, the latter is completely insensitive to the perturbation whilst the former does not display any enhancement that scales with the initial coherent displacement $\alpha$. Indeed, as we later shown in Eq.~(\ref{eqn:Xsensitivity}), we find measuring the $\hat{X}$ quadrature yields a sensitivity even worse than the SQL $(\delta \beta)^2 = 1/4$, as the signal washes out due to residual atom-light entanglement [care of the term $\mathrm{cos}(\chi S_z t)$ which becomes an exponential decay after formally treating the quantum noise]. This contrast to the case when  $\hat{M} = \hat{S}_y$ is measured as  illustrated in Fig.~\ref{fig:IdealSensitivity}(a).

Our discussion is made rigorous by exactly computing the achievable sensitivity for each of these observables. Specifically, the sensitivity $\delta\beta$ achievable by measuring $\hat{M}$ is  operationally defined as
\begin{equation}
    (\delta\beta)^2\equiv\frac{\langle (\Delta\hat{M})^2\rangle}{(\partial_{\beta}\langle\hat{M}\rangle)^2}. \label{eqn:SensDefn}
\end{equation}\\
We can explicitly evaluate the sensitivity with respect to $\hat{M} = \hat{S}_y$ by computing the evolution in the Heisenberg picture. 
The relevant expectation values  required to compute the sensitivity given by Eq.~(\ref{eqn:SensDefn}) are shown in Appendix~\ref{app:IdealExpectationValues}), Eq.~(\ref{eqn:plen}). Here we show the sensitivity for $\hat{M} = \hat{S}_y$ as $\beta \to 0$. For this we need the variance $\braket{(\Delta\hat{S}_y)^2}$ to order $\beta^0$ and the signal $\braket{\hat{S}_y}$ to order $\beta$:
\begin{align}
    \begin{split}
    \braket{\hat{S}_y}&=2\alpha\beta N\sin(\chi\tau/2)\Big[\cos\Big(\frac{\chi\tau}{2}\Big)\Big]^{N-1}+O(\beta^2) ,\\
        \braket{(\Delta\hat{S}_y)^2}&=\frac{N}{4}+O(\beta).
    \end{split}
\end{align}
The sensitivity is then given by
\begin{align}\begin{split}
    (\delta\beta)^2 \equiv \frac{\langle (\Delta\hat{S}_{y})^2 \rangle}{\left\vert \frac{d\langle\hat{S}_{y}\rangle}{d\beta} \right\vert^2}&=\frac{1}{16\alpha^2 N[\sin(\chi \tau/2)\cos(\chi \tau/2)^{N-1}]^{2}} ,\\[5pt]
    &\approx\frac{e^{N\chi^2 \tau^2/4}}{16\alpha^2 N[\sin(\chi \tau/2)]^2}, \label{eqn:SensIdealExact}
\end{split}\end{align}
where the approximation of the cosine as an exponential in the second line is valid for $\chi \tau\ll 1$. In fact, Eq.~(\ref{eqn:SensIdealExact}) is valid for any spin projection on the equatorial plane of the Bloch sphere, $\hat{S}_{\varphi}=\hat{S}_x\cos\varphi+\hat{S}_y\sin\varphi$. However, we highlight that in practice $\varphi = 0$, corresponding to $\hat{S}_x$, should be avoided as at the typical working point of the interferometer ($\beta \approx 0$) both the slope of the expectation value and variance vanish at different rates, so the sensitivity would be dominated by technical noise in any experimental realization. A further discussion of this experimental point is made in Sec.~\ref{sec:DetectionNoise}.

For short times ($\chi\sqrt{N}\tau\ll 1$), we attain a sensitivity 
\begin{equation}\label{AppSensitivity}
   ( \delta\beta)^2\approx \frac{1}{4N\alpha^2\chi^2\tau^2} ,
\end{equation}
which is close to the bound set by the Fisher information, $(\delta\beta)^2 \geq \mathcal{F}_Q^-1 \approx (4+4N\alpha^2\chi^2\tau^2)^{-1}$. The divergence as $\tau\to 0$ reflects the limitations imposed on the sensivity by spin projection noise $\langle (\Delta\hat{S}_y)^2\rangle\propto N/4$. Specifically, the atoms and light must interact for a time  sufficiently long such that the small displacement of the cavity field can be mapped into a resolvable rotation of the collective spin. More rigorously, this requires that the perturbation of the collective spin along $S_y$ satisfies $\delta S_y \equiv (N/2)\phi_{\mathrm{tot}} = \chi\alpha\delta\beta\tau \geq \sqrt{N/4}$ where the RHS of the inequality is the characteristic projection noise of a coherent spin state. In fact, satisfying this inequality can be used to qualitatively derive Eq.~(\ref{AppSensitivity}).

We show this result for the sensitivity in Fig.~\ref{fig:IdealSensitivity}, and compare it to the Cr\`amer-Rao bound, given by the Fisher information of Eq.~(\ref{eqn:IdealFisherInformation}).  For the sake of clarity, throughout this paper we plot the attainable sensitivity as the metrological gain with respect to the SQL:
\begin{equation}
    \mathrm{Met.~ gain}~ (\mathrm{dB})=-10\log_{10}[4(\delta\beta)^2].
\end{equation}

To complete our previous discussion comparing the usefulness of the cavity quadratures as opposed to spin observables, the sensitivity attainable with $\hat{M} = \hat{X}$ is 
\begin{align}\begin{split}
    (\delta\beta)^2 \equiv \frac{\langle (\Delta\hat{X})^2 \rangle}{\left\vert \frac{d\langle\hat{X}\rangle}{d\beta} \right\vert^2}=\frac{1}{4[\cos(\chi \tau/2)]^{2N}}\approx \frac{e^{N\chi^2\tau^2/4}}{4}, \label{eqn:Xsensitivity} 
\end{split}\end{align}
which is never below the SQL.

\section{Engineering the dispersive interaction
}\label{sec:InteractionEngineering}
Our proposed protocol and discussion of the previous section  hinges on the ability to engineer Eq.~(\ref{eqn:Hal}). In this section, we outline two experimentally feasible methods to realize this interaction. Whilst our focus is on implementation in an optical cavity, we point out that the following is readily applicable to other platforms with spin-boson interactions, including trapped ion arrays \cite{Bollinger_2018}.

We begin by assuming the underlying microscopic model describing the coupling of a single bosonic cavity mode to a collection of $N$ two-level systems (atomic transitions) can be written as a Tavis-Cummings Hamiltonian
\begin{equation}
    \hat{H}_{\mathrm{TC}} = g\left( \hat{a}^{\dagger}\hat{S}^- + \hat{a}\hat{S}^+ \right) - \Delta_c \hat{a}^{\dagger}\hat{a} . \label{eqn:HTC}
\end{equation}
Here, $\Delta_c$ is the detuning of the atomic transition from the cavity mode frequency and $2g$ is the single-photon Rabi frequency.

\subsection{Dispersive protocol}\label{sec:InteractionEngineerginDispersive}
Our first proposed scheme assumes that the cavity detuning is large with respect to other relevant scales in the Hamiltonian, specifically $\vert\Delta_c\vert \gg \vert g\vert \sqrt{N}$, and is a collective generalization of the strong dispersive limit in single qubit microwave cavity experiments \cite{Bertet_2002,Schuster_2007,Blais_2004,Girvin_2014}. We shift $\hat{H}_{\mathrm{TC}}$ into the interaction picture generated by $\hat{H}_0 = -\Delta_c\hat{a}^{\dagger}\hat{a}$:
\begin{equation}
    \hat{H}_{\mathrm{I}}=g\big(\hat{S}^+\hat{a}e^{i\Delta_c t}+\hat{a}^{\dagger}\hat{S}^-e^{-i\Delta_c t}\big) .
\end{equation}
Using the approach of Ref.~\cite{James2007} we can compute an effective time-averaged Hamiltonian, which in the original frame is 
\begin{equation}\label{eqn:HamDisp}
    \hat{H}_{\mathrm{D}}=-\Delta_c\hat{a}^{\dagger}\hat{a}+\frac{g^2}{\Delta_c}\hat{S}^+\hat{S}^-+\frac{2g^2}{\Delta_c}\hat{a}^{\dagger}\hat{a}\hat{S}_z.
\end{equation}
For this approximation to be valid, the timescale induced by the first term of $\hat{H}_{\mathrm{D}}$ should be greater than the timescale induced by the corrections (second and third terms). Crudely, the second term can potentially generate a mean field rotation of the spins at a rate $g^2N/\Delta_c$, while the third term rotates the spins at a rate $g^2|\alpha|^2/\Delta_c$ (for an initial coherent state of amplitude $\alpha$) or alternatively rotates the photon distribution in phase space at a rate $g^2N/\Delta_c$. All these timescales should be less than $\Delta_c$. Therefore, we require both $\vert \Delta_c \vert\gg \vert g\vert\sqrt{N}$ and $\vert\Delta_c\vert\gg \vert g\alpha|$.

\subsection{Resonant protocol}\label{sec:InteractionEngineeringResonant}
The second scheme we consider conversely assumes that the cavity is tuned to be resonant with the atomic transition. Though not immediately obvious, injecting a large coherent field leads to a slightly modified version of Eq.~(\ref{eqn:Hal}), where the cavity photon number $\hat{a}^{\dagger}\hat{a}$ couples to the spin projection along $x$, $\hat{S}_x$, instead of along $z$. While the naive expectation is that a large classical field should produce Rabi flopping of the atoms, we demonstrate in this section that the dynamics should be augmented by a dispersive interaction that arises due to quantum fluctuations. This scheme was presented briefly in Ref.~\cite{PRL}, but we make the arguments justifying its validity more rigorously here.

Large coherent cavity fields $\ket{\alpha}$ have  well defined phases, with a phase spread $\delta\phi\sim 1/|\alpha|$. Conversely, they have  large photon number fluctuations $\delta n\sim|\alpha|$. This suggests that, in the presence of such a field, the entangling atom-light dynamics will initially be driven by number fluctuations. To account for them more explicitly, we introduce here the number-phase representation of the bosonic operators \cite{Susskind1964}
\begin{align}\begin{split}
\hat{n}&=\hat{a}^{\dagger}\hat{a} ,\\
\hat{a}&=\sqrt{\hat{n}+1}\,e^{i\hat{\phi}}=e^{i\hat{\phi}}\sqrt{\hat{n}}\,,
\end{split}\end{align}
where the last equality is a consequence of the general relation $e^{i\hat{\phi}}f(\hat{n})=f(\hat{n}+1)e^{i\hat{\phi}}$ for any function $f$.
Substitution of these identities into $\hat{H}_{\mathrm{TC}}$ with $\Delta_c = 0$ yields
\begin{equation}
    \hat{H}=g\big(\hat{S}^+e^{i\hat{\phi}}\sqrt{\hat{n}}+h.c.\big).
\end{equation}
Our previous discussion about phase fluctuations would imply that we can replace the phase operator by a classical number, at least for short times. This would be correct, but the Tavis-Cummings model is sufficiently simple that we do not need to make this approximation. Instead, we notice that the combination $\hat{S}^+e^{i\hat{\phi}}$ has the same matrix elements as the operator $\hat{S}^+$ in the sense that 
\begin{equation}
    \bra{n,m_z}\hat{S}^+\ket{n,m_z'}=\bra{n,m_z}\hat{S}^+ e^{i\hat{\phi}}\ket{n+1,m_z'},
\end{equation}
where $n$ is the photon occupation number and $\ket{m_z}$ is an eigenstate of $\hat{S}_z$ with eigenvalue $m_z$. It would therefore prove useful to find a transformation that implements this mapping. This can be achieved using the operator $\hat{T}=e^{i\hat{\phi}(\hat{S}_z+N/2)}$, so that
\begin{equation}
    \hat{S}^+e^{i\hat{\phi}(\hat{S}_z+N/2)}=e^{i\hat{\phi}(\hat{S}_z+N/2-1)}\hat{S}^+.
\end{equation}
One needs to be cautious about these relations because $\hat{T}$ is only well defined the way we have written it when acting on states with $n>N$. In fact any state $\ket{n,m_z}$ with $n<m_z+N/2$ would transform into a state with negative number of bosonic excitations. The standard way out of this problem is to define the action of $\hat{T}$ in such states to be 0. The downside is that $\hat{T}$ defined this way is not a unitary operator, in the sense that $\hat{T}^{\dagger}\hat{T}\neq 1$. However, it remains true that $\hat{T}\hat{T}^{\dagger}=1$ so we can still perform the transformation $\hat{H}\rightarrow\hat{T}^{\dagger}\hat{H}\hat{T}$ by inserting $\hat{T}\hat{T}^{\dagger}$ in between operators and states in any expression. Furthermore, we will assume further on that we are working with states that have support in photon numbers much greater than $N$ so we will not need to care about the precise definition of $\hat{T}$ close to $n=N$. Under these approximations, then
\begin{equation}
    \hat{T}^{\dagger}\hat{n}\,\hat{T}=\hat{n}-\hat{S}_z-\frac{N}{2},
\end{equation}
and, consequently,
\begin{equation}
    \hat{T}^{\dagger}\hat{H}\hat{T}=g\bigg(\hat{S}^+\sqrt{\hat{n}-\frac{N}{2}-\hat{S}_z}+h.c.\bigg).
\end{equation}
As we have assumed that the cavity mode initially has a large mean occupation $\braket{\hat{n}}\equiv\bar{n}\gg N$ with small fluctuations $\sqrt{\langle(\Delta\hat{n})^2\rangle}\ll\bar{n}$, we can replace $\hat{n}\to\bar{n}+\delta\hat{n}$ and keep only the first order in $\delta{\hat{n}}$. Then the Hamiltonian becomes
\begin{align}\begin{split}\label{AppResonantFirstApproximation}
    \hat{T}^{\dagger}\hat{H}\hat{T}&\approx g\Bigg[\sqrt{\bar{n}}\hat{S}^+\bigg(1+\frac{\delta\hat{n}-\frac{N}{2}-\hat{S}_z}{2\bar{n}}\bigg)+h.c.\Bigg] , \\[5pt] 
    &=g\sqrt{\bar{n}}\bigg(1-\frac{N+1
    }{2\bar{n}}\bigg)\hat{S}_x+\frac{g\hat{n}}{\sqrt{\bar{n}}}\hat{S}_x\\[5pt]
    &\hspace{3.2cm}-\frac{g}{2\sqrt{\bar{n}}}\Big(\hat{S}_x\hat{S}_z+\hat{S}_z\hat{S}_x\Big) .
\end{split}\end{align}
The first term and mean-field contribution $\propto \vert\alpha\vert^2$ of the second term, describe Rabi flopping with frequency $2g\sqrt{\bar{n}}$ (with a small correction), whilst fluctuations in photon number generate evolution of the spins through the second term on a timescale $\frac{g\sqrt{\braket{(\delta\hat{n})^2}}}{\sqrt{\bar{n}}}$, which for the case of an initial coherent state is equal to $g$ and  independent of $\bar{n}$. The third term may appear bigger than the second because of the presence of two collective spin operators. However it is highly non-resonant in the frame of the Rabi flopping and hence it generates evolution of the spins with a timescale $\frac{(g N/\sqrt{\bar{n}})^2}{g\sqrt{\bar{n}}}=gN^2/\bar{n}^{3/2}$, which can be made as small as desired by increasing the amplitude of the initial coherent state. On the other hand, the second term commutes with the Rabi flopping so we can write the relevant Hamiltonian as
\begin{align}\begin{split}
    \hat{T}^{\dagger}\hat{H}\hat{T}&=g\sqrt{\bar{n}}\bigg(1-\frac{N+1}{2\bar{n}}\bigg)\hat{S}_x+\frac{g\hat{n}}{\sqrt{\bar{n}}}\hat{S}_x, \\[5pt]
    &\approx g\sqrt{\bar{n}}\hat{S}_x+\frac{g\,\hat{n}}{\sqrt{\bar{n}}}\hat{S}_x . 
    \label{eqn:HamRes}
\end{split}\end{align}

The transformation defined by $\hat{T}$ also acts on other operators and states, and so to be rigorous we calculate its action on them and show that these corrections can be made small. In the case relevant for this publication and consistent with the notation of Eq.~(\ref{eqn:HamRes}), the system starts with all the atoms in the ground-state of the transition. Then, $\hat{T}^{\dagger}\ket{\tilde{\psi}_0}=\ket{\tilde{\psi}_0}$. Furthermore, 
\begin{equation}
    \hat{T}^{\dagger}\hat{a}\hat{T}=\sqrt{1-\frac{\frac{N}{2}+\hat{S}_z}{\hat{n}+1}}\hat{a}=\hat{a}\Big[1+O\big(N/\bar{n}\big)\Big],
\end{equation}
so the relative corrections to bosonic operators are of the order of $N/\bar{n}$, which is already assumed to be small. Spin operators also transform:
\begin{align}
    \begin{split}
        \hat{T}^{\dagger}\hat{S}_z\hat{T}&=\hat{S}_z ,\\[5pt]
       \hat{T}^{\dagger}\hat{S}^+\hat{T}  &=\hat{S}^+e^{-i\hat{\phi}}.
    \end{split}
\end{align}
In this case, the validity of the approximation relies on the phase spread of the state at the end of the protocol. Considering $\sim\pm\sqrt{N}$ fluctuations in $\hat{S}_x$, Eq.~(\ref{eqn:HamRes}) indicates that the initially coherent state will grow to a size $\sim\sqrt{N}gt$ in phase space at a short time $t$. This distribution subtends a phase spread with respect to $\alpha=0$ given by
\begin{equation}
    \delta\phi=\frac{\sqrt{N}gt}{\alpha}.
\end{equation}
Hence, we can expect that 
\begin{equation}
    \braket{\hat{S}^+e^{-i\hat{\phi}}}\approx \braket{\hat{S}^+}+O(\sqrt{N}gt/\alpha),
\end{equation}
and the corrections can be made smaller by increasing $\alpha$.

Lastly, we note that $\hat{n}$ is coupled to $\hat{S}_x$ instead of $\hat{S}_z$. Since this amounts to a rotation of our basis about $\hat{S}_y$, none of the previous results for the Fisher information and sensitivity are altered, as long as we change the initial state to  $\ket{\tilde{\psi}_0}=\ket{(-N/2)_z}\ket{\alpha}$ and perform  measurements of a spin projection in the $yz$ plane. In that case  the attained   sensitivity is  exactly the one  discussed above.

\begin{figure}
    \centering
    \includegraphics[width=0.48\textwidth,trim={0.3cm 0 0 0},clip]{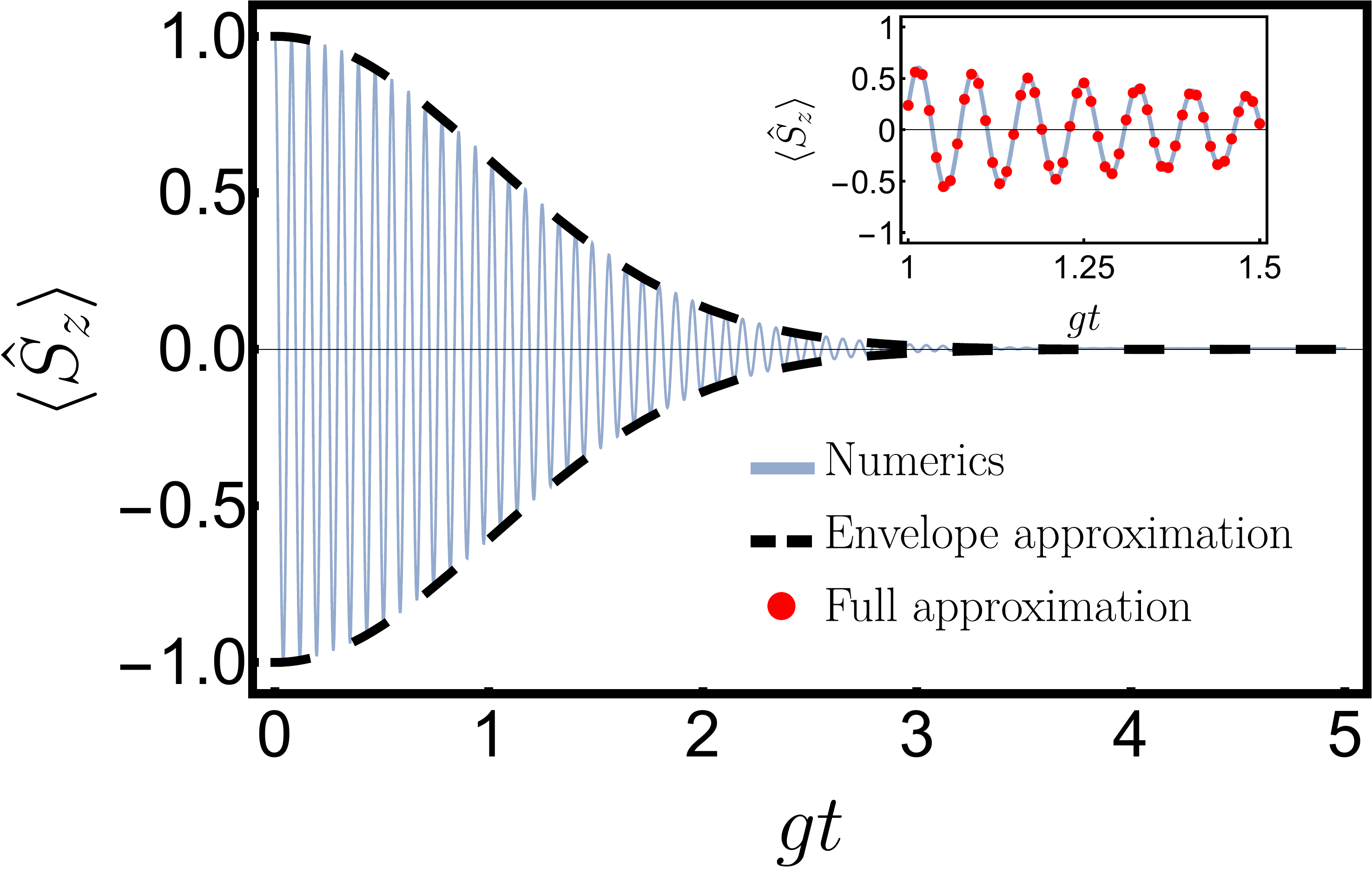}
    \caption{Evolution of $\langle \hat{S}_z \rangle$ predicted by the complete Tavis-Cummings model (solid blue), Eq.~(\ref{eqn:HTC}), for the initial state $\ket{N/2_z}\ket{\alpha}$ with $N=40$ and $\alpha=40$. The decay envelope of the oscillations is compared to that predicted by the effective dispersive interaction Eq.~(\ref{eqn:HamRes}) (dashed black). In the inset we show that the frequency of Rabi oscillations is also captured correctly when including the small correction in the first line of Eq.~(\ref{eqn:HamRes}) (red dots). Though not noticeable, the discrepancy between the exact evolution and the approximation is of about $4\%$, which is consistent with $N/\alpha^2=0.025$.}
    \label{fig:AppFigResonantApproximations}
\end{figure}

\begin{figure}
    \centering
    \includegraphics[width=0.48\textwidth]{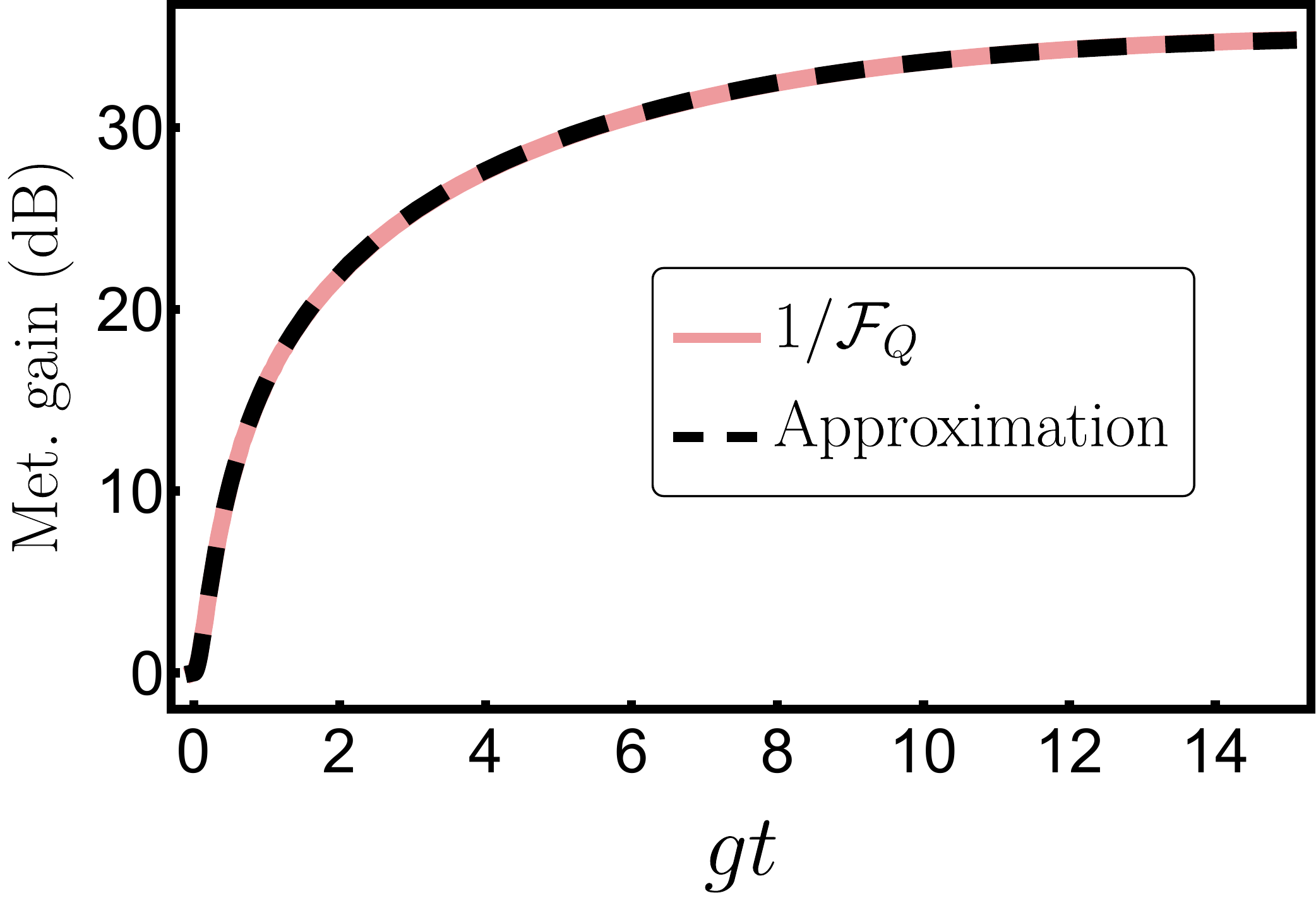}
    \caption{Cr\`amer-Rao bound, $1/\mathcal{F}_Q$, on metrological gain with respect to the SQL, independently calculated using the Tavis-Cummings model (solid red) and effective dispersive interaction Eq.~(\ref{eqn:HamRes}) (dashed black). Calculations are for the initial state $\ket{N/2_z}\ket{\alpha}$ with $N=40$ and $\alpha=40$.}
    \label{fig:ResonantFisherInformation}
\end{figure}

To further support the validity of our approximations, we compare the results of numerical simulations using: (i) the exact Tavis-Cummings Hamiltonian, (ii) the approximation of Eq.~(\ref{eqn:HamRes}). The results are shown in Fig.~\ref{fig:AppFigResonantApproximations}. We also show (see Fig.~\ref{fig:ResonantFisherInformation}) that the Fisher information relevant for our protocol is the same whether it is calculated with the exact Tavis-Cummings Hamiltonian or with Eq.~(\ref{eqn:HamRes}).

Our estimates for the errors introduced by approximating $\hat{H}_{TC}$ with Eq.~(\ref{eqn:HamRes}), further complemented by Fig.~\ref{fig:AppFigResonantApproximations} and Fig.~\ref{fig:ResonantFisherInformation}, show that the approximation is justified for the timescales we are interested in.

Given that we have shown two different ways of obtaining Eq.~(\ref{eqn:Hal}), it is worthwile to point out their differences. Most important of all, the coupling constant $\chi$ takes a different form in both protocols:
\begin{equation}
    \chi_{\text{Dispersive}}=\frac{2g^2}{\Delta_c}, \hspace{1cm}\chi_{\text{Resonant}}=\frac{g}{\alpha},
\end{equation}
and generically
\begin{equation}
    \frac{\chi_{\text{Dispersive}}}{\chi_{\text{Resonant}}}=\frac{2g\alpha}{\Delta_c}\ll 1,
\end{equation}
since $g\alpha/\Delta\ll 1$ is one of the necessary conditions for the dispersive protocol to be valid. Hence, the resonant protocol is faster, which is helpful against dissipation as we will show in the next section. On the other hand, if occupation of the bosonic mode is restricted to be small by other technical reasons (as may happen in other platforms), then it might only be possible to engineer the dispersive protocol.
Another difference is that the spin projection to which photon number fluctuations couple determine what initial spin states are useful. In the dispersive protocol, coupled to $\hat{S}_z$, this means that it is better to start with a state pointing in the $xy$ plane; whereas for the resonant protocol, coupled to $\hat{S}_x$, states pointing in the $yz$ plane are to be preferred. 

Finally, note that $\chi_{\text{Resonant}}=g/\alpha$ depends on $\alpha$ in such a way as to cancel all the $\alpha$ dependence of Eq.~(\ref{AppSensitivity}). For short times, the only effect of increasing $\alpha$ is then to guarantee that the approximations leading into Eq.~(\ref{eqn:HamRes}) are valid. This feature, namely $\alpha$ independence of the sensitivity for short times, will also hold for the resonant protocol in the presence of dissipation, as will be shown in a later section.
\section{Effects of dissipation}\label{sec:Dissipation}
As discussed in the introduction, dissipation and decoherence are a major obstacle that must be overcome in any realistic implementation of a quantum technology. In particular, photons leaked through the mirrors of optical cavities are an intrinsic source of decoherence, with photon loss rates typically much faster than the single-photon Rabi frequency. Another important source of decoherence is spontaneous emission of the atoms, which sets a characteristic time scale within which the atomic coherent dynamics must occur to be useful. In the following we will address each of these sources of intrinsic decoherence separately and use analytic calculations to show that they do not fundamentally limit the attainable sensitivity for reasonable parameter regimes. A combined analytic treatment of both is not possible, and it is also computationally difficult for relevant experiment parameters, but we will argue that cavity decay is the dominant decay process for our protocol.

\subsection{Cavity Decay}\label{sec:CavityLeakage}
In the presence of photon loss, the evolution of the system is given by a master equation with Hamiltonian $\hat{H}=\chi\hat{S}_z\hat{a}^{\dagger}\hat{a}$ and jump operator $\sqrt{\kappa}\hat{a}$:
\begin{equation}\label{CavityDecayMasterEquation}
    \dot{\hat{\rho}}=-i\Big[\chi \hat{S}_z\hat{a}^{\dagger}\hat{a},\hat{\rho}\Big]+\kappa\Big(\hat{a}\hat{\rho}\hat{a}^{\dagger}-\frac{\{\hat{a}^{\dagger}\hat{a},\hat{\rho}\}}{2}\Big)\equiv\mathcal{L}_{\chi}\hat{\rho}.
\end{equation}
We remark again that the results of this section are valid for both the dispersive and resonant protocols as long as the initial states and final measurements are chosen appropriately. 
\subsubsection{Fisher information}
We begin by first seeking to understand how photon decay destroys useful entanglement in the atom-light system, which can be characterized by the Fisher information. Before launching into the complex calculation for the complete model, it is useful to consider a toy model of a simpler bosonic cat state and examine how coherences and entanglement lead to cat death \cite{ONeill2014}.

For our preliminary example, we define the initial cat state as
\begin{equation}\label{eqn:catstate}
    \ket{\psi_{\mathrm{cat}}}=\frac{1}{\mathcal{N}}\bigg(\frac{\ket{\alpha_1}+\ket{\alpha_2}}{\sqrt{2}}\bigg),
\end{equation}
where we take $\alpha_1\neq\alpha_2$, $\mathcal{N}$ is a normalization factor accounting for the 
non-orthogonality of $\ket{\alpha_1}$ and $\ket{\alpha_2}$. We subject the cat state to evolution described by only the dissipative terms in Eq.~(\ref{CavityDecayMasterEquation}).
Rewriting the initially pure state as a density matrix, $\hat{\rho} = \vert \psi_{\mathrm{cat}} \rangle \langle \psi_{\mathrm{cat}} \vert$, the time evolution of the relevant matrix elements is given by
\begin{align}
    \begin{split}
        \ket{\alpha_i}\bra{\alpha_i} &\to \ket{\alpha_i e^{-\kappa t/2}}\bra{\alpha_i e^{-\kappa t/2}}\equiv\ket{\alpha_i^t}\bra{\alpha_i^t} ,\\[5pt]
        \ket{\alpha_1}\bra{\alpha_2}&\to c_t \ket{\alpha_1 e^{-\kappa t/2}}\bra{\alpha_2 e^{-\kappa t/2}} \\[5pt]
        &= c_t \ket{\alpha_1^t}\bra{\alpha_2^t} ,
    \end{split}
\end{align}
where we define $\alpha_i^t\equiv\alpha_i e^{-\frac{\kappa t}{2}}$ for $i=1,2$ and
\begin{align}
    \begin{split}
        c_t&=\exp\bigg[\frac{|\alpha_1|^2+|\alpha_2|^2}{2}(e^{-\kappa t}-1)-\alpha_1\alpha_2^*(1-e^{-\kappa t})\bigg]\\[5pt]
        &\approx\exp\bigg[-\frac{\kappa t}{2}\Big(|\alpha_1|^2+|\alpha_2|^2-2\alpha_1\alpha_2^*\Big)\bigg],
        \end{split}
\end{align}
for which the approximation holds for $\kappa t\ll 1$. For a mixed state and with respect to the generator $\hat{Y}=-i(\hat{a}-\hat{a}^{\dagger})$, the Fisher information is defined as \cite{Braunstein1994}:
\begin{equation}
\mathcal{F}_{Q}=2\sum_{a\neq b}\frac{(\lambda_a-\lambda_b)^2}{\lambda_a+\lambda_b}|\bra{a}\hat{Y}\ket{b}|^2,
\end{equation}
where the $\{\ket{a}\}$ are eigenstates of the time evolved density matrix $\hat{\rho}_t$ and $\{\lambda_a\}$ are their corresponding eigenvalues. Given that the cat-state  only has support in the subspace spanned by $\ket{\alpha_1^t }$ and $\ket{\alpha_2^t}$, which we denote by $I$, we can simplify the Fisher information (see Appendix \ref{app:SimplifiedFisherInformation}) to
\begin{align}\begin{split}\label{eqn:FisherSimplified}
\mathcal{F}_{Q}&=2\sum_{a, b\,\in \,I}\frac{(\lambda_a-\lambda_b)^2}{\lambda_a+\lambda_b}|\bra{a}\hat{Y}\ket{b}|^2\\[5pt]
& \hspace{0.5cm}+4\mathrm{Tr}\big[\hat{P}_I\hat{Y}^{\dagger}\hat{Y}\hat{\rho}\big]-4\mathrm{Tr}\big[\hat{P}_I\hat{Y}^{\dagger}\hat{P}_I\hat{Y}\hat{\rho}\big],
\end{split}\end{align}
where the $\hat{P}_I$ are projectors into $I$. For large enough $|\alpha_1-\alpha_2|^2$ and $\kappa t\ll 1$, $\ket{\alpha_1^t}$ and $\ket{\alpha_2^t}$ are almost orthogonal, so we can consider them to be a basis of $I$. In this basis we define
$\hat{\eta}=\ket{\alpha_1^t}\bra{\alpha_1^t}-\ket{\alpha_2^t}\bra{\alpha_2^t}$ and $\hat{\nu}=\ket{\alpha_1^t}\bra{\alpha_2^t}$ which allows us to express $\hat{\rho}$ and $\hat{X}$ as follows:
\begin{align}\begin{split}
    \hat{\rho}&\approx \frac{\mathbb{I}}{2}+\frac{c_t\hat{\nu} +c_t^*\hat{\nu}^{\dagger}}{2} , \\[5pt]
    P_{I}\hat{Y}P_{I}&\approx -ie^{-\kappa t/2}(\alpha_1-\alpha_1^*)\frac{\mathbb{I}+\hat{\eta}}{2}\\
    &\hspace{0.5cm}-ie^{-\kappa t/2}(\alpha_2-\alpha_2^*)\frac{\mathbb{I}-\hat{\eta}}{2}, \\[5pt]
    P_{I}\hat{Y}^{\dagger}\hat{Y}P_{I}&\approx\big[-e^{-\kappa t}(\alpha_1-\alpha_1^*)^2+1\big]\frac{\mathbb{I}+\hat{\eta}}{2}\\
    &\hspace{0.5cm}+\big[-e^{-\kappa t}(\alpha_2-\alpha_2^*)^2+1\big]\frac{\mathbb{I}-\hat{\eta}}{2},
\end{split}\end{align}
From this representation, the Fisher information of the dying cat state is calculated to be
\begin{equation}
    \mathcal{F}_{Q}\approx 4+4 \Big[\mathrm{Im}(\alpha_1-\alpha_2)\Big]^2 e^{-\kappa t} e^{-\kappa|\alpha_1-\alpha_2|^2 t} . \label{eqn:FisherCat}
\end{equation}
In the absence of dissipation, a large separation in phase space $\propto\alpha_1-\alpha_2$ along the imaginary axis is desirable and leads to a large Fisher information. As has been explained previously in this article and discussed elsewhere \cite{Zurek2001,Toscano_2006}, this is because a large separation leads to fine structure in phase-space which increases the sensitivity of the state to small perturbation. However, for finite $\kappa$ this fine structure is also destroyed very rapidly, illustrated here by the exponential decay of the Fisher information with separation $\propto e^{-\kappa|\alpha_1-\alpha_2|^2t}$.

The example of the bosonic cat state is useful as it can provide powerful intuition into the fragility of the more complex spin-boson cat-state [Eq.~(\ref{eqn:ALcat})]. In particular, it allows us to make a heuristic prediction for the expected scaling of the Fisher information in the presence of photon decay. 

Our toy model consists of approximating the generalized spin-boson cat-state of Eq.~(\ref{eqn:ALcat}) by a simpler superposition involving only the characteristic spin fluctuations $m_z \sim \pm\sqrt{N}$: 
\begin{multline}
 \vert \psi^{\mathrm{SB}}_{\mathrm{cat}} \rangle = \frac{1}{\sqrt{2}}\left( \vert (\sqrt{N})_z\rangle \otimes \vert \alpha e^{-i\chi\sqrt{N}t}\rangle \right. \\
 \left. + \vert (-\sqrt{N})_z\rangle \otimes \vert \alpha e^{i\chi\sqrt{N}t}\rangle \right). 
\end{multline}
The bosonic components of this toy spin-boson cat-state are separated by a characteristic distance $\sim  \chi\alpha\sqrt{N}t$ which dynamically increases. Substituting $\alpha_1 - \alpha_2 \to 2i\chi \alpha \sqrt{N}t$ into Eq.~(\ref{eqn:FisherCat}) and optimisation with respect to $t$ yields a predicted scaling 
\begin{equation}
    \mathcal{F}_Q-4\sim \bigg(\frac{\chi^2 N\alpha^2}{\kappa^2}\bigg)^{1/3} ,
\end{equation}
for the Fisher information.

This apparently simplistic analysis is borne out by more intensive calculations. In particular, we now outline a detailed analysis of the Fisher information for the full spin-boson generalized cat-state, dynamically generated by the dispersive interaction, Eq.~(\ref{eqn:Hal}), and subject to photon loss at rate $\kappa$.

We write the density matrix corresponding to the initial pure state, Eq.~(\ref{eqn:IniState}), as:
\begin{equation}
  \hat{\rho}_0=\Big(\sum_{{m_z},{n_z}} c_{m_z} c^*_{n_z} \ket{m_z}\bra{n_z}\Big)\otimes \ket{\alpha}\bra{\alpha}.
\end{equation}
After evolution for a time $\tau$ under both the coherent and dissipative dynamics, described by  Eq.~(\ref{CavityDecayMasterEquation}), the density matrix is given by (see Appendix \ref{app:ResourceStateinCavity} for more details):
\begin{equation}\label{eqn:FirstStepState}
    \hat{\rho}_{t}=e^{-i\hat{H}t}\Big(\hat{\rho}_{\text{t}}^{\text{spin}}\otimes\ket{\alpha e^{-\kappa t/2}}\bra{\alpha e^{-\kappa t/2}}\Big)e^{i\hat{H}t},
\end{equation}
where
\begin{align}\begin{split}
    \hat{\rho}^{\text{spin}}_{t}&=\sum_{{m_z},{n_z}}c_{m_z} c^*_{n_z} e^{f({m_z}-{n_z},t)}\ket{m_z}\bra{n_z},\\
    f(z,t)&=\frac{\kappa\alpha^2}{\kappa-i\chi z}\Big(1-e^{-\kappa t+i\chi z t}\Big)-\alpha^2(1-e^{\kappa t}).
\end{split}\end{align}

The Fisher information of this state is again obtained via Eq.~(\ref{eqn:FisherSimplified}). The structure of $\hat{\rho}_t$ allows one to re-express $\mathcal{F}_{Q}$ in terms of spin operators alone (see Appendix \ref{app:FisherInfoPhotonLoss}) and we thus obtain:
\begin{equation}\label{eqn:FisherSpin}
    \mathcal{F}_{Q}=4+2\alpha^2e^{-\kappa t}\sum_{r,s}\frac{(\lambda_r-\lambda_s)^2}{\lambda_r+\lambda_s}\Big|\bra{r}\hat{O}\ket{s}\Big|^2, 
\end{equation}
where $\{\ket{r}\}$ are now eigenstates of $\hat{\rho}_t^{\text{spin}}$ only, $\{\lambda_r\}$ their corresponding eigenvalues,
and
\begin{equation}
    \hat{O}=-i\big(e^{-i\chi\hat{S}_z t}-e^{i\chi\hat{S}_z t}\big) .
\end{equation}
In the case that the initial collective spin is large, $N\gg1$, and is prepared in a coherent spin state polarized along the $x$ direction, we can use a Gaussian approximation for the expansion coefficients, $c_{m_z}\propto e^{-\frac{{m_z}^2}{N}}$. For $\chi\sqrt{N}\ll \kappa$, $\chi\sqrt{N}t\ll 1$ we also have that $f(z,t)\approx i\alpha^2(\frac{\chi\kappa  t^2}{2}-\frac{\chi \kappa^2 t^3}{3})z-\frac{\chi^2\kappa \alpha^2t^3 z^2}{6}$. With these approximations we are then able to evaluate Eq.~(\ref{eqn:FisherSpin}) analytically (see Appendix \ref{app:FisherInfoPhotonLoss} for more details), 
\begin{equation}\label{eqn:FisherTime}
    \mathcal{F}_{Q}=4+\frac{4\chi^2N\alpha^2t^2e^{-\kappa t}}{1+\frac{2\chi^2\alpha^2 N}{\kappa^2}\Big[1-e^{-\kappa t}\big(1+\kappa t+\frac{\kappa^2 t^2}{2}\big)\Big]}.
\end{equation}
When $\kappa t\ll 1$, $\mathcal{F}_Q$ becomes
\begin{equation}\label{eqn:FisherTime2}
    \mathcal{F}_Q=4+\frac{4N\alpha^2\chi^2 t^2}{1+\frac{N\chi^2\kappa\alpha^2 t^3}{3}},
\end{equation}
from which the optimal time and $\mathcal{F}_{Q}$ are
\begin{align}\label{eqn:OptimalFisher}
    \begin{split}
        t_{\text{opt}}&=\bigg(\frac{6}{\chi^2\alpha^2\kappa N}\bigg)^{1/3} , \\
        (\mathcal{F}_{Q})_{\text{opt}}&=4+4\bigg(\frac{4\chi^2\alpha^2N}{3\kappa^2}\bigg)^{1/3}\\[5pt]
        &\approx4+4.4\bigg(\frac{\chi^2\alpha^2N}{\kappa^2}\bigg)^{1/3}.
    \end{split}
\end{align}
This result is consistent with the toy model argument up to prefactors. Our results for the optimal Fisher information are valid for 
\begin{equation}\label{eqn:FisherBoundary}
    \frac{\kappa}{\alpha}\ll \chi \sqrt{N}\ll \kappa \alpha^2,
\end{equation}
where the left hand inequality comes from $\kappa t_{\text{opt}}\ll 1$ and the right hand one from $\chi\sqrt{N} t_{\text{opt}}\ll 1$. If the left inequality is not satisfied, then there is no appreciable Fisher information because dissipation is too strong. If the right inequality is not satisfied then Eq.~(\ref{eqn:OptimalFisher}) is no longer valid but for contrary reasons: $(\mathcal{F}_{Q})_{\text{opt}}$ can saturate the value of the ideal case, $(\mathcal{F}_{Q})_{\text{opt}}=4+8\alpha^2$.

\begin{figure*}
    \centering
    \includegraphics[width=0.48\textwidth]{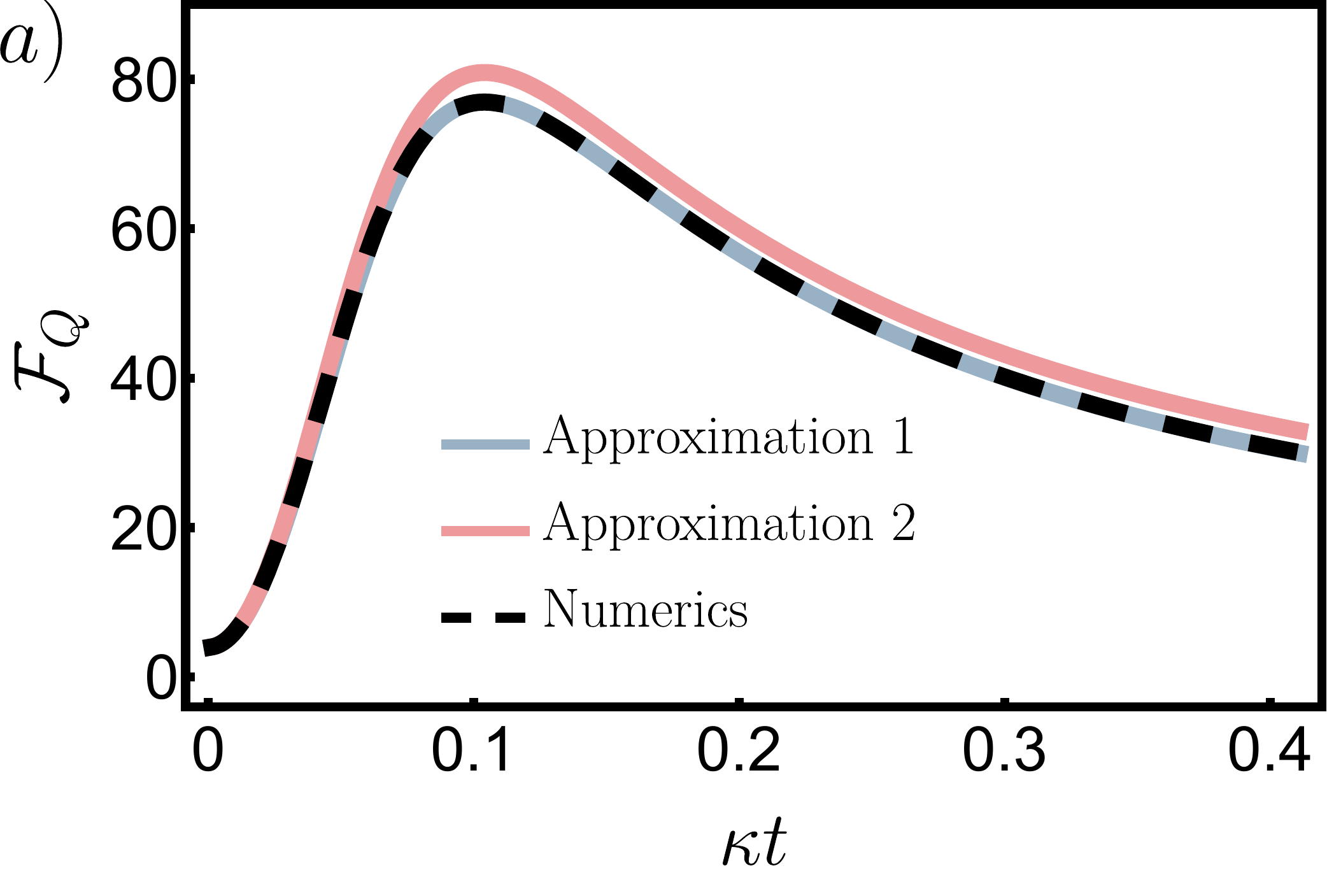}\hspace{0.5cm}
    \includegraphics[width=0.48\textwidth]{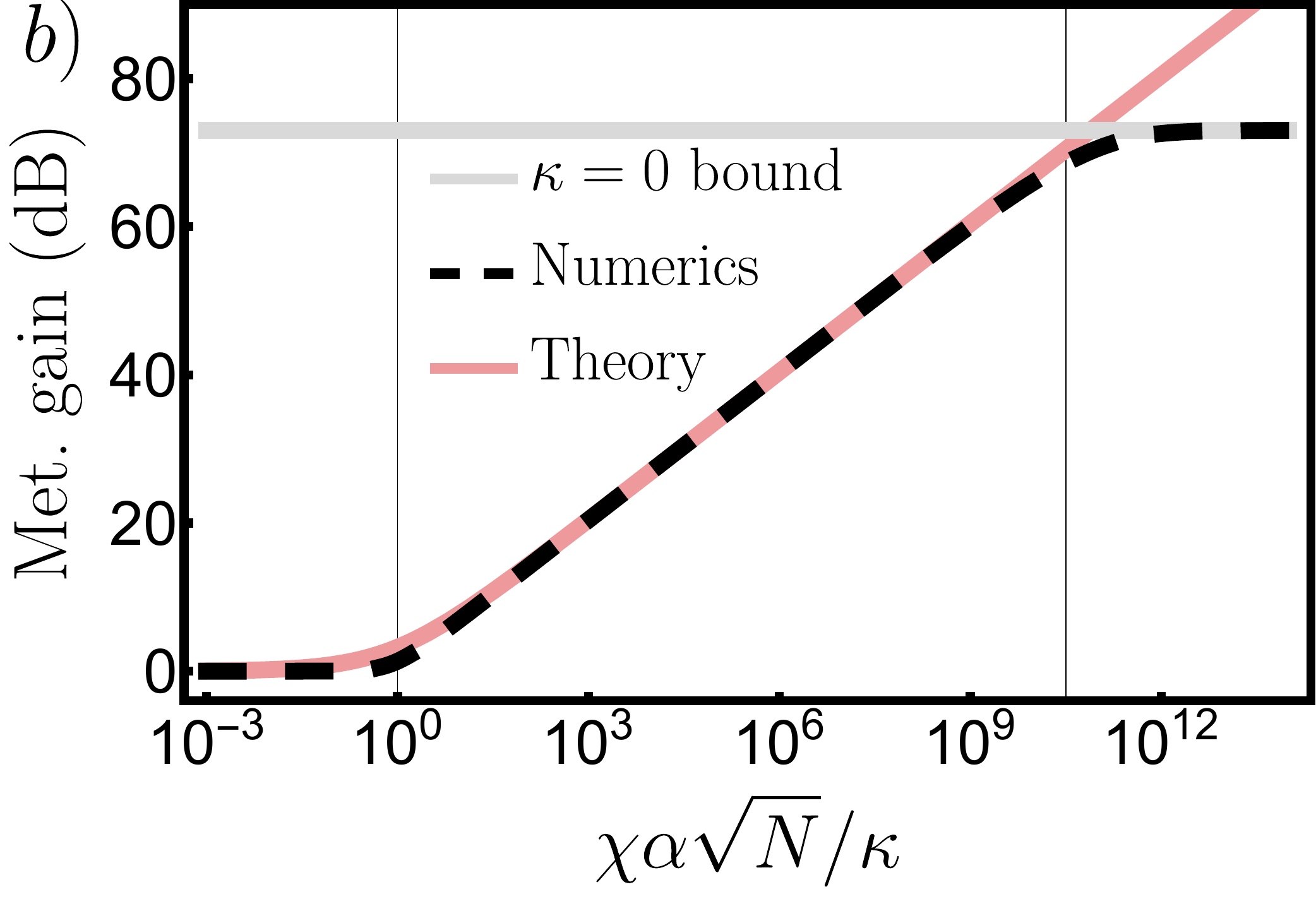}
    \caption{(a) Evolution of QFI with interaction time. Numerical evaluation of $\mathcal{F}_{Q}$ using Eq.~(\ref{eqn:FisherSpin}) (dashed black) is compared to approximate analytic expressions Eq.~(\ref{eqn:FisherTime}) (lower solid, blue) and Eq.~(\ref{eqn:FisherTime2}) (upper solid, red) for $N=1000$, $\alpha=100\sqrt{N}$ and $\chi\alpha\sqrt{N}/\kappa=73\gg 1$. (b) Optimal QFI as a function of $\chi\alpha\sqrt{N}/\kappa$ for $N=1000$ and $\alpha=100\sqrt{N}$. We compare the numerical optimization of Eq.~(\ref{eqn:FisherSpin}) (dashed black) to the approximate analytic expression Eq.~(\ref{eqn:OptimalFisher}) (solid red). The gray horizontal line indicates the optimal QFI for $\kappa=0$, which is attained for very large values of $\chi\alpha\sqrt{N}/\kappa$. The vertical lines mark the region delimited by Eq.~(\ref{eqn:FisherBoundary}), where our results for $(\mathcal{F}_Q)_{\text{opt}}$ are expected to work.}
    \label{fig:FigFisherInformation}
\end{figure*}

We benchmark our analytic calculations by comparison to a full numerical evaluation of Eq.~(\ref{eqn:FisherSpin}), shown in Fig.~\ref{fig:FigFisherInformation}. We choose $N=1000$, $\alpha=100\sqrt{N}$ and $\chi\alpha\sqrt{N}/\kappa=73$. As implied from Eq.(\ref{eqn:FisherTime}), this ratio controls the time development of $\mathcal{F}_Q$ (in units of $\kappa t$). Since it is larger than 1 there should be metrological enhancement, i.e. $\mathcal{F}_Q\gg 4$.  Such a ratio can be experimentally realized, for example, using the parameters in Refs. ~\cite{Norcia_2018,Norcia2016} and applying the resonant interaction described in sec.~\ref{sec:InteractionEngineeringResonant}: $\chi\alpha/2\pi=g/2\pi=11\text{ kHz}$, $N=10^6$ and $\kappa/2\pi=150\text{ kHz}$. We also compare the optimal $\mathcal{F}_Q$ obtained through our analytic expressions against numerical simulations for various values of $\chi\alpha\sqrt{N}/\kappa$ and find that the agreement is excellent in the region where our approximation holds, given by Eq.~(\ref{eqn:FisherBoundary}).

\subsubsection{Achievable sensitivity with collective spin observables}

The effects of photon loss on the time-reversal protocol and the achievable sensitivity $(\delta \beta)^2$ with respect to measurements of collective spin observables can also be analytically evaluated. Specifically, we explicitly calculate the time evolution of relevant operators and evaluate expectations values of collective observables at the end of the time-reversal protocol. 

In the case of nonzero cavity decay, the initial evolution is implemented by $\mathcal{L}_{\chi}$, defined in Eq.~(\ref{CavityDecayMasterEquation}), acting during a time $\tau_1$, and the reversed evolution is implemented by $\mathcal{L}_{-\chi}$. For generality, we assume the second evolution takes time $\tau_2$ which is not neccesarily identical to $\tau_1$. This latter assumption is motivated by the naive expectation that as photons are lost from the cavity the occupation of the cavity field driving the precession of the collective spin is reduced. This will destroy the symmetry of the time-reversal protocol, and thus in our calculation we consider whether choosing $\tau_2 > \tau_1$ may offset this issue and lead to improvements in the achievable sensitivity. 

For simplicity, our calculations are carried out in the Heisenberg picture, for which we have to use the Hilbert-Schmidt adjoints of $\mathcal{L}_{\chi}$ acting in reverse order on the operators of interest. To obtain the sensitivity, we need to calculate the evolution of spin operators and their variances. In particular, we need the evolution of $\hat{S}^+$, $(\hat{S}^+)^2$ and $\hat{S}^+\hat{S}^-$ since they are enough to construct the sensitivity of any spin measurement in the $xy$ plane. The calculations are involved and we quote only the final result (the full derivation can be found in Appendix \ref{app:SensitivityKappa}): 
\begin{widetext}
\begin{align}\begin{split}\label{eqn:AppkappaFinalExpectationValue}
 &\Big\langle(S^+)^m(\tau_1,\tau_2)\Big\rangle=\mathrm{Tr}\Big\{\hat{\rho}_0\big[(\hat{S}^+)^m(\tau_1,\tau_2)\big]\Big\}=\exp\Big[\alpha^2\big(\eta_{\tau_2,m}e^{-\kappa \tau_1+im\chi \tau_1}+\eta_{\tau_1,m}^*\big)+\eta_{\tau_2,m}\beta^2\Big]\Upsilon
\end{split}\end{align}
\end{widetext}
where $\braket{(\hat{S})^m(\tau_1,\tau_2)}$ indicates the expectation value of $\hat{S}^m$ at the end of the protocol and
\begin{align}\begin{split}
    \eta_{\tau,m}&=\frac{i\chi m}{\kappa+i\chi m}(e^{-\kappa\tau-i\chi m\tau}-1)\\[5pt]
   \Upsilon&=\mathrm{Tr}\Big\{(\hat{S}^{+})^m\exp\Big[2\eta_{\tau_2,m}\alpha\beta e^{-\frac{\kappa \tau_1}{2}+i\frac{\chi m \tau_1}{2}}\\
    &\hspace{2.08cm}\cos\big(\chi \tau_1\hat{S}_z+m\chi \tau_1/2\big)\Big]\hat{\rho}_0^{\text{spin}}\Big\}
\end{split}\end{align}

Some general properties of the expectation values can be understood by looking at the exponential prefactor in Eq.~(\ref{eqn:AppkappaFinalExpectationValue}:
\begin{equation}\label{AppExponentialPrefactor}
    \exp\Big[\alpha^2\big(\eta_{\tau_2,n}e^{-\kappa \tau_1+in\chi \tau_1}+\eta_{\tau_1,n}^*\big)+\eta_{\tau_2,n}\beta^2\Big].
\end{equation}
Setting $\tau_1=\tau_2=\tau$ for simplicity and expanding the argument of the exponential in Eq.~(\ref{AppExponentialPrefactor}) for $\kappa t\ll 1$ and $\chi t\ll 1$ we obtain
\begin{equation}
    -i\kappa \alpha^2\chi n \tau^2-\frac{\alpha^2n^2\kappa \tau^3}{3} . \label{eqn:ContrastDecay}
\end{equation}
The first term of Eq.~(\ref{eqn:ContrastDecay}) describes a mismatched overall rotation between the first and last evolution steps, as we foreshadowed. Specifically, the first evolution of the protocol generates a rotation of the spins about $\hat{z}$ through an angle of $\phi_1 \sim \chi\alpha^2 \tau$. In the second evolution period the original coherent state is damped due to photon loss and so the rotation of the spin about $\hat{z}$ is reduced $\phi_2 \sim -\alpha^2 e^{-\kappa \tau}\chi \tau$. Combining these, we then find an overall residual rotation of the spin at the end of the protocol $\phi_{\mathrm{tot}} \sim \chi \alpha^2\kappa \tau^2$ for $\kappa \tau\ll 1$. In principle, this rotation can be corrected by a judicious choice of the measured projection $\hat{S}_{\varphi}$ and so does not affect the sensitivity. On the other hand, the second term of Eq.~(\ref{eqn:ContrastDecay}) arises due to contrast decay of the collective spin induced by decoherence and does modify the sensitivity in an irreversible manner. 

\begin{figure}
    \centering
    \includegraphics[width=0.48\textwidth]{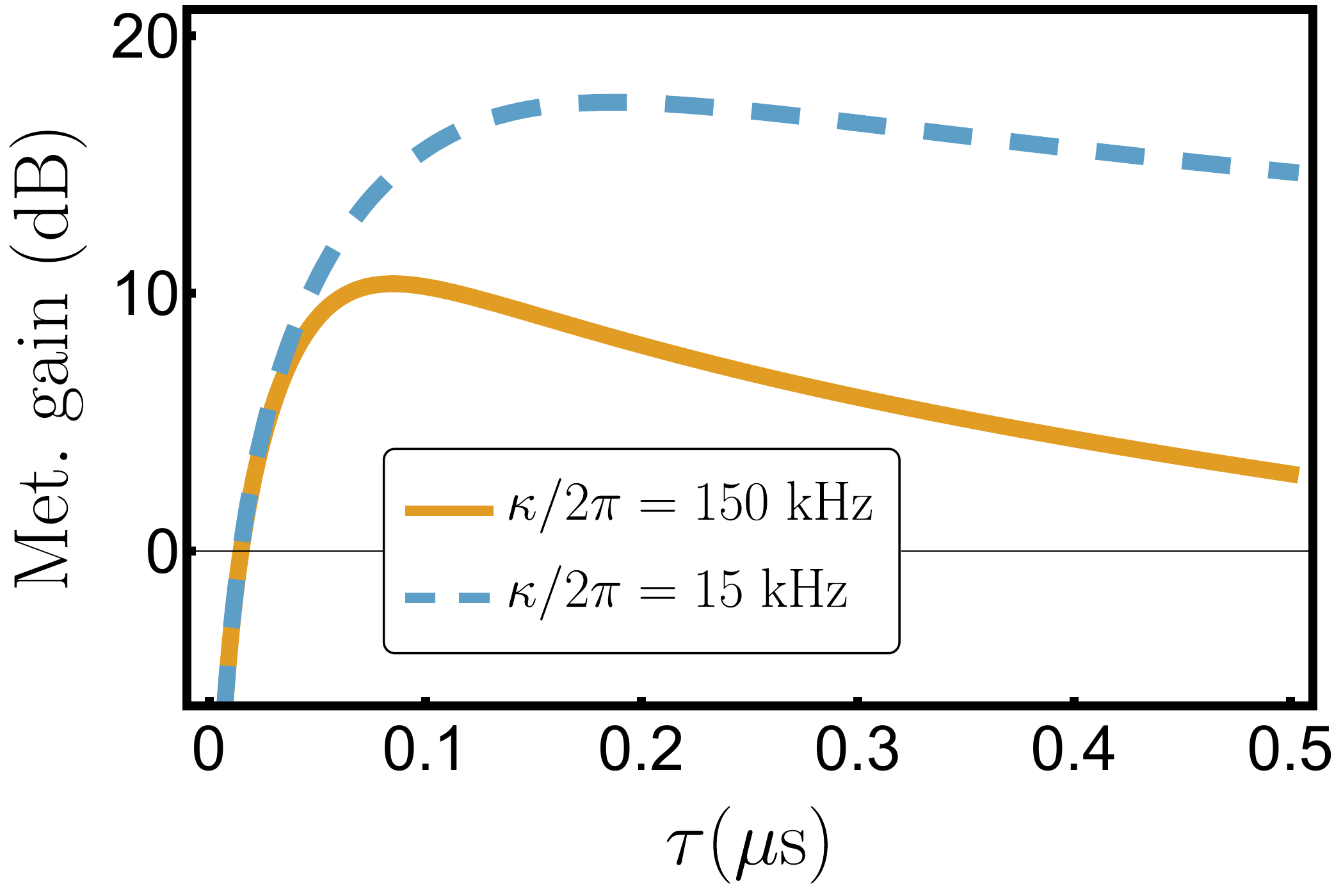}
    \caption{Metrological gain as a function of interaction time $\tau$ (in $\mu$s) when photon leakage from the cavity is accounted for $\chi\alpha=g=2\pi\times 11\text{ kHz}$ and $N=10^6$. 
    We compare two cavity decay rates: $\kappa/2\pi=15\text{ kHz}$ (dashed blue) and $\kappa/2\pi=150\text{ kHz}$ (solid orange).}
    \label{fig:KappaTimeSensitivity}
\end{figure}

From Eq.~(\ref{eqn:AppkappaFinalExpectationValue}) the relevant expectation values can be calculated and they are shown in Appendix~\ref{app:SensitivityKappa}, Eq.~(\ref{eqn:KappaExpectationValues}). They reduce to Eq.~(\ref{eqn:plen}) when $\kappa=0$. With these results we can calculate the sensitivity of a measurement of $\hat{S}_y$, in the limit $\chi \sqrt{N}\tau_1\ll 1$, $\chi \sqrt{N}\tau_2\ll 1$:
\begin{align}
    \begin{split}
    (\delta\beta)^2&=\Bigg(\frac{1+e^{\frac{4\chi^2\alpha^2f(\tau_1,\tau_2)}{\kappa^2}}}{8N}+\frac{1-e^{\frac{4\chi^2\alpha^2f(\tau_1,\tau_2)}{\kappa^2}}}{8}\Bigg)\\
    &\times \frac{e^{\kappa \tau_1}e^{-\frac{\chi^2\alpha^2f(\tau_1,\tau_2)}{\kappa^2}}}{\chi^2\alpha^2(e^{-\kappa \tau_2}-1)/\kappa^2},
\end{split}
\end{align}
where 

\begin{equation}
    f(\tau_1,\tau_2)=e^{-\kappa\tau_1}\big[2\kappa\tau_1+e^{-\kappa\tau_2}+\kappa(\tau_2-\tau_1)e^{-\kappa\tau_2}\big]-1.
\end{equation}
Cavity decay has introduced an $N$ independent summand to $(\delta\beta)^2$ that will ultimately limit the attainable sensitivity as $N$ is increased. Note also that the time development of $\delta\beta$ is parametrized by $N$ and $\chi\alpha/\kappa$. We plot the full sensitivity at $\beta=0$ and $\tau_1=\tau_2=\tau$ as a function of $\tau$ for realistic parameter values~\cite{Norcia_2018,Norcia2016} and using the resonant protocol: $N=10^6$, $\alpha=10^4$, $\chi\alpha=g=2\pi\times 11\text{ kHz}$ and $\kappa/2\pi=15,\, 150\text{ kHz}$ in Fig.~\ref{fig:KappaTimeSensitivity}. Note that, as in the ideal case, the sensitivity for the resonant protocol is $\alpha$ independent. 

\begin{figure}
    \centering
    \includegraphics[width=0.48\textwidth]{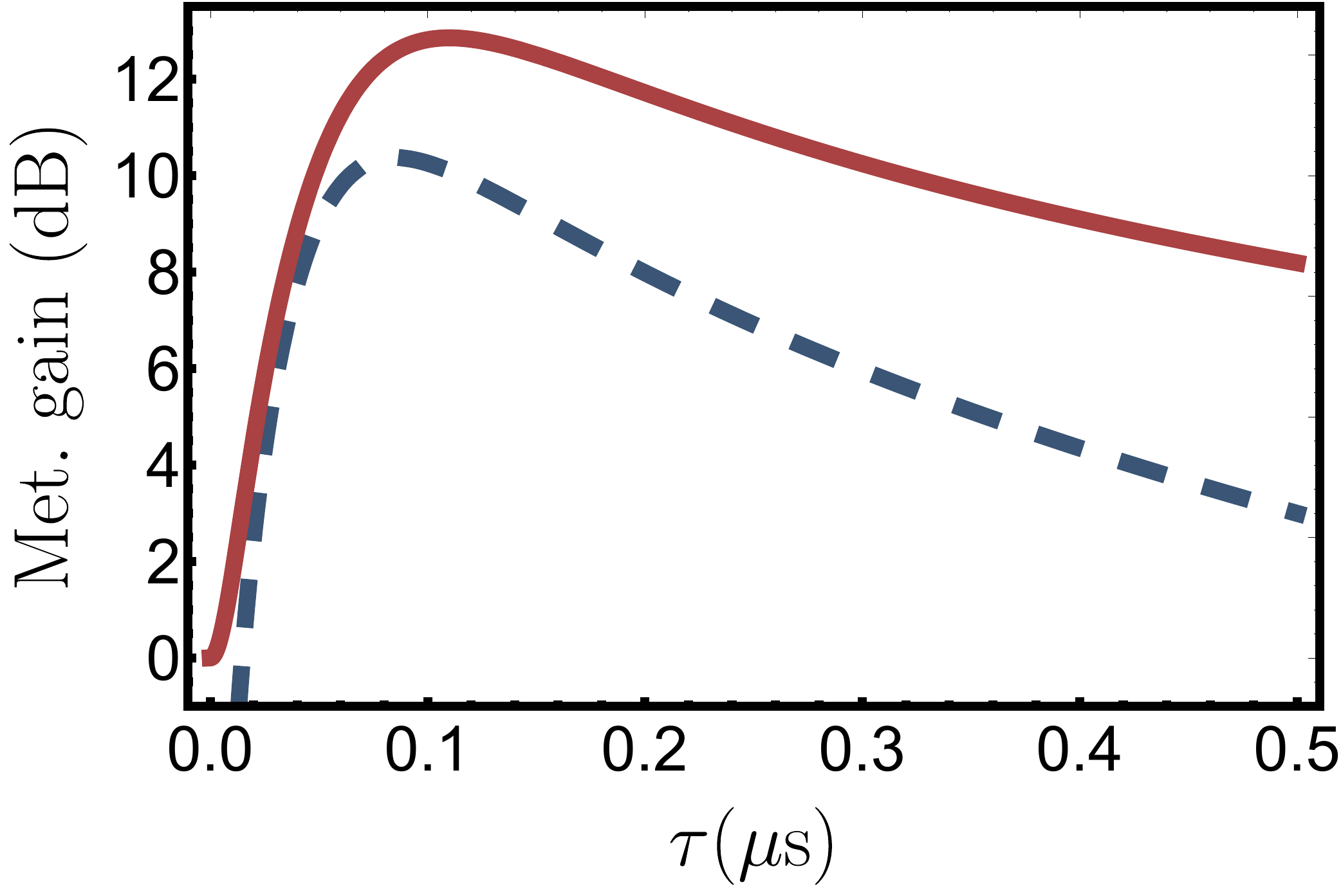}
    \caption{Comparison of metrological gain using the time-reversal protocol and $\hat{M} = \hat{S}_y$ (dashed dark blue) to that predicted from the Cr\`amer-Rao bound (solid red). Calculations are for $N=10^6$, $\chi\alpha=g=2\pi\times 11\text{ kHz}$ and $\kappa/2\pi=150\text{ kHz}$.}
    \label{fig:KappaFisher}
\end{figure}

We also compare $(\delta\beta)^2$ to the Cr\`amer-Rao bound in Fig.~\ref{fig:KappaFisher}. They attain a maximum at roughly the same time and differ by only a few dB.

\begin{figure}
    \centering
    \includegraphics[width=0.48\textwidth]{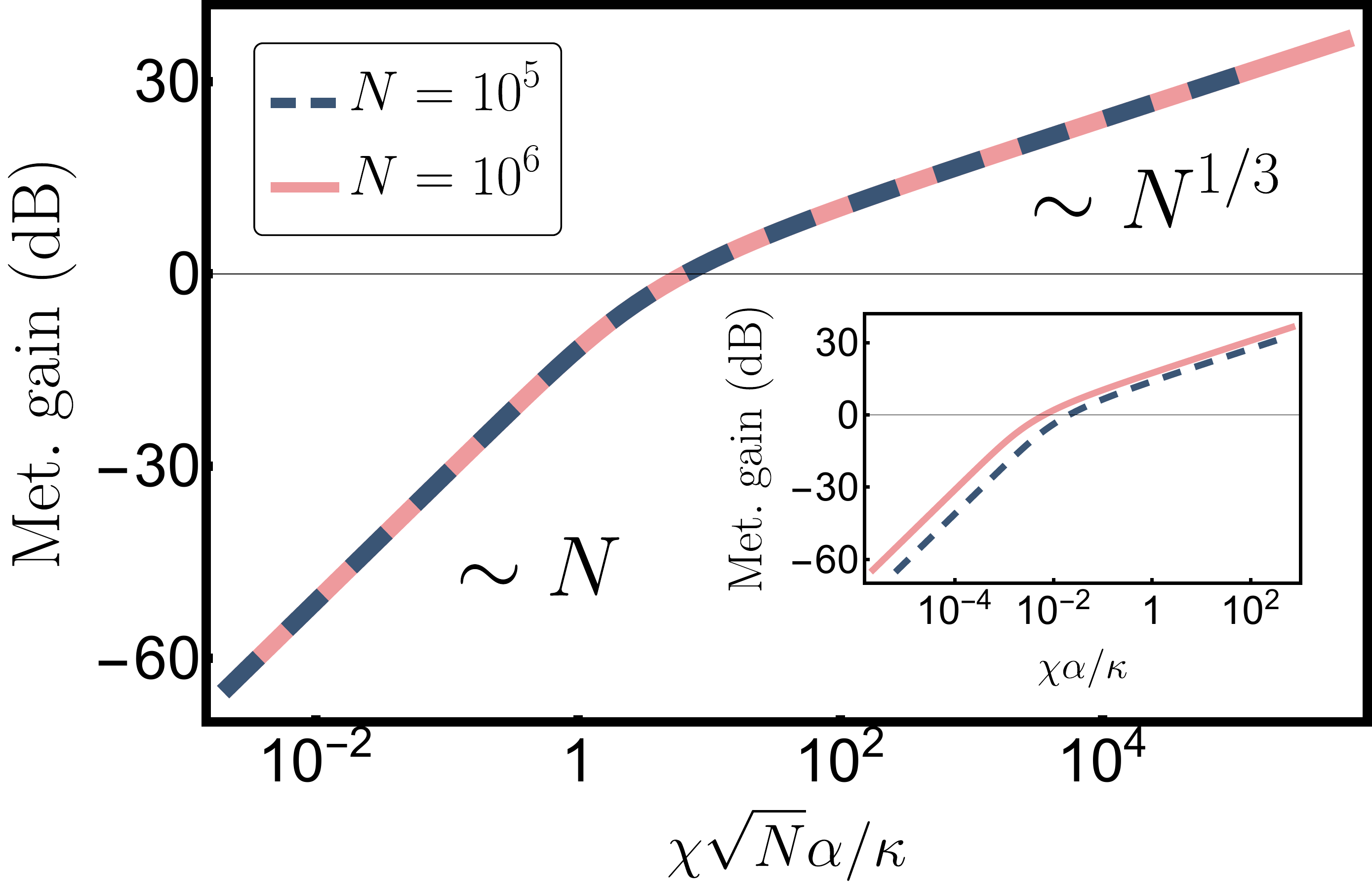}
    \caption{Optimal sensitivity as a ratio of the characteristic interaction scale and cavity decoherence rate, $\chi\sqrt{N}\alpha/\kappa$ for different N. Inset shows $(\delta\beta)^2_{\text{opt}}$ as a function of $\chi\alpha/\kappa$ emphasizing that, overall, larger $N$ is better. When $\chi\alpha\sqrt{N}/\kappa \gtrsim 1$, there is enhanced sensitivity which scales like $N^{-1/3}$. Conversely the protocol does not beat the SQL for $\chi\alpha\sqrt{N}/\kappa \lesssim 1$.}
    \label{fig:KappaOptimalSensitivity}
\end{figure}

Further restricting to $\kappa \tau\ll 1$ the idealized sensitivity of Eq.~(\ref{AppSensitivity}) is modified to
\begin{equation}
     ( \delta\beta)^2\approx \frac{1}{4N\alpha^2\chi^2\tau^2}+\frac{\kappa \tau}{6},
\end{equation}
which upon minimization in time gives
\begin{align}
    \begin{split}\label{eqn:KappaOptimalSensitivity}
        t_{\mathrm{opt}}&=\bigg(\frac{3}{\kappa \chi^2N\alpha^2}\bigg)^{1/3} , \\[5pt]
        (\delta\beta)^2_{\mathrm{opt}}&=\frac{1}{4}\bigg(\frac{3\kappa^2}{\chi^2N\alpha^2}\bigg)^{1/3}.
    \end{split}
\end{align}
Even though the scaling with $N$ is reduced, as compared to Eq.~(\ref{AppSensitivity}), increasing the number of atoms still results in an enhanced sensitivity. Furthermore, the figure of merit quantifying the optimal sensitivity is clearly $\chi\alpha\sqrt{N}/\kappa$, which in the case of the resonant protocol reduces to $g\sqrt{N}/\kappa$ i.e. the ratio between the collectively enhanced coupling and the cavity decay rate. This is further confirmed in Fig.~\ref{fig:KappaOptimalSensitivity}, where we plot $(\delta\beta)^2_{
\text{opt}}$ calculated from the full set of equations Eqs.~(\ref{eqn:KappaExpectationValues}) as a function of $\chi\alpha\sqrt{N}/\kappa$ for different values of $N$.

\begin{figure}
    \centering
    \includegraphics[width=0.48\textwidth]{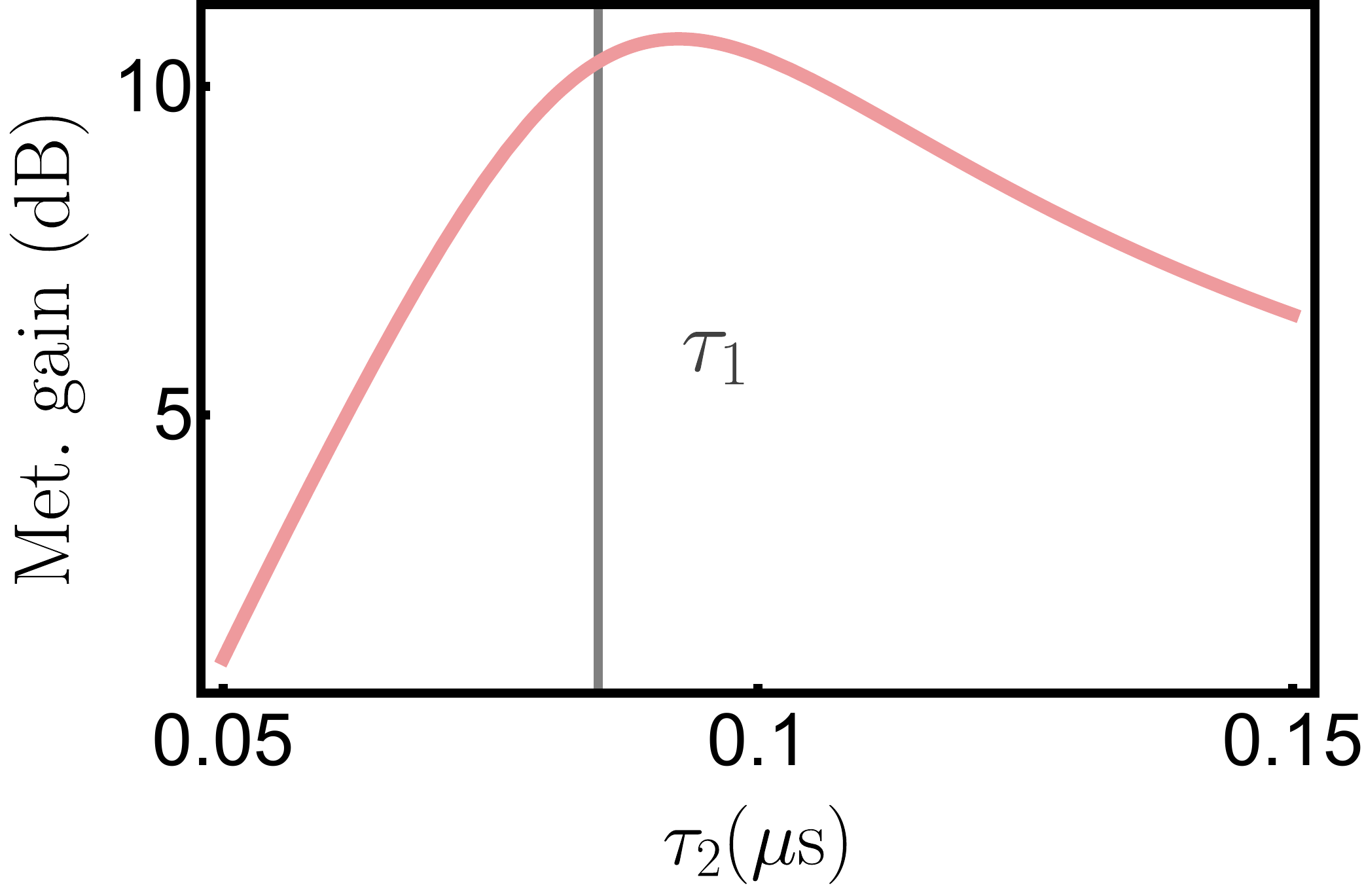}
    \caption{Sensitivity as a function of $\tau_2$ for $\chi\alpha=g=2\pi\times 11\text{ kHz}$, $N=10^6$ and $\kappa/2\pi=150\text{ kHz}$ and fix $\tau_1=85$~ns. Note that the best gain is obtained for $\tau_2$ slightly larger than $\tau_1$.} 
    \label{fig:KappaDifferentTimes}
\end{figure}

As discussed before, the effects of decoherence can be partly compensated by changing the forward ($\tau_1$) and backward ($\tau_2$) evolution times of the protocol. Indeed, as Fig.~\ref{fig:KappaDifferentTimes} shows, the optimal $\tau_2$ is longer than $\tau_1$. However, we found that optimizing over both $\tau_1$ and $\tau_2$ lead at most to a gain of 0.3 dB for the current cavity decay rate of $\kappa/(2\pi) = 150$~kHz. While the optimal result $\tau_2 \geq \tau_1$ can be intuitively understood as offsetting the decreased cavity occupation in the second period of atom-light interaction which generates the rotations of the collective spin, the optimisation of interaction-based readout protocols \cite{Linnemann_2017,Nolan_2017,Haine_2018,Mirkhalaf_2018} in the presence of significant dissipation, such as the case here, remains an interesting open question for future investigation. 


Finally, we compare quantitatively the optimal sensitivities attainable with the resonant and dispersive protocols, thus complementing the discussion at the end of Sec. \ref{sec:InteractionEngineering}. Setting $\chi_{\text{Dispersive}}=2g^2/\Delta_c$ and $\chi_{\text{Resonant}}=g/|\alpha|$, we get that
\begin{equation}
\frac{(\delta\beta)^2_{\text{D}}}{(\delta\beta)^2_{\text{R}}}=\bigg(\frac{\chi_{\text{Resonant}}}{\chi_{\text{Dispersive}}}\bigg)^{2/3}=\bigg(\frac{2g|\alpha|}{\Delta_c}\bigg)^{-2/3} .
\end{equation}
Given that for the dispersive Hamiltonian to work we need that $g|\alpha|/\Delta_c\ll 1$, we conclude that $(\delta\beta)_{\text{D}}\gg(\delta\beta)_{\text{R}}$, so that the resonant protocol will generically be better than the dispersive one.

\subsection{Spontaneous emission}\label{sec:SpontaneousEmission}
Another source of intrinsic decoherence is spontaneous emission of the atoms. Care must be taken when considering the effect of spontaneous emission, particularly in differentiating the dispersive and resonant protocols which generate a dispersive interaction in different (rotated) frames with respect to the spin degree of freedom. Due to this, we present a separate calculation and results for each protocol. Lastly, we note that in this case an analytic result for the Fisher information is not possible and so we focus on evaluating only the achievable sensitivity via the time-reversal protocol and collective measurements.

\subsubsection{Resonant protocol}
The nature of spontaneous emission on the resonant protocol is affected by the presence of a very strong single particle drive term along the $x$ direction. In principle,
the master equation describing the evolution of the atom-light density matrix $\hat{\rho}$ is given by:
\begin{equation}
    \dot{\hat{\rho}}=-ig\Big[\alpha\hat{S}_x+\frac{\hat{a}^{\dagger}\hat{a}}{\alpha}\hat{S}_x,\hat{\rho}\Big]+2\gamma\sum_i\Big(\hat{\sigma}_i^-\hat{\rho} \hat{\sigma}^+_i-\frac{\{\hat{\sigma}^+_i\hat{\sigma}^-_i,\hat{\rho}\}}{2}\Big).
\end{equation}
However, this is modified by the presence of a rotation at Rabi frequency $2g\alpha$, which we assume is fast compared to the spontaneous decay rate, $\gamma$. Upon moving to the rotating frame of the drive $\hat{H}_0 = 2g\alpha\hat{S}_x$, this assumption allows us to perform a RWA on the decay terms and end with the following effective master equation
\begin{align}\begin{split}
    \dot{\hat{\rho}}&=-ig\Big[\frac{\hat{a}^{\dagger}\hat{a}}{\alpha}\hat{S}_x,\hat{\rho}\Big]\\[5pt]
    &+\gamma\sum_i\Big(2\hat{s}_x^i\hat{\rho} \hat{s}_x^i+\hat{s}_z^i\hat{\rho} \hat{s}_z^i++\hat{s}_y^i\hat{\rho} \hat{s}_y^i-\hat{\rho}\Big)\\[5pt]
    &\equiv \mathcal{M}_{\chi}\hat{\rho},
\end{split}\end{align}
where we have set $\chi=g/\alpha$. 

To compute the sensitivity we work in the Heisenberg picture again and evolve $\hat{S}^+$, $(\hat{S}^+)^2$ and $\hat{S}^+\hat{S}^-$ using the Hilbert-Schmidt adjoint of $\mathcal{M}_{\chi}$. Note that in this case $\hat{S}^+$ is a raising operator with respect to the eigenstates of $\hat{S}_x$ since the dispersive interaction is oriented in this direction. As we will show later on, spontaneous emission is not the limiting factor so we will take the forward and backward evolution times to be the same and denote them by $\tau$. 
Both the calculations and the final result are involved, and so we show them in Appendix~\ref{app:SensitivityGamma} and Eq.~(\ref{eqn:AppResonantEV}), respectively.

\subsubsection{Dispersive protocol}
In the dispersive protocol there is no single particle drive, so the dissipative terms are unmodified. On the other hand, the presence of the term proportional to $\hat{S}^+\hat{S}^-$ in the Hamiltonian which we have previously neglected [see Eq.~(\ref{eqn:HamDisp})] must now be accounted for. It is important as it generates additional entanglement between the atoms, thus making the system more susceptible to the effect of spontaneous emission. Taking this into account, the dynamics of the atom-light system is now described by the master equation for the density matrix $\hat{\rho}$,
\begin{align}\begin{split}
    \dot{\hat{\rho}}&=-i\bigg[\frac{\chi}{2}\hat{S}^+\hat{S}^-+\chi\hat{a}^{\dagger}\hat{a}\hat{S}_z,\hat{\rho}\bigg]\\[5pt]
    &\hspace{2cm}+2\gamma\sum_i\Big(\hat{\sigma}_i^-\hat{\rho} \hat{\sigma}^+_i-\frac{\{\hat{\sigma}^+_i\hat{\sigma}^-_i,\hat{\rho}\}}{2}\Big).
\end{split}\end{align}
As in the resonant protocol, the derivation and final results for the relevant expectation values are very involved, so they are shown in Appendix~\ref{app:SensitivityGamma} and Eq.~(\ref{eqn:AppEDispersiveEV}), respectively.

\subsubsection{Timescales and sensitivity}
\begin{figure}
    \centering
    \includegraphics[width=0.48\textwidth]{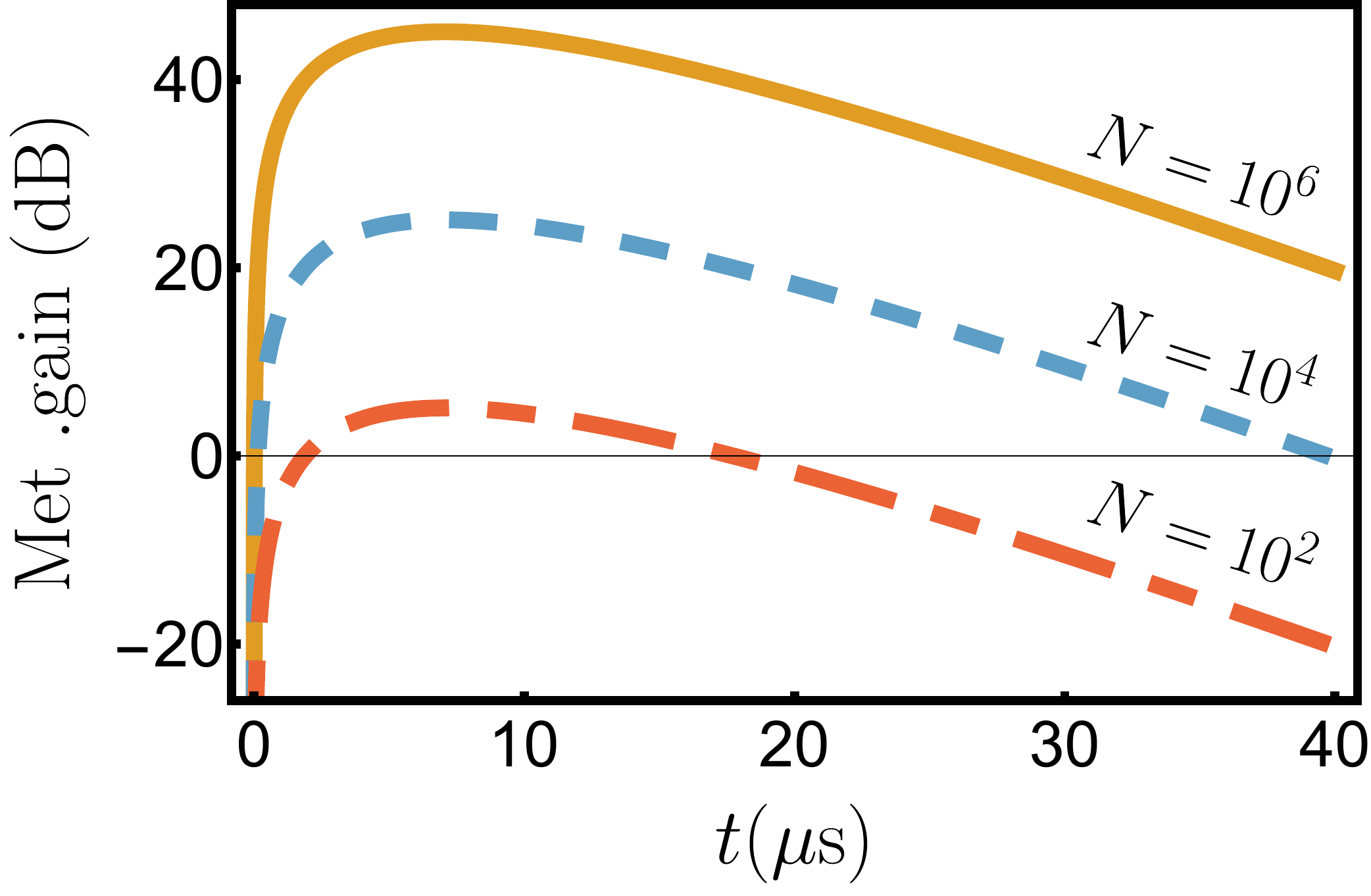}
    \caption{Sensitivity as a function of time using the resonant protocol for $\chi\alpha=g=2\pi\times 11\text{ kHz}$, $\gamma/2\pi=7.5\text{ kHz}$ and $N=10^2$ (lower, dot-dashed red), $10^4$ (middle, dashed blue) and $10^6$ (upper, solid orange). Note that the optimum occurs always at the same time.}
    \label{fig:GammaTimeSensitivity}
\end{figure}

In both the resonant and dispersive protocols we identify that there are two relevant timescales : $t \sim \gamma^{-1}$ which describes single particle decay effects and $t \sim (N\gamma\chi^2)^{-1/3}$ which characterizes entanglement dynamics. The latter arises in a manner analogous to the cavity system since the entangling evolution is creating spin cat states due to photon number fluctuations, with a susceptibility to decoherence similar to that of their bosonic counterparts.
Given that the resonant scheme generally leads to a better sensitivity, we focus in this case in what follows. Calculations for the dispersive case are very similar and give qualitatively similar results. In the resonant protocol, the entanglement timescale is made less relevant by using a large $\alpha$,  as $(N\gamma\chi^2)^{-1/3} \propto \alpha^{2/3}$. As we already require $\alpha \gg \sqrt{N}$ for the resonant protocol to be valid, we assume $\alpha$ can be increased sufficiently so that only single-particle decay is relevant. In this scenario, we have that, for short times
\begin{equation}
   (\delta\beta)^2\approx \frac{e^{6\gamma \tau}}{4\alpha^2 N\chi^2\tau^2} . \label{eqn:SensSpontEmit_Resonant}
\end{equation}
Minimizing Eq.~(\ref{eqn:SensSpontEmit_Resonant}) with respect to time indicates that the optimal sensitivity is determined by the ratio $\chi\alpha\sqrt{N}/\gamma$ and is attained at $3\gamma t_{\text{opt}}=1$, as shown in Fig.~\ref{fig:GammaTimeSensitivity}. For the experimental parameters discussed in Refs.  \cite{Norcia_2018,Norcia2016} ($g=2\pi\times 11\text{kHz}$, $\gamma=2\pi\times 7.5\text{kHz}$ and $N=10^6$) and in the case of the resonant protocol ($\chi=g/\alpha$), it follows that $\chi\alpha\sqrt{N}/\gamma=g\sqrt{N}/\gamma\approx 1500$. Looking at Fig.~(\ref{fig:GammaTimeSensitivity}) we conclude that spontaneous emission alone is not a limiting factor for the protocol.

To understand the importance of spontaneous emission relative to cavity leakage, we can compare the timescales to reach optimal sensitivity in both cases. Examining Eq.~(\ref{eqn:SensSpontEmit_Resonant}) the optimal squeezing is reached at $t \sim \gamma^{-1}$. This is to be compared with the optimal squezing time in the presence of photon loss, which is $t \sim (\kappa g^2 N)^{-1/3}$. Considering the same parameter regime from Refs.~\cite{Norcia2016,Norcia_2018} as previously, we have that $\gamma^{-1} \gg (\kappa g^2 N)^{-1/3}$ and thus we expect cavity decay to be far and away the dominant limitation of the protocol. To be more concrete, substituting the optimal time $t = (3\kappa g^2 N)^{-1/3}$ [Eq.~(\ref{eqn:KappaOptimalSensitivity})] into Eq.~(\ref{eqn:SensSpontEmit_Resonant}) we note spontaneous emission leads to a correction of $\approx 6\%$ to the sensitivity, which is negligible and justifies the detailed calculations we presented in Ref.~\cite{PRL}.

\subsection{Detection noise}\label{sec:DetectionNoise}
Prior work discussing the implementation of time-reversal and related interaction-based readout schemes have highlighted their utility in suppressing issues associated with detection noise \cite{Davis_2016,Nolan_2017,Haine_2018,Hosten_2016,Fabian_2018,Mirkhalaf_2018,Huang_2018}. In the case of time-reversal, this robustness can be associated with the fact that characterization of the metrological sensitivity only requires measurement of simple observables such as mean spin-projections \cite{Hosten_2016,Davis_2016}. On the other hand, work on the more general interaction-based readout schemes has demonstrated that the robustness to detection noise is preserved even when full distribution functions of observables are used \cite{Nolan_2017,Haine_2018}. 

In the absence of decoherence, our time-reversal protocol is robust to detection noise up to the level of the fundamental quantum noise. Specifically, we can model detection noise in an observable $\hat{M}$ as a Gaussian fluctuation of standard deviation $\sigma^{\hat{M}}_{\mathrm{det}}$ which does not contribute to the observed mean $\langle \hat{M} \rangle$ but does to the variance $\langle (\Delta\hat{M})^2\rangle \to \langle (\Delta\hat{M})^2\rangle + (\sigma^{\hat{M}}_{\mathrm{det}})^2$. For a measurement of $\hat{M} = \hat{S}_{\varphi} = \mathrm{cos}(\varphi)\hat{S}_x + \mathrm{sin}(\varphi)\hat{S}_y$ the achievable sensitivity then generalizes to:
\begin{equation}
    (\delta\beta)^2 \approx \frac{e^{N\chi^2\tau^2/4}}{4N\alpha^2\chi^2\tau^2} \left[ 1 + 4\mathrm{csc}^2(\varphi)\frac{\sigma^2_{\mathrm{det}}}{N} \right] . \label{eqn:SensDetIdeal}
\end{equation}
The optimal robustness occurs for $\varphi = \pi/2$ (i.e., $\hat{M} = \hat{S}_y$), for which detection noise $\sigma_{\mathrm{det}} \lesssim \sqrt{N}$ only leads to a numeric prefactor correction to the ideal sensitivity. We point out that the $\mathrm{csc}^2(\varphi)$ dependence implies this feature is not overly sensitive to the choice of $\varphi$. 

\begin{figure}
    \centering
    \includegraphics[width=0.48\textwidth]{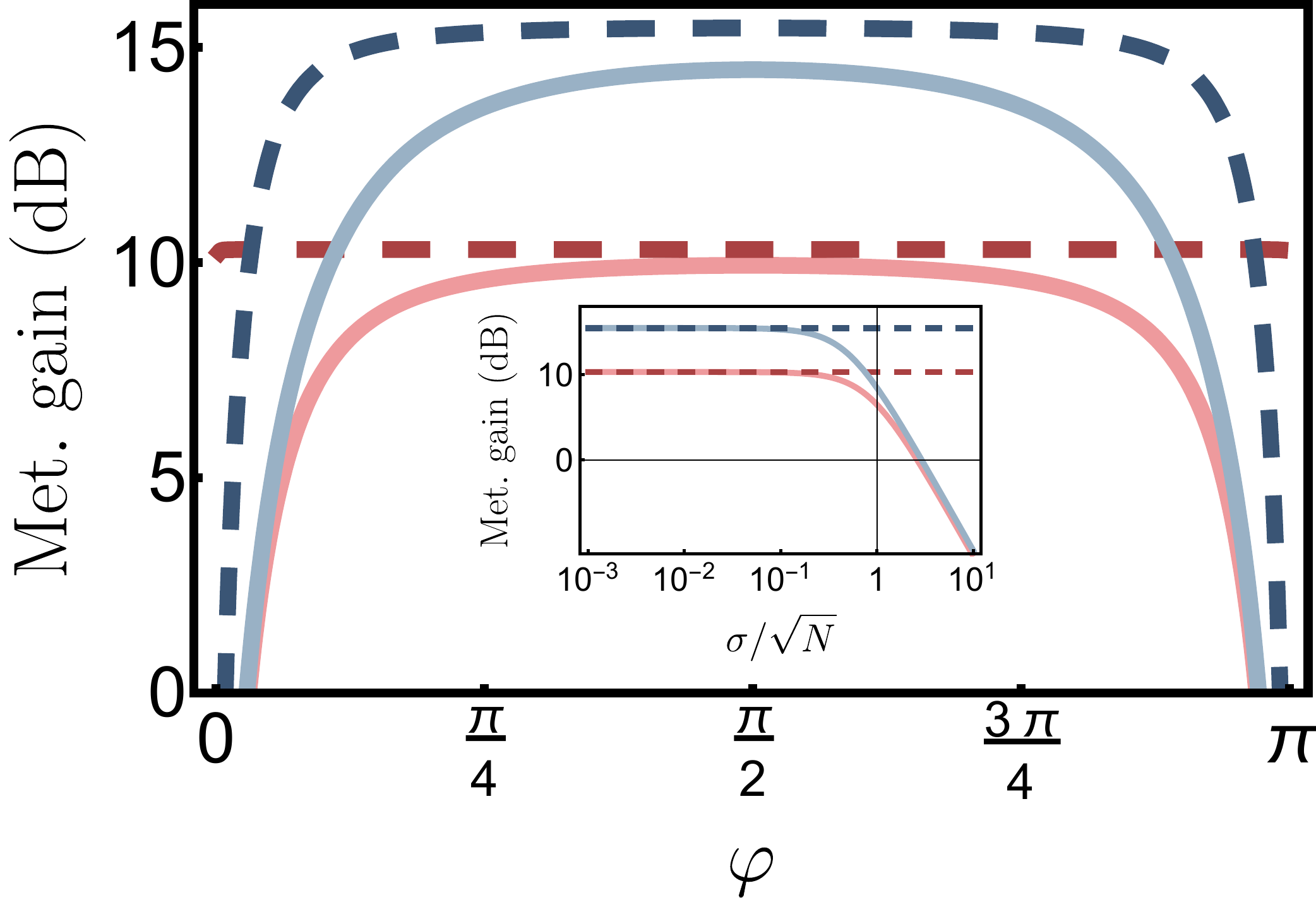}
    \caption{Robustness to detection noise as a function of measurement basis $\hat{M} = \hat{S}_{\varphi} = \mathrm{cos}(\varphi)\hat{S}_x + \mathrm{sin}(\varphi)\hat{S}_y$. Detection noise of  $\sigma_{\mathrm{det}} = \sqrt{N}/4$ is included for the case with spontaneous emission [$\gamma/(2\pi)=7.5$~kHz, upper solid line, in blue], and with photon loss [$\kappa/(2\pi) = 150$~kHz, lower solid line, in red], leading to an optimal sensitivity for $\varphi = \pi/2$. For comparison we also plot the relevant results for $\sigma_{\mathrm{det}} = 0$ (dashed lines). Inset shows scaling of sensitivity with $\sigma_{\mathrm{det}}$ for $\varphi = \pi/2$. Dashed lines in this case represent  the $\sigma_{\text{det}}=0$ result. All calculations are for $N = 10^6$, $t = 85~$ns and other parameters used in  Refs.  \cite{Norcia2016,Norcia_2018}.}
    \label{fig:DetectionNoise}
\end{figure}

The robustness is preserved when photon leakage and atomic spontaneous emission are included. Specifically, in the former case and for $\kappa \tau, \chi\sqrt{N}\tau \ll 1$: 
\begin{multline}
    (\delta\beta)^2 \approx \frac{1}{4\alpha^2 N \chi^2\tau^2}\left[ 1 + 4\mathrm{csc}^2(\varphi)\frac{\sigma^2_{\mathrm{det}}}{N} \right] \\
    + \frac{\kappa \tau}{6}\left[ \frac{N-1 + \mathrm{csc}^2(\varphi)}{N} + 4\mathrm{csc}^2(\varphi)\frac{\sigma^2_{\mathrm{det}}}{N} \right] , \label{eqn:SensDetKappa}
\end{multline}
while in the latter and for $\chi\sqrt{N}\tau \ll 1$: 
\begin{equation}
    (\delta\beta)^2\approx  \frac{1}{4N\chi^2\alpha^2\tau^2}\bigg[\frac{e^{6\gamma\tau}-\cos(\varphi)^2}{\sin(\varphi)^2}\bigg]+\frac{\sigma_{\mathrm{det}}^2\csc(\varphi)^2}{N^2\alpha^2\chi^2\tau^2} .
\end{equation}
We plot example results for the parameters of Refs.~\cite{Norcia2016,Norcia_2018} and $N= 10^6$ in Fig.~(\ref{fig:DetectionNoise}) using Eqs.~(\ref{eqn:SensDetIdeal}) and (\ref{eqn:SensDetKappa}) and as a function of $\varphi$. The inset shows the scaling of the metrological gain for fixed $\varphi = \pi/2$ with varying $\sigma_{\mathrm{det}}$, and confirms the protocol is robust to detection noise $\sigma_{\mathrm{det}} \lesssim \sqrt{N}$.

\section{Conclusions}
In this paper we have described a protocol for quantum enhanced sensing in an optical QED cavity which leverages the ability to work in a strong collective coupling limit due to the large atom number accessible in such systems. We demonstrated that a dispersive light-matter interaction can be engineered by either detuning the cavity, or operating it on resonance with the atomic transition and injecting a large coherent field, and used to generate metrologically useful entangled atom-light states. Our detailed analysis of intrinsic decoherence, particularly photon loss through the cavity mirrors and spontaneous emission of the atoms, predicts that entangled states of the cavity field can be generated for sensing of optical electromagnetic fields below the SQL by up to $10-20$~dB in realistic experimental conditions. 

Our protocol and results are not exclusive to the optical cavity platform, and could be readily implemented in a range of AMO platforms and frequency regimes. These include microwave cavities \cite{Delglise_2008}, circuit-QED \cite{Wallraff_2009,Viennot_2018,Fink2009}, trapped ion arrays \cite{Bollinger_2018} and other hybrid quantum systems systems \cite{Kolkowitz_2012,Aspelmeyer_2014}, particularly in the context of sensing weak forces or small mechanical displacements \cite{Gilmore2017,Burd2019}.  

\begin{acknowledgements}
We acknowledge helpful discussions with Kevin Gilmore and Michael Perlin during the preparation of this manuscript. This work is supported by the AFOSR grant FA9550-18-1-0319, by the DARPA and ARO grant W911NF-16-1-0576, the ARO single investigator award W911NF-19-1-0210,  the NSF PHY1820885, NSF JILA-PFC PHY-1734006 grants, and by NIST.
\end{acknowledgements}
\bibliography{library}

\begin{thebibliography}{65}%
\makeatletter
\providecommand \@ifxundefined [1]{%
 \@ifx{#1\undefined}
}%
\providecommand \@ifnum [1]{%
 \ifnum #1\expandafter \@firstoftwo
 \else \expandafter \@secondoftwo
 \fi
}%
\providecommand \@ifx [1]{%
 \ifx #1\expandafter \@firstoftwo
 \else \expandafter \@secondoftwo
 \fi
}%
\providecommand \natexlab [1]{#1}%
\providecommand \enquote  [1]{``#1''}%
\providecommand \bibnamefont  [1]{#1}%
\providecommand \bibfnamefont [1]{#1}%
\providecommand \citenamefont [1]{#1}%
\providecommand \href@noop [0]{\@secondoftwo}%
\providecommand \href [0]{\begingroup \@sanitize@url \@href}%
\providecommand \@href[1]{\@@startlink{#1}\@@href}%
\providecommand \@@href[1]{\endgroup#1\@@endlink}%
\providecommand \@sanitize@url [0]{\catcode `\\12\catcode `\$12\catcode
  `\&12\catcode `\#12\catcode `\^12\catcode `\_12\catcode `\%12\relax}%
\providecommand \@@startlink[1]{}%
\providecommand \@@endlink[0]{}%
\providecommand \url  [0]{\begingroup\@sanitize@url \@url }%
\providecommand \@url [1]{\endgroup\@href {#1}{\urlprefix }}%
\providecommand \urlprefix  [0]{URL }%
\providecommand \Eprint [0]{\href }%
\providecommand \doibase [0]{http://dx.doi.org/}%
\providecommand \selectlanguage [0]{\@gobble}%
\providecommand \bibinfo  [0]{\@secondoftwo}%
\providecommand \bibfield  [0]{\@secondoftwo}%
\providecommand \translation [1]{[#1]}%
\providecommand \BibitemOpen [0]{}%
\providecommand \bibitemStop [0]{}%
\providecommand \bibitemNoStop [0]{.\EOS\space}%
\providecommand \EOS [0]{\spacefactor3000\relax}%
\providecommand \BibitemShut  [1]{\csname bibitem#1\endcsname}%
\let\auto@bib@innerbib\@empty
\bibitem [{\citenamefont {Lewis-Swan}\ \emph {et~al.}(2020)\citenamefont
  {Lewis-Swan}, \citenamefont {Barberena}, \citenamefont {Muniz}, \citenamefont
  {Cline}, \citenamefont {Young}, \citenamefont {Thompson},\ and\ \citenamefont
  {Rey}}]{PRL}%
  \BibitemOpen
  \bibfield  {author} {\bibinfo {author} {\bibfnamefont {R.~J.}\ \bibnamefont
  {Lewis-Swan}}, \bibinfo {author} {\bibfnamefont {D.}~\bibnamefont
  {Barberena}}, \bibinfo {author} {\bibfnamefont {J.~A.}\ \bibnamefont
  {Muniz}}, \bibinfo {author} {\bibfnamefont {J.~R.~K.}\ \bibnamefont {Cline}},
  \bibinfo {author} {\bibfnamefont {D.}~\bibnamefont {Young}}, \bibinfo
  {author} {\bibfnamefont {J.~K.}\ \bibnamefont {Thompson}}, \ and\ \bibinfo
  {author} {\bibfnamefont {A.~M.}\ \bibnamefont {Rey}},\ }\href {\doibase
  10.1103/PhysRevLett.124.193602} {\bibfield  {journal} {\bibinfo  {journal}
  {Phys. Rev. Lett.}\ }\textbf {\bibinfo {volume} {124}},\ \bibinfo {pages}
  {193602} (\bibinfo {year} {2020})}\BibitemShut {NoStop}%
\bibitem [{\citenamefont {Walls}\ and\ \citenamefont
  {Milburn}(2008)}]{walls_quantum_2008}%
  \BibitemOpen
  \bibfield  {author} {\bibinfo {author} {\bibfnamefont {D.~F.}\ \bibnamefont
  {Walls}}\ and\ \bibinfo {author} {\bibfnamefont {G.~J.}\ \bibnamefont
  {Milburn}},\ }\href@noop {} {\emph {\bibinfo {title} {{Quantum Optics}}}},\
  \bibinfo {edition} {2nd}\ ed.\ (\bibinfo  {publisher} {Springer},\ \bibinfo
  {year} {2008})\BibitemShut {NoStop}%
\bibitem [{\citenamefont {Caves}(1981)}]{Caves1981}%
  \BibitemOpen
  \bibfield  {author} {\bibinfo {author} {\bibfnamefont {C.~M.}\ \bibnamefont
  {Caves}},\ }\href {\doibase 10.1103/PhysRevD.23.1693} {\bibfield  {journal}
  {\bibinfo  {journal} {Phys. Rev. D}\ }\textbf {\bibinfo {volume} {23}},\
  \bibinfo {pages} {1693} (\bibinfo {year} {1981})}\BibitemShut {NoStop}%
\bibitem [{\citenamefont {{Abbott, B. P., \emph{et.
  al.}}}(2016)}]{AdvancedLigo2016}%
  \BibitemOpen
  \bibfield  {author} {\bibinfo {author} {\bibnamefont {{Abbott, B. P.,
  \emph{et. al.}}}} (\bibinfo {collaboration} {LIGO Scientific Collaboration
  and Virgo Collaboration}),\ }\href {\doibase 10.1103/PhysRevLett.116.131103}
  {\bibfield  {journal} {\bibinfo  {journal} {Phys. Rev. Lett.}\ }\textbf
  {\bibinfo {volume} {116}},\ \bibinfo {pages} {131103} (\bibinfo {year}
  {2016})}\BibitemShut {NoStop}%
\bibitem [{\citenamefont {Aasi}(2013)}]{Aasi2013}%
  \BibitemOpen
  \bibfield  {author} {\bibinfo {author} {\bibfnamefont {J.~{\it et al }.}\
  \bibnamefont {Aasi}},\ }\href@noop {} {\bibfield  {journal} {\bibinfo
  {journal} {Nature Photonics}\ }\textbf {\bibinfo {volume} {7}},\ \bibinfo
  {pages} {613} (\bibinfo {year} {2013})}\BibitemShut {NoStop}%
\bibitem [{\citenamefont {Malnou}\ \emph {et~al.}(2019)\citenamefont {Malnou},
  \citenamefont {Palken}, \citenamefont {Brubaker}, \citenamefont {Vale},
  \citenamefont {Hilton},\ and\ \citenamefont {Lehnert}}]{Malnou2019}%
  \BibitemOpen
  \bibfield  {author} {\bibinfo {author} {\bibfnamefont {M.}~\bibnamefont
  {Malnou}}, \bibinfo {author} {\bibfnamefont {D.~A.}\ \bibnamefont {Palken}},
  \bibinfo {author} {\bibfnamefont {B.~M.}\ \bibnamefont {Brubaker}}, \bibinfo
  {author} {\bibfnamefont {L.~R.}\ \bibnamefont {Vale}}, \bibinfo {author}
  {\bibfnamefont {G.~C.}\ \bibnamefont {Hilton}}, \ and\ \bibinfo {author}
  {\bibfnamefont {K.~W.}\ \bibnamefont {Lehnert}},\ }\href {\doibase
  10.1103/PhysRevX.9.021023} {\bibfield  {journal} {\bibinfo  {journal} {Phys.
  Rev. X}\ }\textbf {\bibinfo {volume} {9}},\ \bibinfo {pages} {021023}
  (\bibinfo {year} {2019})}\BibitemShut {NoStop}%
\bibitem [{\citenamefont {Burd}\ \emph {et~al.}(2019)\citenamefont {Burd},
  \citenamefont {Srinivas}, \citenamefont {Bollinger}, \citenamefont {Wilson},
  \citenamefont {Wineland}, \citenamefont {Leibfried}, \citenamefont
  {Slichter},\ and\ \citenamefont {Allcock}}]{Burd2019}%
  \BibitemOpen
  \bibfield  {author} {\bibinfo {author} {\bibfnamefont {S.~C.}\ \bibnamefont
  {Burd}}, \bibinfo {author} {\bibfnamefont {R.}~\bibnamefont {Srinivas}},
  \bibinfo {author} {\bibfnamefont {J.~J.}\ \bibnamefont {Bollinger}}, \bibinfo
  {author} {\bibfnamefont {A.~C.}\ \bibnamefont {Wilson}}, \bibinfo {author}
  {\bibfnamefont {D.~J.}\ \bibnamefont {Wineland}}, \bibinfo {author}
  {\bibfnamefont {D.}~\bibnamefont {Leibfried}}, \bibinfo {author}
  {\bibfnamefont {D.~H.}\ \bibnamefont {Slichter}}, \ and\ \bibinfo {author}
  {\bibfnamefont {D.~T.~C.}\ \bibnamefont {Allcock}},\ }\href {\doibase
  10.1126/science.aaw2884} {\bibfield  {journal} {\bibinfo  {journal}
  {Science}\ }\textbf {\bibinfo {volume} {364}},\ \bibinfo {pages} {1163}
  (\bibinfo {year} {2019})}\BibitemShut {NoStop}%
\bibitem [{\citenamefont {Penasa}\ \emph {et~al.}(2016)\citenamefont {Penasa},
  \citenamefont {Gerlich}, \citenamefont {Rybarczyk}, \citenamefont
  {M\'etillon}, \citenamefont {Brune}, \citenamefont {Raimond}, \citenamefont
  {Haroche}, \citenamefont {Davidovich},\ and\ \citenamefont
  {Dotsenko}}]{Penasa_2016}%
  \BibitemOpen
  \bibfield  {author} {\bibinfo {author} {\bibfnamefont {M.}~\bibnamefont
  {Penasa}}, \bibinfo {author} {\bibfnamefont {S.}~\bibnamefont {Gerlich}},
  \bibinfo {author} {\bibfnamefont {T.}~\bibnamefont {Rybarczyk}}, \bibinfo
  {author} {\bibfnamefont {V.}~\bibnamefont {M\'etillon}}, \bibinfo {author}
  {\bibfnamefont {M.}~\bibnamefont {Brune}}, \bibinfo {author} {\bibfnamefont
  {J.~M.}\ \bibnamefont {Raimond}}, \bibinfo {author} {\bibfnamefont
  {S.}~\bibnamefont {Haroche}}, \bibinfo {author} {\bibfnamefont
  {L.}~\bibnamefont {Davidovich}}, \ and\ \bibinfo {author} {\bibfnamefont
  {I.}~\bibnamefont {Dotsenko}},\ }\href {\doibase 10.1103/PhysRevA.94.022313}
  {\bibfield  {journal} {\bibinfo  {journal} {Phys. Rev. A}\ }\textbf {\bibinfo
  {volume} {94}},\ \bibinfo {pages} {022313} (\bibinfo {year}
  {2016})}\BibitemShut {NoStop}%
\bibitem [{\citenamefont {Vlastakis}\ \emph {et~al.}(2013)\citenamefont
  {Vlastakis}, \citenamefont {Kirchmair}, \citenamefont {Leghtas},
  \citenamefont {Nigg}, \citenamefont {Frunzio}, \citenamefont {Girvin},
  \citenamefont {Mirrahimi}, \citenamefont {Devoret},\ and\ \citenamefont
  {Schoelkopf}}]{Vlastakis2013}%
  \BibitemOpen
  \bibfield  {author} {\bibinfo {author} {\bibfnamefont {B.}~\bibnamefont
  {Vlastakis}}, \bibinfo {author} {\bibfnamefont {G.}~\bibnamefont
  {Kirchmair}}, \bibinfo {author} {\bibfnamefont {Z.}~\bibnamefont {Leghtas}},
  \bibinfo {author} {\bibfnamefont {S.~E.}\ \bibnamefont {Nigg}}, \bibinfo
  {author} {\bibfnamefont {L.}~\bibnamefont {Frunzio}}, \bibinfo {author}
  {\bibfnamefont {S.~M.}\ \bibnamefont {Girvin}}, \bibinfo {author}
  {\bibfnamefont {M.}~\bibnamefont {Mirrahimi}}, \bibinfo {author}
  {\bibfnamefont {M.~H.}\ \bibnamefont {Devoret}}, \ and\ \bibinfo {author}
  {\bibfnamefont {R.~J.}\ \bibnamefont {Schoelkopf}},\ }\href {\doibase
  10.1126/science.1243289} {\bibfield  {journal} {\bibinfo  {journal}
  {Science}\ }\textbf {\bibinfo {volume} {342}},\ \bibinfo {pages} {607}
  (\bibinfo {year} {2013})}\BibitemShut {NoStop}%
\bibitem [{\citenamefont {McCormick}\ \emph {et~al.}(2019)\citenamefont
  {McCormick}, \citenamefont {Keller}, \citenamefont {Burd}, \citenamefont
  {Wineland}, \citenamefont {Wilson},\ and\ \citenamefont
  {Leibfried}}]{Didi_2018}%
  \BibitemOpen
  \bibfield  {author} {\bibinfo {author} {\bibfnamefont {K.~C.}\ \bibnamefont
  {McCormick}}, \bibinfo {author} {\bibfnamefont {J.}~\bibnamefont {Keller}},
  \bibinfo {author} {\bibfnamefont {S.~C.}\ \bibnamefont {Burd}}, \bibinfo
  {author} {\bibfnamefont {D.~J.}\ \bibnamefont {Wineland}}, \bibinfo {author}
  {\bibfnamefont {A.~C.}\ \bibnamefont {Wilson}}, \ and\ \bibinfo {author}
  {\bibfnamefont {D.}~\bibnamefont {Leibfried}},\ }\href {\doibase
  10.1038/s41586-019-1421-y} {\bibfield  {journal} {\bibinfo  {journal}
  {Nature}\ }\textbf {\bibinfo {volume} {572}},\ \bibinfo {pages} {86}
  (\bibinfo {year} {2019})}\BibitemShut {NoStop}%
\bibitem [{\citenamefont {Huelga}\ \emph {et~al.}(1997)\citenamefont {Huelga},
  \citenamefont {Macchiavello}, \citenamefont {Pellizzari}, \citenamefont
  {Ekert}, \citenamefont {Plenio},\ and\ \citenamefont {Cirac}}]{Huelga1997}%
  \BibitemOpen
  \bibfield  {author} {\bibinfo {author} {\bibfnamefont {S.~F.}\ \bibnamefont
  {Huelga}}, \bibinfo {author} {\bibfnamefont {C.}~\bibnamefont
  {Macchiavello}}, \bibinfo {author} {\bibfnamefont {T.}~\bibnamefont
  {Pellizzari}}, \bibinfo {author} {\bibfnamefont {A.~K.}\ \bibnamefont
  {Ekert}}, \bibinfo {author} {\bibfnamefont {M.~B.}\ \bibnamefont {Plenio}}, \
  and\ \bibinfo {author} {\bibfnamefont {J.~I.}\ \bibnamefont {Cirac}},\ }\href
  {\doibase 10.1103/PhysRevLett.79.3865} {\bibfield  {journal} {\bibinfo
  {journal} {Phys. Rev. Lett.}\ }\textbf {\bibinfo {volume} {79}},\ \bibinfo
  {pages} {3865} (\bibinfo {year} {1997})}\BibitemShut {NoStop}%
\bibitem [{\citenamefont {Schuster}\ \emph {et~al.}(2007)\citenamefont
  {Schuster}, \citenamefont {Houck}, \citenamefont {Schreier}, \citenamefont
  {Wallraff}, \citenamefont {Gambetta}, \citenamefont {Blais}, \citenamefont
  {Frunzio}, \citenamefont {Majer}, \citenamefont {Johnson}, \citenamefont
  {Devoret}, \citenamefont {Girvin},\ and\ \citenamefont
  {Schoelkopf}}]{Schuster_2007}%
  \BibitemOpen
  \bibfield  {author} {\bibinfo {author} {\bibfnamefont {D.~I.}\ \bibnamefont
  {Schuster}}, \bibinfo {author} {\bibfnamefont {A.~A.}\ \bibnamefont {Houck}},
  \bibinfo {author} {\bibfnamefont {J.~A.}\ \bibnamefont {Schreier}}, \bibinfo
  {author} {\bibfnamefont {A.}~\bibnamefont {Wallraff}}, \bibinfo {author}
  {\bibfnamefont {J.~M.}\ \bibnamefont {Gambetta}}, \bibinfo {author}
  {\bibfnamefont {A.}~\bibnamefont {Blais}}, \bibinfo {author} {\bibfnamefont
  {L.}~\bibnamefont {Frunzio}}, \bibinfo {author} {\bibfnamefont
  {J.}~\bibnamefont {Majer}}, \bibinfo {author} {\bibfnamefont
  {B.}~\bibnamefont {Johnson}}, \bibinfo {author} {\bibfnamefont {M.~H.}\
  \bibnamefont {Devoret}}, \bibinfo {author} {\bibfnamefont {S.~M.}\
  \bibnamefont {Girvin}}, \ and\ \bibinfo {author} {\bibfnamefont {R.~J.}\
  \bibnamefont {Schoelkopf}},\ }\href {\doibase 10.1038/nature05461} {\bibfield
   {journal} {\bibinfo  {journal} {Nature}\ }\textbf {\bibinfo {volume}
  {445}},\ \bibinfo {pages} {515} (\bibinfo {year} {2007})}\BibitemShut
  {NoStop}%
\bibitem [{\citenamefont {Girvin}(2014)}]{Girvin_2014}%
  \BibitemOpen
  \bibfield  {author} {\bibinfo {author} {\bibfnamefont {S.~M.}\ \bibnamefont
  {Girvin}},\ }in\ \href {\doibase 10.1093/acprof:oso/9780199681181.003.0003}
  {\emph {\bibinfo {booktitle} {Quantum Machines: Measurement and Control of
  Engineered Quantum Systems}}}\ (\bibinfo  {publisher} {Oxford University
  Press},\ \bibinfo {year} {2014})\ pp.\ \bibinfo {pages}
  {113--256}\BibitemShut {NoStop}%
\bibitem [{\citenamefont {Hacker}\ \emph {et~al.}(2019)\citenamefont {Hacker},
  \citenamefont {Welte}, \citenamefont {Daiss}, \citenamefont {Shaukat},
  \citenamefont {Ritter}, \citenamefont {Li},\ and\ \citenamefont
  {Rempe}}]{Hacker_2019}%
  \BibitemOpen
  \bibfield  {author} {\bibinfo {author} {\bibfnamefont {B.}~\bibnamefont
  {Hacker}}, \bibinfo {author} {\bibfnamefont {S.}~\bibnamefont {Welte}},
  \bibinfo {author} {\bibfnamefont {S.}~\bibnamefont {Daiss}}, \bibinfo
  {author} {\bibfnamefont {A.}~\bibnamefont {Shaukat}}, \bibinfo {author}
  {\bibfnamefont {S.}~\bibnamefont {Ritter}}, \bibinfo {author} {\bibfnamefont
  {L.}~\bibnamefont {Li}}, \ and\ \bibinfo {author} {\bibfnamefont
  {G.}~\bibnamefont {Rempe}},\ }\href {\doibase 10.1038/s41566-018-0339-5}
  {\bibfield  {journal} {\bibinfo  {journal} {Nature Photonics}\ }\textbf
  {\bibinfo {volume} {13}},\ \bibinfo {pages} {110} (\bibinfo {year}
  {2019})}\BibitemShut {NoStop}%
\bibitem [{\citenamefont {Leroux}\ \emph {et~al.}(2010)\citenamefont {Leroux},
  \citenamefont {Schleier-Smith},\ and\ \citenamefont
  {Vuleti{\'{c}}}}]{Leroux2010c}%
  \BibitemOpen
  \bibfield  {author} {\bibinfo {author} {\bibfnamefont {I.~D.}\ \bibnamefont
  {Leroux}}, \bibinfo {author} {\bibfnamefont {M.~H.}\ \bibnamefont
  {Schleier-Smith}}, \ and\ \bibinfo {author} {\bibfnamefont {V.}~\bibnamefont
  {Vuleti{\'{c}}}},\ }\href {\doibase 10.1103/PhysRevLett.104.073602}
  {\bibfield  {journal} {\bibinfo  {journal} {Phys. Rev. Lett.}\ }\textbf
  {\bibinfo {volume} {104}},\ \bibinfo {pages} {073602} (\bibinfo {year}
  {2010})}\BibitemShut {NoStop}%
\bibitem [{\citenamefont {Schleier-Smith}\ \emph {et~al.}(2010)\citenamefont
  {Schleier-Smith}, \citenamefont {Leroux},\ and\ \citenamefont
  {Vuleti{\'{c}}}}]{Schleier-Smith2010b}%
  \BibitemOpen
  \bibfield  {author} {\bibinfo {author} {\bibfnamefont {M.~H.}\ \bibnamefont
  {Schleier-Smith}}, \bibinfo {author} {\bibfnamefont {I.~D.}\ \bibnamefont
  {Leroux}}, \ and\ \bibinfo {author} {\bibfnamefont {V.}~\bibnamefont
  {Vuleti{\'{c}}}},\ }\href {\doibase 10.1103/PhysRevLett.104.073604}
  {\bibfield  {journal} {\bibinfo  {journal} {Phys. Rev. Lett.}\ }\textbf
  {\bibinfo {volume} {104}},\ \bibinfo {pages} {1} (\bibinfo {year}
  {2010})}\BibitemShut {NoStop}%
\bibitem [{\citenamefont {Hosten}\ \emph
  {et~al.}(2016{\natexlab{a}})\citenamefont {Hosten}, \citenamefont
  {Krishnakumar}, \citenamefont {Engelsen},\ and\ \citenamefont
  {Kasevich}}]{Hosten_2016}%
  \BibitemOpen
  \bibfield  {author} {\bibinfo {author} {\bibfnamefont {O.}~\bibnamefont
  {Hosten}}, \bibinfo {author} {\bibfnamefont {R.}~\bibnamefont
  {Krishnakumar}}, \bibinfo {author} {\bibfnamefont {N.~J.}\ \bibnamefont
  {Engelsen}}, \ and\ \bibinfo {author} {\bibfnamefont {M.~A.}\ \bibnamefont
  {Kasevich}},\ }\href {\doibase 10.1126/science.aaf3397} {\bibfield  {journal}
  {\bibinfo  {journal} {Science}\ }\textbf {\bibinfo {volume} {352}},\ \bibinfo
  {pages} {1552} (\bibinfo {year} {2016}{\natexlab{a}})}\BibitemShut {NoStop}%
\bibitem [{\citenamefont {Hosten}\ \emph
  {et~al.}(2016{\natexlab{b}})\citenamefont {Hosten}, \citenamefont {Engelsen},
  \citenamefont {Krishnakumar},\ and\ \citenamefont {Kasevich}}]{Hosten_2016b}%
  \BibitemOpen
  \bibfield  {author} {\bibinfo {author} {\bibfnamefont {O.}~\bibnamefont
  {Hosten}}, \bibinfo {author} {\bibfnamefont {N.~J.}\ \bibnamefont
  {Engelsen}}, \bibinfo {author} {\bibfnamefont {R.}~\bibnamefont
  {Krishnakumar}}, \ and\ \bibinfo {author} {\bibfnamefont {M.~A.}\
  \bibnamefont {Kasevich}},\ }\href {\doibase 10.1038/nature16176} {\bibfield
  {journal} {\bibinfo  {journal} {Nature}\ }\textbf {\bibinfo {volume} {529}},\
  \bibinfo {pages} {505} (\bibinfo {year} {2016}{\natexlab{b}})}\BibitemShut
  {NoStop}%
\bibitem [{\citenamefont {Cox}\ \emph {et~al.}(2016)\citenamefont {Cox},
  \citenamefont {Greve}, \citenamefont {Weiner},\ and\ \citenamefont
  {Thompson}}]{Cox_2016}%
  \BibitemOpen
  \bibfield  {author} {\bibinfo {author} {\bibfnamefont {K.~C.}\ \bibnamefont
  {Cox}}, \bibinfo {author} {\bibfnamefont {G.~P.}\ \bibnamefont {Greve}},
  \bibinfo {author} {\bibfnamefont {J.~M.}\ \bibnamefont {Weiner}}, \ and\
  \bibinfo {author} {\bibfnamefont {J.~K.}\ \bibnamefont {Thompson}},\ }\href
  {\doibase 10.1103/PhysRevLett.116.093602} {\bibfield  {journal} {\bibinfo
  {journal} {Phys. Rev. Lett.}\ }\textbf {\bibinfo {volume} {116}},\ \bibinfo
  {pages} {093602} (\bibinfo {year} {2016})}\BibitemShut {NoStop}%
\bibitem [{\citenamefont {Norcia}\ \emph {et~al.}(2018)\citenamefont {Norcia},
  \citenamefont {Lewis-Swan}, \citenamefont {Cline}, \citenamefont {Zhu},
  \citenamefont {Rey},\ and\ \citenamefont {Thompson}}]{Norcia_2018}%
  \BibitemOpen
  \bibfield  {author} {\bibinfo {author} {\bibfnamefont {M.~A.}\ \bibnamefont
  {Norcia}}, \bibinfo {author} {\bibfnamefont {R.~J.}\ \bibnamefont
  {Lewis-Swan}}, \bibinfo {author} {\bibfnamefont {J.~R.~K.}\ \bibnamefont
  {Cline}}, \bibinfo {author} {\bibfnamefont {B.}~\bibnamefont {Zhu}}, \bibinfo
  {author} {\bibfnamefont {A.~M.}\ \bibnamefont {Rey}}, \ and\ \bibinfo
  {author} {\bibfnamefont {J.~K.}\ \bibnamefont {Thompson}},\ }\href {\doibase
  10.1126/science.aar3102} {\bibfield  {journal} {\bibinfo  {journal}
  {Science}\ }\textbf {\bibinfo {volume} {361}},\ \bibinfo {pages} {259}
  (\bibinfo {year} {2018})}\BibitemShut {NoStop}%
\bibitem [{\citenamefont {Hu}\ \emph {et~al.}(2017)\citenamefont {Hu},
  \citenamefont {Chen}, \citenamefont {Vendeiro}, \citenamefont {Urvoy},
  \citenamefont {Braverman},\ and\ \citenamefont {Vuleti\ifmmode~\acute{c}\else
  \'{c}\fi{}}}]{Hu_2017}%
  \BibitemOpen
  \bibfield  {author} {\bibinfo {author} {\bibfnamefont {J.}~\bibnamefont
  {Hu}}, \bibinfo {author} {\bibfnamefont {W.}~\bibnamefont {Chen}}, \bibinfo
  {author} {\bibfnamefont {Z.}~\bibnamefont {Vendeiro}}, \bibinfo {author}
  {\bibfnamefont {A.}~\bibnamefont {Urvoy}}, \bibinfo {author} {\bibfnamefont
  {B.}~\bibnamefont {Braverman}}, \ and\ \bibinfo {author} {\bibfnamefont
  {V.}~\bibnamefont {Vuleti\ifmmode~\acute{c}\else \'{c}\fi{}}},\ }\href
  {\doibase 10.1103/PhysRevA.96.050301} {\bibfield  {journal} {\bibinfo
  {journal} {Phys. Rev. A}\ }\textbf {\bibinfo {volume} {96}},\ \bibinfo
  {pages} {050301} (\bibinfo {year} {2017})}\BibitemShut {NoStop}%
\bibitem [{\citenamefont {Lewis-Swan}\ \emph {et~al.}(2018)\citenamefont
  {Lewis-Swan}, \citenamefont {Norcia}, \citenamefont {Cline}, \citenamefont
  {Thompson},\ and\ \citenamefont {Rey}}]{RLS_2018}%
  \BibitemOpen
  \bibfield  {author} {\bibinfo {author} {\bibfnamefont {R.~J.}\ \bibnamefont
  {Lewis-Swan}}, \bibinfo {author} {\bibfnamefont {M.~A.}\ \bibnamefont
  {Norcia}}, \bibinfo {author} {\bibfnamefont {J.~R.~K.}\ \bibnamefont
  {Cline}}, \bibinfo {author} {\bibfnamefont {J.~K.}\ \bibnamefont {Thompson}},
  \ and\ \bibinfo {author} {\bibfnamefont {A.~M.}\ \bibnamefont {Rey}},\ }\href
  {\doibase 10.1103/PhysRevLett.121.070403} {\bibfield  {journal} {\bibinfo
  {journal} {Phys. Rev. Lett.}\ }\textbf {\bibinfo {volume} {121}},\ \bibinfo
  {pages} {070403} (\bibinfo {year} {2018})}\BibitemShut {NoStop}%
\bibitem [{\citenamefont {Ludlow}\ \emph {et~al.}(2015)\citenamefont {Ludlow},
  \citenamefont {Boyd}, \citenamefont {Ye}, \citenamefont {Peik},\ and\
  \citenamefont {Schmidt}}]{Ludlow2015}%
  \BibitemOpen
  \bibfield  {author} {\bibinfo {author} {\bibfnamefont {A.~D.}\ \bibnamefont
  {Ludlow}}, \bibinfo {author} {\bibfnamefont {M.~M.}\ \bibnamefont {Boyd}},
  \bibinfo {author} {\bibfnamefont {J.}~\bibnamefont {Ye}}, \bibinfo {author}
  {\bibfnamefont {E.}~\bibnamefont {Peik}}, \ and\ \bibinfo {author}
  {\bibfnamefont {P.~O.}\ \bibnamefont {Schmidt}},\ }\href {\doibase
  10.1103/RevModPhys.87.637} {\bibfield  {journal} {\bibinfo  {journal} {Rev.
  Mod. Phys.}\ }\textbf {\bibinfo {volume} {87}},\ \bibinfo {pages} {637}
  (\bibinfo {year} {2015})}\BibitemShut {NoStop}%
\bibitem [{\citenamefont {Braunstein}\ and\ \citenamefont
  {Caves}(1994)}]{Braunstein1994}%
  \BibitemOpen
  \bibfield  {author} {\bibinfo {author} {\bibfnamefont {S.~L.}\ \bibnamefont
  {Braunstein}}\ and\ \bibinfo {author} {\bibfnamefont {C.~M.}\ \bibnamefont
  {Caves}},\ }\href {\doibase 10.1103/PhysRevLett.72.3439} {\bibfield
  {journal} {\bibinfo  {journal} {Phys. Rev. Lett.}\ }\textbf {\bibinfo
  {volume} {72}},\ \bibinfo {pages} {3439} (\bibinfo {year}
  {1994})}\BibitemShut {NoStop}%
\bibitem [{\citenamefont {Jaekel}\ and\ \citenamefont
  {Reynaud}(1990)}]{Jaekel_1990}%
  \BibitemOpen
  \bibfield  {author} {\bibinfo {author} {\bibfnamefont {M.~T.}\ \bibnamefont
  {Jaekel}}\ and\ \bibinfo {author} {\bibfnamefont {S.}~\bibnamefont
  {Reynaud}},\ }\href {\doibase 10.1209/0295-5075/13/4/003} {\bibfield
  {journal} {\bibinfo  {journal} {Europhysics Letters ({EPL})}\ }\textbf
  {\bibinfo {volume} {13}},\ \bibinfo {pages} {301} (\bibinfo {year}
  {1990})}\BibitemShut {NoStop}%
\bibitem [{\citenamefont {Zurek}(2001)}]{Zurek2001}%
  \BibitemOpen
  \bibfield  {author} {\bibinfo {author} {\bibfnamefont {W.~H.}\ \bibnamefont
  {Zurek}},\ }\href {\doibase 10.1038/35089017} {\bibfield  {journal} {\bibinfo
   {journal} {Nature}\ }\textbf {\bibinfo {volume} {412}},\ \bibinfo {pages}
  {712} (\bibinfo {year} {2001})}\BibitemShut {NoStop}%
\bibitem [{\citenamefont {Pezz\`e}\ \emph {et~al.}(2018)\citenamefont
  {Pezz\`e}, \citenamefont {Smerzi}, \citenamefont {Oberthaler}, \citenamefont
  {Schmied},\ and\ \citenamefont {Treutlein}}]{Pezze_2018}%
  \BibitemOpen
  \bibfield  {author} {\bibinfo {author} {\bibfnamefont {L.}~\bibnamefont
  {Pezz\`e}}, \bibinfo {author} {\bibfnamefont {A.}~\bibnamefont {Smerzi}},
  \bibinfo {author} {\bibfnamefont {M.~K.}\ \bibnamefont {Oberthaler}},
  \bibinfo {author} {\bibfnamefont {R.}~\bibnamefont {Schmied}}, \ and\
  \bibinfo {author} {\bibfnamefont {P.}~\bibnamefont {Treutlein}},\ }\href
  {\doibase 10.1103/RevModPhys.90.035005} {\bibfield  {journal} {\bibinfo
  {journal} {Rev. Mod. Phys.}\ }\textbf {\bibinfo {volume} {90}},\ \bibinfo
  {pages} {035005} (\bibinfo {year} {2018})}\BibitemShut {NoStop}%
\bibitem [{\citenamefont {Giovannetti}\ \emph {et~al.}(2006)\citenamefont
  {Giovannetti}, \citenamefont {Lloyd},\ and\ \citenamefont
  {Maccone}}]{Giovannetti_2006}%
  \BibitemOpen
  \bibfield  {author} {\bibinfo {author} {\bibfnamefont {V.}~\bibnamefont
  {Giovannetti}}, \bibinfo {author} {\bibfnamefont {S.}~\bibnamefont {Lloyd}},
  \ and\ \bibinfo {author} {\bibfnamefont {L.}~\bibnamefont {Maccone}},\ }\href
  {\doibase 10.1103/PhysRevLett.96.010401} {\bibfield  {journal} {\bibinfo
  {journal} {Phys. Rev. Lett.}\ }\textbf {\bibinfo {volume} {96}},\ \bibinfo
  {pages} {010401} (\bibinfo {year} {2006})}\BibitemShut {NoStop}%
\bibitem [{\citenamefont {Yurke}\ \emph {et~al.}(1986)\citenamefont {Yurke},
  \citenamefont {McCall},\ and\ \citenamefont {Klauder}}]{Yurke_1986}%
  \BibitemOpen
  \bibfield  {author} {\bibinfo {author} {\bibfnamefont {B.}~\bibnamefont
  {Yurke}}, \bibinfo {author} {\bibfnamefont {S.~L.}\ \bibnamefont {McCall}}, \
  and\ \bibinfo {author} {\bibfnamefont {J.~R.}\ \bibnamefont {Klauder}},\
  }\href {\doibase 10.1103/PhysRevA.33.4033} {\bibfield  {journal} {\bibinfo
  {journal} {Phys. Rev. A}\ }\textbf {\bibinfo {volume} {33}},\ \bibinfo
  {pages} {4033} (\bibinfo {year} {1986})}\BibitemShut {NoStop}%
\bibitem [{\citenamefont {Holland}\ and\ \citenamefont
  {Burnett}(1993)}]{Holland_1993}%
  \BibitemOpen
  \bibfield  {author} {\bibinfo {author} {\bibfnamefont {M.~J.}\ \bibnamefont
  {Holland}}\ and\ \bibinfo {author} {\bibfnamefont {K.}~\bibnamefont
  {Burnett}},\ }\href {\doibase 10.1103/PhysRevLett.71.1355} {\bibfield
  {journal} {\bibinfo  {journal} {Phys. Rev. Lett.}\ }\textbf {\bibinfo
  {volume} {71}},\ \bibinfo {pages} {1355} (\bibinfo {year}
  {1993})}\BibitemShut {NoStop}%
\bibitem [{\citenamefont {Toscano}\ \emph {et~al.}(2006)\citenamefont
  {Toscano}, \citenamefont {Dalvit}, \citenamefont {Davidovich},\ and\
  \citenamefont {Zurek}}]{Toscano_2006}%
  \BibitemOpen
  \bibfield  {author} {\bibinfo {author} {\bibfnamefont {F.}~\bibnamefont
  {Toscano}}, \bibinfo {author} {\bibfnamefont {D.~A.~R.}\ \bibnamefont
  {Dalvit}}, \bibinfo {author} {\bibfnamefont {L.}~\bibnamefont {Davidovich}},
  \ and\ \bibinfo {author} {\bibfnamefont {W.~H.}\ \bibnamefont {Zurek}},\
  }\href {\doibase 10.1103/PhysRevA.73.023803} {\bibfield  {journal} {\bibinfo
  {journal} {Phys. Rev. A}\ }\textbf {\bibinfo {volume} {73}},\ \bibinfo
  {pages} {023803} (\bibinfo {year} {2006})}\BibitemShut {NoStop}%
\bibitem [{\citenamefont {Schneider}\ \emph {et~al.}(1998)\citenamefont
  {Schneider}, \citenamefont {Wiseman}, \citenamefont {Munro},\ and\
  \citenamefont {Milburn}}]{Schneider_1998}%
  \BibitemOpen
  \bibfield  {author} {\bibinfo {author} {\bibfnamefont {S.}~\bibnamefont
  {Schneider}}, \bibinfo {author} {\bibfnamefont {H.}~\bibnamefont {Wiseman}},
  \bibinfo {author} {\bibfnamefont {W.}~\bibnamefont {Munro}}, \ and\ \bibinfo
  {author} {\bibfnamefont {G.}~\bibnamefont {Milburn}},\ }\href {\doibase
  10.1002/(SICI)1521-3978(199806)46:4/5<391::AID-PROP391>3.0.CO;2-0} {\bibfield
   {journal} {\bibinfo  {journal} {Fortschritte der Physik}\ }\textbf {\bibinfo
  {volume} {46}},\ \bibinfo {pages} {391} (\bibinfo {year} {1998})}\BibitemShut
  {NoStop}%
\bibitem [{\citenamefont {Munro}\ \emph {et~al.}(2002)\citenamefont {Munro},
  \citenamefont {Nemoto}, \citenamefont {Milburn},\ and\ \citenamefont
  {Braunstein}}]{Munro_2002}%
  \BibitemOpen
  \bibfield  {author} {\bibinfo {author} {\bibfnamefont {W.~J.}\ \bibnamefont
  {Munro}}, \bibinfo {author} {\bibfnamefont {K.}~\bibnamefont {Nemoto}},
  \bibinfo {author} {\bibfnamefont {G.~J.}\ \bibnamefont {Milburn}}, \ and\
  \bibinfo {author} {\bibfnamefont {S.~L.}\ \bibnamefont {Braunstein}},\ }\href
  {\doibase 10.1103/PhysRevA.66.023819} {\bibfield  {journal} {\bibinfo
  {journal} {Phys. Rev. A}\ }\textbf {\bibinfo {volume} {66}},\ \bibinfo
  {pages} {023819} (\bibinfo {year} {2002})}\BibitemShut {NoStop}%
\bibitem [{\citenamefont {Hillery}\ \emph {et~al.}(1984)\citenamefont
  {Hillery}, \citenamefont {O'Connell}, \citenamefont {Scully},\ and\
  \citenamefont {Wigner}}]{HILLERY_1984}%
  \BibitemOpen
  \bibfield  {author} {\bibinfo {author} {\bibfnamefont {M.}~\bibnamefont
  {Hillery}}, \bibinfo {author} {\bibfnamefont {R.}~\bibnamefont {O'Connell}},
  \bibinfo {author} {\bibfnamefont {M.}~\bibnamefont {Scully}}, \ and\ \bibinfo
  {author} {\bibfnamefont {E.}~\bibnamefont {Wigner}},\ }\href {\doibase
  https://doi.org/10.1016/0370-1573(84)90160-1} {\bibfield  {journal} {\bibinfo
   {journal} {Physics Reports}\ }\textbf {\bibinfo {volume} {106}},\ \bibinfo
  {pages} {121 } (\bibinfo {year} {1984})}\BibitemShut {NoStop}%
\bibitem [{\citenamefont {Agarwal}\ and\ \citenamefont
  {Pathak}(2004)}]{Agarwal_2004}%
  \BibitemOpen
  \bibfield  {author} {\bibinfo {author} {\bibfnamefont {G.~S.}\ \bibnamefont
  {Agarwal}}\ and\ \bibinfo {author} {\bibfnamefont {P.~K.}\ \bibnamefont
  {Pathak}},\ }\href {\doibase 10.1103/PhysRevA.70.053813} {\bibfield
  {journal} {\bibinfo  {journal} {Phys. Rev. A}\ }\textbf {\bibinfo {volume}
  {70}},\ \bibinfo {pages} {053813} (\bibinfo {year} {2004})}\BibitemShut
  {NoStop}%
\bibitem [{\citenamefont {Helstrom}(1976)}]{Helstrom_1976}%
  \BibitemOpen
  \bibfield  {author} {\bibinfo {author} {\bibfnamefont {C.~W.}\ \bibnamefont
  {Helstrom}},\ }\href {https://nla.gov.au/nla.cat-vn617918} {\emph {\bibinfo
  {title} {Quantum detection and estimation theory}}}\ (\bibinfo  {publisher}
  {Academic Press New York},\ \bibinfo {year} {1976})\ pp.\ \bibinfo {pages}
  {ix, 309 p. :}\BibitemShut {NoStop}%
\bibitem [{\citenamefont {Jarzyna}\ and\ \citenamefont
  {Demkowicz-Dobrza\ifmmode~\acute{n}\else
  \'{n}\fi{}ski}(2012)}]{Jarzyna_2012}%
  \BibitemOpen
  \bibfield  {author} {\bibinfo {author} {\bibfnamefont {M.}~\bibnamefont
  {Jarzyna}}\ and\ \bibinfo {author} {\bibfnamefont {R.}~\bibnamefont
  {Demkowicz-Dobrza\ifmmode~\acute{n}\else \'{n}\fi{}ski}},\ }\href {\doibase
  10.1103/PhysRevA.85.011801} {\bibfield  {journal} {\bibinfo  {journal} {Phys.
  Rev. A}\ }\textbf {\bibinfo {volume} {85}},\ \bibinfo {pages} {011801}
  (\bibinfo {year} {2012})}\BibitemShut {NoStop}%
\bibitem [{\citenamefont {Bollinger}\ \emph {et~al.}(1996)\citenamefont
  {Bollinger}, \citenamefont {Itano}, \citenamefont {Wineland},\ and\
  \citenamefont {Heinzen}}]{Bollinger_1996}%
  \BibitemOpen
  \bibfield  {author} {\bibinfo {author} {\bibfnamefont {J.~J.~.}\ \bibnamefont
  {Bollinger}}, \bibinfo {author} {\bibfnamefont {W.~M.}\ \bibnamefont
  {Itano}}, \bibinfo {author} {\bibfnamefont {D.~J.}\ \bibnamefont {Wineland}},
  \ and\ \bibinfo {author} {\bibfnamefont {D.~J.}\ \bibnamefont {Heinzen}},\
  }\href {\doibase 10.1103/PhysRevA.54.R4649} {\bibfield  {journal} {\bibinfo
  {journal} {Phys. Rev. A}\ }\textbf {\bibinfo {volume} {54}},\ \bibinfo
  {pages} {R4649} (\bibinfo {year} {1996})}\BibitemShut {NoStop}%
\bibitem [{\citenamefont {Macr\`{\i}}\ \emph {et~al.}(2016)\citenamefont
  {Macr\`{\i}}, \citenamefont {Smerzi},\ and\ \citenamefont
  {Pezz\`e}}]{Macri_2016}%
  \BibitemOpen
  \bibfield  {author} {\bibinfo {author} {\bibfnamefont {T.}~\bibnamefont
  {Macr\`{\i}}}, \bibinfo {author} {\bibfnamefont {A.}~\bibnamefont {Smerzi}},
  \ and\ \bibinfo {author} {\bibfnamefont {L.}~\bibnamefont {Pezz\`e}},\ }\href
  {\doibase 10.1103/PhysRevA.94.010102} {\bibfield  {journal} {\bibinfo
  {journal} {Phys. Rev. A}\ }\textbf {\bibinfo {volume} {94}},\ \bibinfo
  {pages} {010102} (\bibinfo {year} {2016})}\BibitemShut {NoStop}%
\bibitem [{\citenamefont {Strobel}\ \emph {et~al.}(2014)\citenamefont
  {Strobel}, \citenamefont {Muessel}, \citenamefont {Linnemann}, \citenamefont
  {Zibold}, \citenamefont {Hume}, \citenamefont {Pezze}, \citenamefont
  {Smerzi},\ and\ \citenamefont {Oberthaler}}]{Strobel2014}%
  \BibitemOpen
  \bibfield  {author} {\bibinfo {author} {\bibfnamefont {H.}~\bibnamefont
  {Strobel}}, \bibinfo {author} {\bibfnamefont {W.}~\bibnamefont {Muessel}},
  \bibinfo {author} {\bibfnamefont {D.}~\bibnamefont {Linnemann}}, \bibinfo
  {author} {\bibfnamefont {T.}~\bibnamefont {Zibold}}, \bibinfo {author}
  {\bibfnamefont {D.~B.}\ \bibnamefont {Hume}}, \bibinfo {author}
  {\bibfnamefont {L.}~\bibnamefont {Pezze}}, \bibinfo {author} {\bibfnamefont
  {A.}~\bibnamefont {Smerzi}}, \ and\ \bibinfo {author} {\bibfnamefont {M.~K.}\
  \bibnamefont {Oberthaler}},\ }\href {\doibase 10.1126/science.1250147}
  {\bibfield  {journal} {\bibinfo  {journal} {Science}\ }\textbf {\bibinfo
  {volume} {345}},\ \bibinfo {pages} {424} (\bibinfo {year}
  {2014})}\BibitemShut {NoStop}%
\bibitem [{\citenamefont {Hudelist}\ \emph {et~al.}(2014)\citenamefont
  {Hudelist}, \citenamefont {Kong}, \citenamefont {Liu}, \citenamefont {Jing},
  \citenamefont {Ou},\ and\ \citenamefont {Zhang}}]{Hudelist_2014}%
  \BibitemOpen
  \bibfield  {author} {\bibinfo {author} {\bibfnamefont {F.}~\bibnamefont
  {Hudelist}}, \bibinfo {author} {\bibfnamefont {J.}~\bibnamefont {Kong}},
  \bibinfo {author} {\bibfnamefont {C.}~\bibnamefont {Liu}}, \bibinfo {author}
  {\bibfnamefont {J.}~\bibnamefont {Jing}}, \bibinfo {author} {\bibfnamefont
  {Z.}~\bibnamefont {Ou}}, \ and\ \bibinfo {author} {\bibfnamefont
  {W.}~\bibnamefont {Zhang}},\ }\href {\doibase 10.1038/ncomms4049} {\bibfield
  {journal} {\bibinfo  {journal} {Nature Communications}\ }\textbf {\bibinfo
  {volume} {5}} (\bibinfo {year} {2014}),\ 10.1038/ncomms4049}\BibitemShut
  {NoStop}%
\bibitem [{\citenamefont {Linnemann}\ \emph {et~al.}(2016)\citenamefont
  {Linnemann}, \citenamefont {Strobel}, \citenamefont {Muessel}, \citenamefont
  {Schulz}, \citenamefont {Lewis-Swan}, \citenamefont {Kheruntsyan},\ and\
  \citenamefont {Oberthaler}}]{Linnemann_2016}%
  \BibitemOpen
  \bibfield  {author} {\bibinfo {author} {\bibfnamefont {D.}~\bibnamefont
  {Linnemann}}, \bibinfo {author} {\bibfnamefont {H.}~\bibnamefont {Strobel}},
  \bibinfo {author} {\bibfnamefont {W.}~\bibnamefont {Muessel}}, \bibinfo
  {author} {\bibfnamefont {J.}~\bibnamefont {Schulz}}, \bibinfo {author}
  {\bibfnamefont {R.~J.}\ \bibnamefont {Lewis-Swan}}, \bibinfo {author}
  {\bibfnamefont {K.~V.}\ \bibnamefont {Kheruntsyan}}, \ and\ \bibinfo {author}
  {\bibfnamefont {M.~K.}\ \bibnamefont {Oberthaler}},\ }\href {\doibase
  10.1103/PhysRevLett.117.013001} {\bibfield  {journal} {\bibinfo  {journal}
  {Phys. Rev. Lett.}\ }\textbf {\bibinfo {volume} {117}},\ \bibinfo {pages}
  {013001} (\bibinfo {year} {2016})}\BibitemShut {NoStop}%
\bibitem [{\citenamefont {Davis}\ \emph {et~al.}(2016)\citenamefont {Davis},
  \citenamefont {Bentsen},\ and\ \citenamefont {Schleier-Smith}}]{Davis_2016}%
  \BibitemOpen
  \bibfield  {author} {\bibinfo {author} {\bibfnamefont {E.}~\bibnamefont
  {Davis}}, \bibinfo {author} {\bibfnamefont {G.}~\bibnamefont {Bentsen}}, \
  and\ \bibinfo {author} {\bibfnamefont {M.}~\bibnamefont {Schleier-Smith}},\
  }\href {\doibase 10.1103/PhysRevLett.116.053601} {\bibfield  {journal}
  {\bibinfo  {journal} {Phys. Rev. Lett.}\ }\textbf {\bibinfo {volume} {116}},\
  \bibinfo {pages} {053601} (\bibinfo {year} {2016})}\BibitemShut {NoStop}%
\bibitem [{\citenamefont {Szigeti}\ \emph {et~al.}(2017)\citenamefont
  {Szigeti}, \citenamefont {Lewis-Swan},\ and\ \citenamefont
  {Haine}}]{Szigeti_2017}%
  \BibitemOpen
  \bibfield  {author} {\bibinfo {author} {\bibfnamefont {S.~S.}\ \bibnamefont
  {Szigeti}}, \bibinfo {author} {\bibfnamefont {R.~J.}\ \bibnamefont
  {Lewis-Swan}}, \ and\ \bibinfo {author} {\bibfnamefont {S.~A.}\ \bibnamefont
  {Haine}},\ }\href {\doibase 10.1103/PhysRevLett.118.150401} {\bibfield
  {journal} {\bibinfo  {journal} {Phys. Rev. Lett.}\ }\textbf {\bibinfo
  {volume} {118}},\ \bibinfo {pages} {150401} (\bibinfo {year}
  {2017})}\BibitemShut {NoStop}%
\bibitem [{\citenamefont {Wrubel}\ \emph {et~al.}(2018)\citenamefont {Wrubel},
  \citenamefont {Schwettmann}, \citenamefont {Fahey}, \citenamefont {Glassman},
  \citenamefont {Pechkis}, \citenamefont {Griffin}, \citenamefont {Barnett},
  \citenamefont {Tiesinga},\ and\ \citenamefont {Lett}}]{Wrubel_2018}%
  \BibitemOpen
  \bibfield  {author} {\bibinfo {author} {\bibfnamefont {J.~P.}\ \bibnamefont
  {Wrubel}}, \bibinfo {author} {\bibfnamefont {A.}~\bibnamefont {Schwettmann}},
  \bibinfo {author} {\bibfnamefont {D.~P.}\ \bibnamefont {Fahey}}, \bibinfo
  {author} {\bibfnamefont {Z.}~\bibnamefont {Glassman}}, \bibinfo {author}
  {\bibfnamefont {H.~K.}\ \bibnamefont {Pechkis}}, \bibinfo {author}
  {\bibfnamefont {P.~F.}\ \bibnamefont {Griffin}}, \bibinfo {author}
  {\bibfnamefont {R.}~\bibnamefont {Barnett}}, \bibinfo {author} {\bibfnamefont
  {E.}~\bibnamefont {Tiesinga}}, \ and\ \bibinfo {author} {\bibfnamefont
  {P.~D.}\ \bibnamefont {Lett}},\ }\href {\doibase 10.1103/PhysRevA.98.023620}
  {\bibfield  {journal} {\bibinfo  {journal} {Phys. Rev. A}\ }\textbf {\bibinfo
  {volume} {98}},\ \bibinfo {pages} {023620} (\bibinfo {year}
  {2018})}\BibitemShut {NoStop}%
\bibitem [{\citenamefont {Huang}\ \emph {et~al.}(2018)\citenamefont {Huang},
  \citenamefont {Zhuang}, \citenamefont {Lu}, \citenamefont {Ke},\ and\
  \citenamefont {Lee}}]{Huang_2018}%
  \BibitemOpen
  \bibfield  {author} {\bibinfo {author} {\bibfnamefont {J.}~\bibnamefont
  {Huang}}, \bibinfo {author} {\bibfnamefont {M.}~\bibnamefont {Zhuang}},
  \bibinfo {author} {\bibfnamefont {B.}~\bibnamefont {Lu}}, \bibinfo {author}
  {\bibfnamefont {Y.}~\bibnamefont {Ke}}, \ and\ \bibinfo {author}
  {\bibfnamefont {C.}~\bibnamefont {Lee}},\ }\href {\doibase
  10.1103/PhysRevA.98.012129} {\bibfield  {journal} {\bibinfo  {journal} {Phys.
  Rev. A}\ }\textbf {\bibinfo {volume} {98}},\ \bibinfo {pages} {012129}
  (\bibinfo {year} {2018})}\BibitemShut {NoStop}%
\bibitem [{\citenamefont {Haine}(2018)}]{Haine_2018}%
  \BibitemOpen
  \bibfield  {author} {\bibinfo {author} {\bibfnamefont {S.~A.}\ \bibnamefont
  {Haine}},\ }\href {\doibase 10.1103/PhysRevA.98.030303} {\bibfield  {journal}
  {\bibinfo  {journal} {Phys. Rev. A}\ }\textbf {\bibinfo {volume} {98}},\
  \bibinfo {pages} {030303} (\bibinfo {year} {2018})}\BibitemShut {NoStop}%
\bibitem [{\citenamefont {Nolan}\ \emph {et~al.}(2017)\citenamefont {Nolan},
  \citenamefont {Szigeti},\ and\ \citenamefont {Haine}}]{Nolan_2017}%
  \BibitemOpen
  \bibfield  {author} {\bibinfo {author} {\bibfnamefont {S.~P.}\ \bibnamefont
  {Nolan}}, \bibinfo {author} {\bibfnamefont {S.~S.}\ \bibnamefont {Szigeti}},
  \ and\ \bibinfo {author} {\bibfnamefont {S.~A.}\ \bibnamefont {Haine}},\
  }\href {\doibase 10.1103/PhysRevLett.119.193601} {\bibfield  {journal}
  {\bibinfo  {journal} {Phys. Rev. Lett.}\ }\textbf {\bibinfo {volume} {119}},\
  \bibinfo {pages} {193601} (\bibinfo {year} {2017})}\BibitemShut {NoStop}%
\bibitem [{\citenamefont {Mirkhalaf}\ \emph {et~al.}(2018)\citenamefont
  {Mirkhalaf}, \citenamefont {Nolan},\ and\ \citenamefont
  {Haine}}]{Mirkhalaf_2018}%
  \BibitemOpen
  \bibfield  {author} {\bibinfo {author} {\bibfnamefont {S.~S.}\ \bibnamefont
  {Mirkhalaf}}, \bibinfo {author} {\bibfnamefont {S.~P.}\ \bibnamefont
  {Nolan}}, \ and\ \bibinfo {author} {\bibfnamefont {S.~A.}\ \bibnamefont
  {Haine}},\ }\href {\doibase 10.1103/PhysRevA.97.053618} {\bibfield  {journal}
  {\bibinfo  {journal} {Phys. Rev. A}\ }\textbf {\bibinfo {volume} {97}},\
  \bibinfo {pages} {053618} (\bibinfo {year} {2018})}\BibitemShut {NoStop}%
\bibitem [{\citenamefont {Safavi-Naini}\ \emph {et~al.}(2018)\citenamefont
  {Safavi-Naini}, \citenamefont {Lewis-Swan}, \citenamefont {Bohnet},
  \citenamefont {G\"arttner}, \citenamefont {Gilmore}, \citenamefont {Jordan},
  \citenamefont {Cohn}, \citenamefont {Freericks}, \citenamefont {Rey},\ and\
  \citenamefont {Bollinger}}]{Bollinger_2018}%
  \BibitemOpen
  \bibfield  {author} {\bibinfo {author} {\bibfnamefont {A.}~\bibnamefont
  {Safavi-Naini}}, \bibinfo {author} {\bibfnamefont {R.~J.}\ \bibnamefont
  {Lewis-Swan}}, \bibinfo {author} {\bibfnamefont {J.~G.}\ \bibnamefont
  {Bohnet}}, \bibinfo {author} {\bibfnamefont {M.}~\bibnamefont {G\"arttner}},
  \bibinfo {author} {\bibfnamefont {K.~A.}\ \bibnamefont {Gilmore}}, \bibinfo
  {author} {\bibfnamefont {J.~E.}\ \bibnamefont {Jordan}}, \bibinfo {author}
  {\bibfnamefont {J.}~\bibnamefont {Cohn}}, \bibinfo {author} {\bibfnamefont
  {J.~K.}\ \bibnamefont {Freericks}}, \bibinfo {author} {\bibfnamefont {A.~M.}\
  \bibnamefont {Rey}}, \ and\ \bibinfo {author} {\bibfnamefont {J.~J.}\
  \bibnamefont {Bollinger}},\ }\href {\doibase 10.1103/PhysRevLett.121.040503}
  {\bibfield  {journal} {\bibinfo  {journal} {Phys. Rev. Lett.}\ }\textbf
  {\bibinfo {volume} {121}},\ \bibinfo {pages} {040503} (\bibinfo {year}
  {2018})}\BibitemShut {NoStop}%
\bibitem [{\citenamefont {Bertet}\ \emph {et~al.}(2002)\citenamefont {Bertet},
  \citenamefont {Auffeves}, \citenamefont {Maioli}, \citenamefont {Osnaghi},
  \citenamefont {Meunier}, \citenamefont {Brune}, \citenamefont {Raimond},\
  and\ \citenamefont {Haroche}}]{Bertet_2002}%
  \BibitemOpen
  \bibfield  {author} {\bibinfo {author} {\bibfnamefont {P.}~\bibnamefont
  {Bertet}}, \bibinfo {author} {\bibfnamefont {A.}~\bibnamefont {Auffeves}},
  \bibinfo {author} {\bibfnamefont {P.}~\bibnamefont {Maioli}}, \bibinfo
  {author} {\bibfnamefont {S.}~\bibnamefont {Osnaghi}}, \bibinfo {author}
  {\bibfnamefont {T.}~\bibnamefont {Meunier}}, \bibinfo {author} {\bibfnamefont
  {M.}~\bibnamefont {Brune}}, \bibinfo {author} {\bibfnamefont {J.~M.}\
  \bibnamefont {Raimond}}, \ and\ \bibinfo {author} {\bibfnamefont
  {S.}~\bibnamefont {Haroche}},\ }\href {\doibase
  10.1103/PhysRevLett.89.200402} {\bibfield  {journal} {\bibinfo  {journal}
  {Phys. Rev. Lett.}\ }\textbf {\bibinfo {volume} {89}},\ \bibinfo {pages}
  {200402} (\bibinfo {year} {2002})}\BibitemShut {NoStop}%
\bibitem [{\citenamefont {Blais}\ \emph {et~al.}(2004)\citenamefont {Blais},
  \citenamefont {Huang}, \citenamefont {Wallraff}, \citenamefont {Girvin},\
  and\ \citenamefont {Schoelkopf}}]{Blais_2004}%
  \BibitemOpen
  \bibfield  {author} {\bibinfo {author} {\bibfnamefont {A.}~\bibnamefont
  {Blais}}, \bibinfo {author} {\bibfnamefont {R.-S.}\ \bibnamefont {Huang}},
  \bibinfo {author} {\bibfnamefont {A.}~\bibnamefont {Wallraff}}, \bibinfo
  {author} {\bibfnamefont {S.~M.}\ \bibnamefont {Girvin}}, \ and\ \bibinfo
  {author} {\bibfnamefont {R.~J.}\ \bibnamefont {Schoelkopf}},\ }\href
  {\doibase 10.1103/PhysRevA.69.062320} {\bibfield  {journal} {\bibinfo
  {journal} {Phys. Rev. A}\ }\textbf {\bibinfo {volume} {69}},\ \bibinfo
  {pages} {062320} (\bibinfo {year} {2004})}\BibitemShut {NoStop}%
\bibitem [{\citenamefont {James}\ and\ \citenamefont
  {Jerke}(2007)}]{James2007}%
  \BibitemOpen
  \bibfield  {author} {\bibinfo {author} {\bibfnamefont {D.~F.}\ \bibnamefont
  {James}}\ and\ \bibinfo {author} {\bibfnamefont {J.}~\bibnamefont {Jerke}},\
  }\href {\doibase 10.1139/p07-060} {\bibfield  {journal} {\bibinfo  {journal}
  {Canadian Journal of Physics}\ }\textbf {\bibinfo {volume} {85}},\ \bibinfo
  {pages} {625} (\bibinfo {year} {2007})},\ \Eprint
  {http://arxiv.org/abs/https://doi.org/10.1139/p07-060}
  {https://doi.org/10.1139/p07-060} \BibitemShut {NoStop}%
\bibitem [{\citenamefont {Susskind}\ and\ \citenamefont
  {Glogower}(1964)}]{Susskind1964}%
  \BibitemOpen
  \bibfield  {author} {\bibinfo {author} {\bibfnamefont {L.}~\bibnamefont
  {Susskind}}\ and\ \bibinfo {author} {\bibfnamefont {J.}~\bibnamefont
  {Glogower}},\ }\href {\doibase 10.1103/PhysicsPhysiqueFizika.1.49} {\bibfield
   {journal} {\bibinfo  {journal} {Physics Physique Fizika}\ }\textbf {\bibinfo
  {volume} {1}},\ \bibinfo {pages} {49} (\bibinfo {year} {1964})}\BibitemShut
  {NoStop}%
\bibitem [{\citenamefont {O'Neill}\ \emph {et~al.}(2014)\citenamefont
  {O'Neill}, \citenamefont {Church}, \citenamefont {McGreevy}, \citenamefont
  {Thomson},\ and\ \citenamefont {Brodbelt}}]{ONeill2014}%
  \BibitemOpen
  \bibfield  {author} {\bibinfo {author} {\bibfnamefont {D.~G.}\ \bibnamefont
  {O'Neill}}, \bibinfo {author} {\bibfnamefont {D.~B.}\ \bibnamefont {Church}},
  \bibinfo {author} {\bibfnamefont {P.~D.}\ \bibnamefont {McGreevy}}, \bibinfo
  {author} {\bibfnamefont {P.~C.}\ \bibnamefont {Thomson}}, \ and\ \bibinfo
  {author} {\bibfnamefont {D.~C.}\ \bibnamefont {Brodbelt}},\ }\href {\doibase
  10.1177/1098612x14536176} {\bibfield  {journal} {\bibinfo  {journal} {Journal
  of Feline Medicine and Surgery}\ }\textbf {\bibinfo {volume} {17}},\ \bibinfo
  {pages} {125} (\bibinfo {year} {2014})}\BibitemShut {NoStop}%
\bibitem [{\citenamefont {Norcia}\ and\ \citenamefont
  {Thompson}(2016)}]{Norcia2016}%
  \BibitemOpen
  \bibfield  {author} {\bibinfo {author} {\bibfnamefont {M.~A.}\ \bibnamefont
  {Norcia}}\ and\ \bibinfo {author} {\bibfnamefont {J.~K.}\ \bibnamefont
  {Thompson}},\ }\href {\doibase 10.1103/PhysRevA.93.023804} {\bibfield
  {journal} {\bibinfo  {journal} {Phys. Rev. A}\ }\textbf {\bibinfo {volume}
  {93}},\ \bibinfo {pages} {023804} (\bibinfo {year} {2016})}\BibitemShut
  {NoStop}%
\bibitem [{\citenamefont {Linnemann}\ \emph {et~al.}(2017)\citenamefont
  {Linnemann}, \citenamefont {Schulz}, \citenamefont {Muessel}, \citenamefont
  {Kunkel}, \citenamefont {Prüfer}, \citenamefont {Frölian}, \citenamefont
  {Strobel},\ and\ \citenamefont {Oberthaler}}]{Linnemann_2017}%
  \BibitemOpen
  \bibfield  {author} {\bibinfo {author} {\bibfnamefont {D.}~\bibnamefont
  {Linnemann}}, \bibinfo {author} {\bibfnamefont {J.}~\bibnamefont {Schulz}},
  \bibinfo {author} {\bibfnamefont {W.}~\bibnamefont {Muessel}}, \bibinfo
  {author} {\bibfnamefont {P.}~\bibnamefont {Kunkel}}, \bibinfo {author}
  {\bibfnamefont {M.}~\bibnamefont {Prüfer}}, \bibinfo {author} {\bibfnamefont
  {A.}~\bibnamefont {Frölian}}, \bibinfo {author} {\bibfnamefont
  {H.}~\bibnamefont {Strobel}}, \ and\ \bibinfo {author} {\bibfnamefont
  {M.~K.}\ \bibnamefont {Oberthaler}},\ }\href {\doibase
  10.1088/2058-9565/aa802c} {\bibfield  {journal} {\bibinfo  {journal} {Quantum
  Science and Technology}\ }\textbf {\bibinfo {volume} {2}},\ \bibinfo {pages}
  {044009} (\bibinfo {year} {2017})}\BibitemShut {NoStop}%
\bibitem [{\citenamefont {Anders}\ \emph {et~al.}(2018)\citenamefont {Anders},
  \citenamefont {Pezz\`e}, \citenamefont {Smerzi},\ and\ \citenamefont
  {Klempt}}]{Fabian_2018}%
  \BibitemOpen
  \bibfield  {author} {\bibinfo {author} {\bibfnamefont {F.}~\bibnamefont
  {Anders}}, \bibinfo {author} {\bibfnamefont {L.}~\bibnamefont {Pezz\`e}},
  \bibinfo {author} {\bibfnamefont {A.}~\bibnamefont {Smerzi}}, \ and\ \bibinfo
  {author} {\bibfnamefont {C.}~\bibnamefont {Klempt}},\ }\href {\doibase
  10.1103/PhysRevA.97.043813} {\bibfield  {journal} {\bibinfo  {journal} {Phys.
  Rev. A}\ }\textbf {\bibinfo {volume} {97}},\ \bibinfo {pages} {043813}
  (\bibinfo {year} {2018})}\BibitemShut {NoStop}%
\bibitem [{\citenamefont {Del{\'{e}}glise}\ \emph {et~al.}(2008)\citenamefont
  {Del{\'{e}}glise}, \citenamefont {Dotsenko}, \citenamefont {Sayrin},
  \citenamefont {Bernu}, \citenamefont {Brune}, \citenamefont {Raimond},\ and\
  \citenamefont {Haroche}}]{Delglise_2008}%
  \BibitemOpen
  \bibfield  {author} {\bibinfo {author} {\bibfnamefont {S.}~\bibnamefont
  {Del{\'{e}}glise}}, \bibinfo {author} {\bibfnamefont {I.}~\bibnamefont
  {Dotsenko}}, \bibinfo {author} {\bibfnamefont {C.}~\bibnamefont {Sayrin}},
  \bibinfo {author} {\bibfnamefont {J.}~\bibnamefont {Bernu}}, \bibinfo
  {author} {\bibfnamefont {M.}~\bibnamefont {Brune}}, \bibinfo {author}
  {\bibfnamefont {J.-M.}\ \bibnamefont {Raimond}}, \ and\ \bibinfo {author}
  {\bibfnamefont {S.}~\bibnamefont {Haroche}},\ }\href {\doibase
  10.1038/nature07288} {\bibfield  {journal} {\bibinfo  {journal} {Nature}\
  }\textbf {\bibinfo {volume} {455}},\ \bibinfo {pages} {510} (\bibinfo {year}
  {2008})}\BibitemShut {NoStop}%
\bibitem [{\citenamefont {Fink}\ \emph
  {et~al.}(2009{\natexlab{a}})\citenamefont {Fink}, \citenamefont {Bianchetti},
  \citenamefont {Baur}, \citenamefont {G\"oppl}, \citenamefont {Steffen},
  \citenamefont {Filipp}, \citenamefont {Leek}, \citenamefont {Blais},\ and\
  \citenamefont {Wallraff}}]{Wallraff_2009}%
  \BibitemOpen
  \bibfield  {author} {\bibinfo {author} {\bibfnamefont {J.~M.}\ \bibnamefont
  {Fink}}, \bibinfo {author} {\bibfnamefont {R.}~\bibnamefont {Bianchetti}},
  \bibinfo {author} {\bibfnamefont {M.}~\bibnamefont {Baur}}, \bibinfo {author}
  {\bibfnamefont {M.}~\bibnamefont {G\"oppl}}, \bibinfo {author} {\bibfnamefont
  {L.}~\bibnamefont {Steffen}}, \bibinfo {author} {\bibfnamefont
  {S.}~\bibnamefont {Filipp}}, \bibinfo {author} {\bibfnamefont {P.~J.}\
  \bibnamefont {Leek}}, \bibinfo {author} {\bibfnamefont {A.}~\bibnamefont
  {Blais}}, \ and\ \bibinfo {author} {\bibfnamefont {A.}~\bibnamefont
  {Wallraff}},\ }\href {\doibase 10.1103/PhysRevLett.103.083601} {\bibfield
  {journal} {\bibinfo  {journal} {Phys. Rev. Lett.}\ }\textbf {\bibinfo
  {volume} {103}},\ \bibinfo {pages} {083601} (\bibinfo {year}
  {2009}{\natexlab{a}})}\BibitemShut {NoStop}%
\bibitem [{\citenamefont {Viennot}\ \emph {et~al.}(2018)\citenamefont
  {Viennot}, \citenamefont {Ma},\ and\ \citenamefont {Lehnert}}]{Viennot_2018}%
  \BibitemOpen
  \bibfield  {author} {\bibinfo {author} {\bibfnamefont {J.~J.}\ \bibnamefont
  {Viennot}}, \bibinfo {author} {\bibfnamefont {X.}~\bibnamefont {Ma}}, \ and\
  \bibinfo {author} {\bibfnamefont {K.~W.}\ \bibnamefont {Lehnert}},\ }\href
  {\doibase 10.1103/PhysRevLett.121.183601} {\bibfield  {journal} {\bibinfo
  {journal} {Phys. Rev. Lett.}\ }\textbf {\bibinfo {volume} {121}},\ \bibinfo
  {pages} {183601} (\bibinfo {year} {2018})}\BibitemShut {NoStop}%
\bibitem [{\citenamefont {Fink}\ \emph
  {et~al.}(2009{\natexlab{b}})\citenamefont {Fink}, \citenamefont {Bianchetti},
  \citenamefont {Baur}, \citenamefont {G\"oppl}, \citenamefont {Steffen},
  \citenamefont {Filipp}, \citenamefont {Leek}, \citenamefont {Blais},\ and\
  \citenamefont {Wallraff}}]{Fink2009}%
  \BibitemOpen
  \bibfield  {author} {\bibinfo {author} {\bibfnamefont {J.~M.}\ \bibnamefont
  {Fink}}, \bibinfo {author} {\bibfnamefont {R.}~\bibnamefont {Bianchetti}},
  \bibinfo {author} {\bibfnamefont {M.}~\bibnamefont {Baur}}, \bibinfo {author}
  {\bibfnamefont {M.}~\bibnamefont {G\"oppl}}, \bibinfo {author} {\bibfnamefont
  {L.}~\bibnamefont {Steffen}}, \bibinfo {author} {\bibfnamefont
  {S.}~\bibnamefont {Filipp}}, \bibinfo {author} {\bibfnamefont {P.~J.}\
  \bibnamefont {Leek}}, \bibinfo {author} {\bibfnamefont {A.}~\bibnamefont
  {Blais}}, \ and\ \bibinfo {author} {\bibfnamefont {A.}~\bibnamefont
  {Wallraff}},\ }\href {\doibase 10.1103/PhysRevLett.103.083601} {\bibfield
  {journal} {\bibinfo  {journal} {Phys. Rev. Lett.}\ }\textbf {\bibinfo
  {volume} {103}},\ \bibinfo {pages} {083601} (\bibinfo {year}
  {2009}{\natexlab{b}})}\BibitemShut {NoStop}%
\bibitem [{\citenamefont {Kolkowitz}\ \emph {et~al.}(2012)\citenamefont
  {Kolkowitz}, \citenamefont {Bleszynski~Jayich}, \citenamefont
  {Unterreithmeier}, \citenamefont {Bennett}, \citenamefont {Rabl},
  \citenamefont {Harris},\ and\ \citenamefont {Lukin}}]{Kolkowitz_2012}%
  \BibitemOpen
  \bibfield  {author} {\bibinfo {author} {\bibfnamefont {S.}~\bibnamefont
  {Kolkowitz}}, \bibinfo {author} {\bibfnamefont {A.~C.}\ \bibnamefont
  {Bleszynski~Jayich}}, \bibinfo {author} {\bibfnamefont {Q.~P.}\ \bibnamefont
  {Unterreithmeier}}, \bibinfo {author} {\bibfnamefont {S.~D.}\ \bibnamefont
  {Bennett}}, \bibinfo {author} {\bibfnamefont {P.}~\bibnamefont {Rabl}},
  \bibinfo {author} {\bibfnamefont {J.~G.~E.}\ \bibnamefont {Harris}}, \ and\
  \bibinfo {author} {\bibfnamefont {M.~D.}\ \bibnamefont {Lukin}},\ }\href
  {\doibase 10.1126/science.1216821} {\bibfield  {journal} {\bibinfo  {journal}
  {Science}\ }\textbf {\bibinfo {volume} {335}},\ \bibinfo {pages} {1603}
  (\bibinfo {year} {2012})}\BibitemShut {NoStop}%
\bibitem [{\citenamefont {Aspelmeyer}\ \emph {et~al.}(2014)\citenamefont
  {Aspelmeyer}, \citenamefont {Kippenberg},\ and\ \citenamefont
  {Marquardt}}]{Aspelmeyer_2014}%
  \BibitemOpen
  \bibfield  {author} {\bibinfo {author} {\bibfnamefont {M.}~\bibnamefont
  {Aspelmeyer}}, \bibinfo {author} {\bibfnamefont {T.~J.}\ \bibnamefont
  {Kippenberg}}, \ and\ \bibinfo {author} {\bibfnamefont {F.}~\bibnamefont
  {Marquardt}},\ }\href {\doibase 10.1103/RevModPhys.86.1391} {\bibfield
  {journal} {\bibinfo  {journal} {Rev. Mod. Phys.}\ }\textbf {\bibinfo {volume}
  {86}},\ \bibinfo {pages} {1391} (\bibinfo {year} {2014})}\BibitemShut
  {NoStop}%
\bibitem [{\citenamefont {Gilmore}\ \emph {et~al.}(2017)\citenamefont
  {Gilmore}, \citenamefont {Bohnet}, \citenamefont {Sawyer}, \citenamefont
  {Britton},\ and\ \citenamefont {Bollinger}}]{Gilmore2017}%
  \BibitemOpen
  \bibfield  {author} {\bibinfo {author} {\bibfnamefont {K.~A.}\ \bibnamefont
  {Gilmore}}, \bibinfo {author} {\bibfnamefont {J.~G.}\ \bibnamefont {Bohnet}},
  \bibinfo {author} {\bibfnamefont {B.~C.}\ \bibnamefont {Sawyer}}, \bibinfo
  {author} {\bibfnamefont {J.~W.}\ \bibnamefont {Britton}}, \ and\ \bibinfo
  {author} {\bibfnamefont {J.~J.}\ \bibnamefont {Bollinger}},\ }\href {\doibase
  10.1103/PhysRevLett.118.263602} {\bibfield  {journal} {\bibinfo  {journal}
  {Phys. Rev. Lett.}\ }\textbf {\bibinfo {volume} {118}},\ \bibinfo {pages}
  {263602} (\bibinfo {year} {2017})}\BibitemShut {NoStop}%
\end{thebibliography}%
\appendix

\section{Ideal expectation values}\label{app:IdealExpectationValues}

In this appendix we investigate the sensitivity attainable by measurement of only simple observables. First, we will show how the perturbation of the resource state can be inferred from the quadratures of the cavity field, but not in a way that demonstrates a quantum advantage. Then, we will consider the full time-reversal protocol and show that the second period of atom-light interaction allows the perturbation to be mapped into the spin projection and leads to a quantum-enhancement below the SQL. 

\subsection{Direct measurement}\label{appsub:IdealDirectMeasurement}
We consider here the resource state directly after the perturbation by a coherent displacement. As discussed in the main text, the perturbed state is
\begin{equation}
    \ket{\psi_{\beta}}=\mathcal{D}(\beta)e^{-i\chi t\hat{S}_z\hat{a}^{\dagger}\hat{a}}\ket{\psi_0},
\end{equation}
where $\ket{\psi_0} = \ket{\alpha} \otimes \ket{N/2_x}$ is the initial state, with $\alpha \in \mathbb{R}$. The initial entangling evolution which creates the resource state is described by $e^{-i\chi t\hat{S}_z\hat{a}^{\dagger}\hat{a}}$, and $\mathcal{D}(\beta)$ implements the displacement we are trying to infer. For clarity, we take $\beta\in\mathds{R}$.

First, we will demonstrate that the spin observables are completely insensitive probes of the perturbaton. To be general, let us consider an arbitrary spin operator $\hat{O}_s$. Then
\begin{equation}
    \braket{\hat{O}_s}_{\beta}\equiv\bra{\psi_0}e^{i\chi t\hat{S}_z\hat{a}^{\dagger}\hat{a}}\mathcal{D}(\beta)^{\dagger}\hat{O}_s\,\mathcal{D}(\beta)e^{-i\chi t\hat{S}_z\hat{a}^{\dagger}\hat{a}}\ket{\psi_0},
\end{equation}
where $\braket{...}_{\beta}$ indicates the expectation value is taken after the perturbation $\beta$ has occurred. Since $\hat{O}_s$ is a purely spin operator, it commutes with $\mathcal{D}(\beta)$, which is constructed out of only bosonic operators. Hence
\begin{equation}
    \braket{\hat{O}_s}_{\beta}=\bra{\psi_0}e^{i\chi t\hat{S}_z\hat{a}^{\dagger}\hat{a}}\hat{O}_s\,e^{-i\chi t\hat{S}_z\hat{a}^{\dagger}\hat{a}}\ket{\psi_0}.
\end{equation}
Thus, no information about $\beta$ is presented in the spin observables. 

Next, we instead consider the result of measuring a quadrature of the cavity field. To be general, we will consider an arbitrary quadrature 
\begin{equation}
    \hat{X}^{\phi}=\hat{a}e^{i\phi}+\hat{a}^{\dagger}e^{-i\phi}
\end{equation}
characterized by the phase $\phi$.

For simplicity, we perform the calculation in the Heisenberg picture. Thus the operator $\hat{a}$ evolves as:
\begin{align}
    \begin{split}
        \hat{a}&\to e^{i\chi t\hat{S}_z\hat{a}^{\dagger}\hat{a}}\mathcal{D}(\beta)^{\dagger}\hat{a}\,\mathcal{D}(\beta)e^{-i\chi t\hat{S}_z\hat{a}^{\dagger}\hat{a}}\\[8pt]
        &=e^{i\chi t\hat{S}_z\hat{a}^{\dagger}\hat{a}}(\hat{a}+\beta)e^{-i\chi t\hat{S}_z\hat{a}^{\dagger}\hat{a}}\\[8pt]
        &=\hat{a}e^{-i\chi t\hat{S}_z}+\beta.
    \end{split}
\end{align}
Noting that $\bra{N/2_x}e^{-i\chi t\hat{S}_z}\ket{N/2_x}=\cos(\chi t/2)^{N}$, we proceed to compute the following expectation values
\begin{align}
    \begin{split}
        \braket{\hat{a}}_{\beta}&=\alpha \cos(\chi t/2)^N+\beta\\[8pt]
        \braket{\hat{a}^2}_{\beta}&=\alpha^2\cos(\chi t)^N+2\alpha\beta\cos(\chi t/2)^N+\beta^2\\[8pt]
        \braket{\hat{a}^{\dagger}\hat{a}}_{\beta}&=\alpha^2+2\alpha\beta\cos(\chi t/2)^N+\beta^2.
    \end{split}
\end{align}
With this, we can evaluate the relevant expectations of the quadrature $\hat{X}^{\phi}$:
\begin{align}\begin{split}
    \braket{\hat{X}^{\phi}}_{\beta}&=\alpha\cos(\chi t/2)^N\cos(\phi)+\beta\cos(\phi)\\[8pt]
    \braket{(\Delta\hat{X}^{\phi})^2}&=1+2\alpha^2\Big\{\big[\cos(\chi t)^N-\cos(\chi t/2)^{2N}\big]\cos(2\phi)\\[5pt]
    &\hspace{2cm}+1-\cos(\chi t/2)^{2N}\Big\}\\[5pt]
    &\approx 1+\frac{N\alpha^2\chi^2t^2\sin(\phi)^2}{2}\geq 1,
\end{split}\end{align}
The last line demonstrates that the variance is never reduced below the level of vacuum noise [$\braket{(\Delta\hat{X}^{\phi})^2} = 1$]. Moreover, the signal $\langle \hat{X}^{\phi} \rangle_{\beta}$ does not demonstrate an amplified response to the perturbation scaling with atom number $N$ or the initial cavity amplitude $\vert \alpha \vert$. Specifically, 
\begin{equation}
    \partial_{\beta}\braket{\hat{X}^{\phi}}_{\beta}=\cos(\phi).
\end{equation}
As a result, measuring a cavity quadrature does not give an enhanced sensitivity with respect to the SQL: i.e. $(\delta\beta)^2 = \langle (\Delta\hat{X}^{\phi})^2\rangle/\vert \partial\langle \hat{X}^{\phi} \rangle/\partial\beta \vert^2 \geq 1$ for any $\phi$ and $t$.

\subsection{Time reversal protocol}\label{appsub:IdealTimeReversal}
We now move on to the case of the full time-reversal sequence and present a detailed derivation of the achievable sensitivity and related expectation values in the absence of any dissipation. The results are pertinent to the discussion of Sec.~\ref{sec:IdealProtocol}. 

We briefly recap that the evolution in the time-reversal protocol is composed of three key steps: (i) evolution with the dispersive interaction $\hat{H} = \chi \hat{a}^{\dagger}\hat{a}\hat{S}_z$ [Eq.~(\ref{eqn:Hal})] for a time $\tau$, (ii) a coherent displacement of amplitude $\beta$ of the cavity, and (iii) reversed evolution with Hamiltonian $-\hat{H}$ for another duration $\tau$. Collectively, the evolution in the Schrodinger picture is described by $\vert \psi_{2\tau} \rangle = \hat{U}\vert \psi_0 \rangle$ with the unitary operator
\begin{equation}
    \hat{U}=e^{i\chi\hat{S}_z\hat{a}^{\dagger}\hat{a}\, \tau}\mathcal{D}(\beta)e^{-i\chi\hat{S}_z\hat{a}^{\dagger}\hat{a}\, \tau}, \label{eqn:Uoperator}
\end{equation}
where $\mathcal{D}(\beta)=e^{\beta(\hat{a}^{\dagger}-\hat{a})}$, with $\beta\in\mathds{R}$ is a displacement in the $x$ direction in bosonic phase space. It is easier in this instance to evaluate the dynamics in the protocol in the Heisenberg picture, where the evolution of an operator $\hat{O}$ is given by $\hat{O}(t) = \hat{U}^{\dagger}\hat{O}(0)\hat{U}$. To characterize the achievable sensitivity $(\delta\beta)^2$ we only require to compute observables of the form $\langle (\hat{S}_x)^m \rangle$ and $\langle (\hat{S}_y)^m \rangle$ for $m=1,2$, and so it is sufficient to evaluate only the evolution of the operator $(\hat{S}^+)^m$. Lastly, we note that any function of $\hat{S}_z$ (such as $\hat{S}^+\hat{S}^-$) commutes with $\hat{U}$, and thus will not change under time evolution. 

Breaking apart the unitary evolution described by Eq.~(\ref{eqn:Uoperator}), we begin by applying the reverse evolution [step (iii)]:
\begin{align}\begin{split}
    (\hat{S}^+)^m(\tau)&\equiv e^{-i\chi\hat{S}_z \hat{a}^{\dagger}\hat{a}\, \tau}(\hat{S}^+)^me^{i\chi\hat{S}_z \hat{a}^{\dagger}\hat{a}\, \tau}\\[5pt]
    &=(\hat{S}^+)^m e^{-i\chi m\hat{a}^{\dagger}\hat{a}\,\tau}\\[5pt]
    &=(\hat{S}^+)^m: e^{(e^{-i\chi m \tau}-1)\hat{a}^{\dagger}\hat{a}}:\,,
\end{split}\end{align}
where we have used $\hat{S}^+f(\hat{S}_z)=f(\hat{S}_z-1)\hat{S}^+$ and the well known relation $(1+x)^{\hat{a}^{\dagger}\hat{a}}=:e^{x\hat{a}^{\dagger}\hat{a}}:$ to leave the result in a form that will be useful in the remaining steps. Next, we act the displacement operator [step (ii)]
\begin{align}\begin{split}
    (\hat{S}^+)^m(\tau)'&\equiv \mathcal{D}(\beta)^{\dagger}[(\hat{S}^+)^m(\tau)]\mathcal{D}(\beta)\\[5pt]
    &=\mathcal{D}(\beta)^{\dagger}\Big[(\hat{S}^+)^m: e^{(e^{-i\chi m \tau}-1)\hat{a}^{\dagger}\hat{a}}:\Big]\mathcal{D}(\beta)\\[5pt]
    &=(\hat{S}^+)^m: e^{(e^{-i\chi m \tau}-1)(\hat{a}^{\dagger}+\beta)(\hat{a}+\beta)}: .
\end{split}\end{align}

Finally, to evaluate the initial evolution [step (i)] we use the results
\begin{align}
    \begin{split}
        e^{i\chi\hat{S}_z \hat{a}^{\dagger}\hat{a}\, \tau}(\hat{S}^+)^m e^{-i\chi\hat{S}_z \hat{a}^{\dagger}\hat{a}\, \tau}&=(\hat{S}^+)^m e^{i\chi m\hat{a}^{\dagger}\hat{a} \tau}\\[5pt]
        e^{i\chi\hat{S}_z \hat{a}^{\dagger}\hat{a}\, \tau}\hat{a}e^{-i\chi\hat{S}_z \hat{a}^{\dagger}\hat{a}\, \tau}&=\hat{a} e^{-i\chi \hat{S}_z \tau},
    \end{split}
\end{align}
so that
\begin{align}\begin{split}
    (\hat{S}^+)^m(2\tau)&\equiv  e^{i\chi\hat{S}_z \hat{a}^{\dagger}\hat{a}\, \tau}\big[(\hat{S}^+)^m(\tau)'\big] e^{-i\chi\hat{S}_z \hat{a}^{\dagger}\hat{a}\, \tau}\\[5pt]
    &=(\hat{S}^+)^m e^{i\chi m \hat{a}^{\dagger}\hat{a} \tau}\\[5pt]
    &*: e^{(e^{-i\chi m \tau}-1)(\hat{a}^{\dagger} e^{i\chi \hat{S}_z \tau}+\beta)(\hat{a} e^{-i\chi \hat{S}_z \tau}+\beta)}: .
\end{split}\end{align}

Expectation values of this operator can then be taken with respect to the initial state $\vert\psi_0\rangle = \vert N/2_x\rangle \otimes \vert \alpha \rangle$ [Eq.~(\ref{eqn:IniState})]. This is done in two stages: We evaluate the expectation value with respect to $\ket{\alpha}$ and then the spin degree of freedom. 

First, using the fact $\bra{\alpha}e^{i\chi m \hat{a}^{\dagger}\hat{a} \tau}=\bra{\alpha e^{-i\chi m \tau}}$ we find that we need to evaluate the following expression
\begin{align}\begin{split}
    &\bra{\alpha e^{-i\chi m \tau}}: e^{(e^{-i\chi m \tau}-1)(\hat{a}^{\dagger} e^{i\chi \hat{S}_z \tau}+\beta)(\hat{a} e^{-i\chi \hat{S}_z \tau}+\beta)}:\ket{\alpha}\\[5pt]
    &= e^{(e^{-i\chi m \tau}-1)(\alpha e^{im \chi \tau}e^{i\chi \hat{S}_z \tau}+\beta)(\alpha e^{-i\chi \hat{S}_z \tau}+\beta)} e^{\alpha^2(e^{i\chi m \tau}-1)}\\[5pt]
    &=e^{\beta^2(e^{-i\chi m \tau}-1)}\\[5pt]
    &\hspace{1cm}*\exp\Big[-4i\alpha\beta \sin(\chi m \tau/2)\cos(\chi\hat{S}_z \tau+\chi m \tau/2)\Big]
\end{split}\end{align}
where we have used the relation for normal ordered expressions $\bra{\gamma}:f(a^{\dagger},a):\ket{\alpha}=f(\gamma^*,\alpha)$ and computed the overlap $\braket{\alpha e^{-i\chi m \tau}|\alpha}=e^{\alpha^2(e^{i\chi m \tau}-1)}$. Defining $g(\tau)=-4\alpha\beta\sin(\chi m \tau/2)$ we can perform a Jacobi-Anger expansion of the exponential
\begin{equation}
    e^{ig(\tau)\cos(\chi\hat{S}_z \tau+\chi m \tau/2)}=\sum_{n}i^n J_n\big[g(\tau)\big]e^{in\chi(\hat{S}_z+m/2)\tau},
\end{equation}
where $J_n$ is the nth Bessel function of the first kind. Thus to evaluate the expectation with respect to state $\ket{N/2_x}$, we have to calculate 
\begin{align}\begin{split}
    &\bra{N/2_x}(\hat{S}^+)^m e^{in\chi \tau\hat{S}_z}\ket{N/2_x}\\[5pt]
    &=\braket{(\hat{S}^+)^m}_0\cos(n\chi \tau/2)^{N-m} e^{-i n m \chi \tau/2},
\end{split}\end{align}
where $\braket{(\hat{S}^+)^m}_0=\bra{N/2_x}(\hat{S}^+)^m\ket{N/2_x}$. Putting all the pieces together, we find that
\begin{align}\begin{split}\label{AppA:IdealExpectationValues}
    \braket{(\hat{S}^+)^m(2\tau)}&=\braket{(\hat{S}^+)^m}_0e^{\beta^2(e^{-i\chi m \tau}-1)}\\[5pt]
    &\hspace{1cm}*\sum_n i^n J_n\big[g(\tau)\big]\cos(n\chi \tau/2)^{N-m},
\end{split}\end{align}
By taking $m=1,2$ and using the initial conditions $\braket{\hat{S}^+}_0=N/2$ and $\braket{(\hat{S}^+)^2}_0=N(N-1)/4$ we can get the relevant expectation values needed to compute the sensitivity:

\begin{widetext}
\begin{align}\begin{split}\label{eqn:plen}
    \braket{\hat{S}^+(2\tau)}&=\frac{N}{2}\exp\Big[(e^{-i\chi \tau}-1)\beta^2\Big]\sum_{n=-\infty}^{\infty}i^nJ_n\Big(-4\alpha\beta \sin(\chi \tau/2)\Big)\Big[\cos\Big(\frac{n\chi \tau}{2}\Big)\Big]^{N-1},\\
    \braket{\hat{S}^{+2}(2\tau)}&=\frac{N(N-1)}{4}\exp\Big[(e^{-i2\chi \tau}-1)\beta^2\Big]\sum_{n=-\infty}^{\infty}i^nJ_n\Big(-4\alpha\beta \sin(\chi \tau)\Big)\Big[\cos\Big(\frac{n\chi \tau}{2}\Big)\Big]^{N-2},\\
    \braket{(\hat{S}^+\hat{S}^-)(2\tau)}&=\frac{N(N+1)}{4} .
\end{split}\end{align}
\end{widetext}

\section{Resource state in presence of photon loss}\label{app:ResourceStateinCavity}
In this appendix we present a more detailed treatment of the dispersive dynamics in the presence of photon loss. In particular, we focus on the generated state after the first period of atom-light interaction, Eq.~(\ref{eqn:FirstStepState}).

As the dissipative dynamics will generate a mixed state in general, we begin our calculation from the density matrix representing the initial state $\vert \psi_0 \rangle$ [Eq.~(\ref{eqn:IniState})],
\begin{equation}\label{eqn:AppRhoZero}
    \hat{\rho}_0= |N/2_x\rangle\langle N/2_x|\otimes |\alpha\rangle\langle \alpha|.
\end{equation}
Evolution is described by the Liouvillian
\begin{equation}\label{eqn:AppLiouv}
  \partial_t\hat{\rho}_t=-i \chi\Big[\hat{a}^{\dagger}\hat{a}\hat{S}_z,\hat{\rho}_t\Big]+\kappa \Big(\hat{a}\hat{\rho}_t\hat{a}^{\dagger}-\frac{\{\hat{a}^{\dagger}\hat{a},\hat{\rho}_t\}}{2}\Big)\equiv   \mathcal{L}_{\chi}\hat{\rho}_t .
\end{equation}

To simplify the treatment of the evolution, we perform a change of frame and define the transformed density operator as 
\begin{equation}
    \hat{\xi}_t=e^{i \hat{H} t}e^{\frac{\kappa \hat{a}^{\dagger}\hat{a} t}{2}}\,\hat{\rho}_t\,e^{-i\hat{H} t}e^{\frac{\kappa \hat{a}^{\dagger}\hat{a} t}{2}} ,
\end{equation}
where $\hat{H}=\chi \hat{a}^{\dagger}\hat{a}\hat{S}_z$. This transformation is designed to strip away the commutator and anticommutator parts of $\mathcal{L}_{\chi}$. Some straightforward yet tedious manipulations using the identity $\hat{a}g(\hat{a}^{\dagger}\hat{a})=g(\hat{a}^{\dagger}\hat{a}+1)\hat{a}$, where $g$ is any function, then result in a simplified equation of motion for $\hat{\xi}_t$:
\begin{equation}\label{eqn:AppCSimplEqn}
    \partial_t\hat{\xi}_t=\kappa e^{-\kappa t}e^{-i\chi\hat{S}_z t}\hat{a}\,\hat{\xi}_t\hat{a}^{\dagger}e^{i\chi\hat{S}_z t},
\end{equation}
with the the initial condition $\hat{\xi}_0=\hat{\rho}_0$. 

The superoperator acting on $\hat{\rho}$ in the right hand side of Eq.~(\ref{eqn:AppCSimplEqn}) leaves bosonic coherent states invariant. Furthermore, the initial condition of the system involves precisely a coherent state factor. This suggests an ansatz for the density matrix which is a tensor product: $\hat{\xi}_t=\hat{\xi}^{\text{spin}}_t\otimes|\alpha\rangle\langle\alpha|$. Substitution of this ansatz into Eq.~(\ref{eqn:AppCSimplEqn}) then yields an equation for the spin degree of freedom
\begin{equation}
     \partial_t\hat{\xi}^{\text{spin}}_t=\kappa\alpha^2 e^{-\kappa t}e^{-i\chi\hat{S}_z t}\hat{\xi}^{\text{spin}}_te^{i\chi\hat{S}_z t} . \label{eqn:AppCspinEqn}
\end{equation}
If $\hat{\xi}^{\text{spin}}_t$ is expanded in the $\hat{S}_z$ basis, $\hat{S}_z \vert m_z \rangle = m \vert m_z \rangle$, an analytic form for $\hat{\xi}^{\text{spin}}_t$ can be obtained in a straightforward manner. In particular, we can express $\hat{\xi}^{\text{spin}}_t=\sum_{m,n}d_{mn}(t)\ketbra{m}{n}$, and  inserting this into Eq.~(\ref{eqn:AppCspinEqn}) leads to decoupled equations for each coefficient $d_{mn}$:
\begin{equation}
    \partial_td_{mn}=\kappa\alpha^2 e^{-\kappa t}e^{i\chi(n-m) t}d_{mn} . 
\end{equation}
These can be solved exactly,
\begin{equation}
    d_{mn}(t)=\exp\bigg[\frac{\kappa\alpha^2\big(1-e^{-\kappa t+i\chi(n-m)t}\big)}{\kappa-i\chi(n-m)}\bigg]c_mc_n^*,
\end{equation}
where we have used the initial condition $\hat{\xi}_0^{\text{spin}}=\ketbra{N/2_x}{N/2_x}$ and expressed the state in the $\hat{S}_z$ basis, i.e. $\ket{N/2_x}=\sum_{m}c_m\ket{m_z}$.

Transforming back to the original frame, the solution $\hat{\xi}_t$ yields an analytic form for the density matrix of the complete atom-light system:
\begin{widetext}
\begin{align}
    \begin{split}
         \hat{\rho}_t&=e^{-i \hat{H} t}e^{-\frac{\kappa \hat{a}^{\dagger}\hat{a} t}{2}}\bigg(\sum_{m,n}d_{mn}(t)\ketbra{m}{n}\otimes\ketbra{\alpha}{\alpha}\bigg)
         e^{i \hat{H} t}e^{-\frac{\kappa \hat{a}^{\dagger}\hat{a} t}{2}}\\[5pt]
         &=e^{-i\hat{H}t}\Bigg(\sum_{m,n}d_{mn}(t)\ketbra{m}{n}
         \otimes e^{\kappa\alpha^2(e^{-\kappa t}-1)}\ketbra{\alpha e^{-\kappa t/2}}{\alpha e^{-\kappa t/2}}\Bigg)e^{i\hat{H}t},
    \end{split}
\end{align}
\end{widetext}
where we have used $e^{-\frac{\kappa \hat{a}^{\dagger}\hat{a} t}{2}}\ket{\alpha}=e^{\frac{\kappa\alpha^2(e^{-\kappa t}-1)}{2}}\ket{\alpha e^{-\kappa t/2}}$. A more compact form of the final result is then
\begin{equation}\label{eqn:AppFirstStepState}
    \hat{\rho}_t=e^{-i\hat{H}t}\bigg(\hat{\rho}_t^{\text{spin}}\otimes\ketbra{\alpha e^{-\kappa t/2}}{\alpha e^{-\kappa t/2}}\bigg)e^{i\hat{H} t},
\end{equation}
with
\begin{align}
    \begin{split}\label{eqn:AppFirstStetStateb}
        \hat{\rho}_t^{\text{spin}}&=\sum_{m,n}c_n^*c_m e^{f(n-m,t)}\ketbra{m}{n},\\[5pt]
        f(z,t)&=\frac{\kappa\alpha^2(1-e^{-\kappa t+i\chi z t})}{\kappa-i\chi z}+\alpha^2(e^{-\kappa t}-1)
    \end{split}
\end{align}
which are used in Sec.~\ref{sec:CavityLeakage} of the main text.

\section{Quantum Fisher information with photon loss}\label{app:FisherInfoPhotonLoss}
Having derived the complete density matrix of the atom-light state generated by the dispersive interaction in the presence of photon loss [Eqs.~(\ref{eqn:FirstStepState}) and (\ref{eqn:AppFirstStepState})], we can use this result to evaluate the quantum Fisher information and establish the metrological utility of the state. 

The quantum Fisher information of a generic quantum state $\hat{\rho}$ can be computed via the expression \cite{Braunstein1994}:
\begin{equation}
\mathcal{F}_{Q}=2\sum_{a\neq b}\frac{(\lambda_a-\lambda_b)^2}{\lambda_a+\lambda_b}|\bra{a}\hat{G}\ket{b}|^2,
\end{equation}
where $\hat{G}$ is the generator of the perturbation,  $\{\ket{a}\}$ are the eigenstates of the density matrix $\hat{\rho}$ and $\{\lambda_a\}$ the corresponding eigenvalues. As we are characterizing the sensitivity of the generated state to small displacements, we take $\hat{G} = \hat{Y} = -i(\hat{a}-\hat{a}^{\dagger})$. 

Computation of $\mathcal{F}_{Q}$ thus requires us to calculate the eigensystem of $\hat{\rho}_t$ and subsequently compute the matrix elements $\bra{a}\hat{Y}\ket{b}$. Given the structure of $\hat{\rho}_t$, we find it convenient to strip away the $e^{\pm i\hat{H} t}$ factors (which amounts to a unitary transformation), so that $\ket{\tilde{a}}\equiv e^{i\hat{H}t}\ket{a}$, $\lambda_a$ remain unchanged and $\bra{a}\hat{Y}\ket{b}\equiv \bra{\tilde{a}}e^{i\hat{H}t}\hat{Y}e^{-i\hat{H}t}\ket{\tilde{b}}$. Under this transformation
\begin{equation}
    e^{i\hat{H}t}\hat{Y}e^{-i\hat{H}t}=-i(\hat{a}e^{-i\chi\hat{S}_zt}-\hat{a}^{\dagger}e^{i\chi\hat{S}_zt}).
\end{equation}
Alternatively, $\ket{\tilde{a}}$ can also be characterized as eigenstates of
\begin{equation}
    \hat{\rho}_t^{\text{spin}}\otimes\ketbra{\alpha e^{-\kappa t/2}}{\alpha e^{-\kappa t/2}},
\end{equation}
where $\hat{\rho}_t^{\text{spin}}$ is defined in Eq.~(\ref{eqn:AppFirstStetStateb}). We highlight that the bosonic component is diagonal in the coherent state basis, and so we expect most of its eigenvalues are zero. This allows us to greatly simplify the calculation of the Fisher information, following the procedure outlined in Appendix \ref{app:SimplifiedFisherInformation}, by picking $P_I$ (as defined in Appendix \ref{app:SimplifiedFisherInformation}) to be $\mathds{1}_{\text{spin}}\otimes \ketbra{\alpha e^{-\kappa t/2}}{\alpha e^{-\kappa t/2}}$. In particular, this feature allows us to reduce the computation to
\begin{equation}\label{eqn:AppFisherSpin}
    \mathcal{F}_{Q}=4+2\alpha^2e^{-\kappa t}\sum_{r,s}\frac{(\lambda_r-\lambda_s)^2}{\lambda_r+\lambda_s}\Big|\bra{r}\hat{O}\ket{s}\Big|^2, 
\end{equation}
where $\{\ket{r}\}$ are now eigenstates of $\hat{\rho}_t^{\text{spin}}$ only, $\{\lambda_r\}$ are their corresponding eigenvalues, and
\begin{equation}
    \hat{O}=-i(e^{-i\chi\hat{S}_zt}-e^{i\chi\hat{S}_zt})
\end{equation}
This is precisely Eq.~(\ref{eqn:FisherSpin}) given in the main text.

A simpler analytic expression can be computed by way of further reasonable approximations. As mentioned in the main text, for $\chi\sqrt{N}\ll \kappa$, $\chi\sqrt{N}t\ll 1$ and large $N$ the matrix elements of $\hat{\rho}_t^{\mathrm{spin}}$ can be approximated as
\begin{align}\begin{split}
   & c_n^*c_m e^{f(n-m,t)}\ketbra{m}{n}\\[5pt]
   & \sim \exp\Big(-\frac{m^2+n^2}{N}+ic_1(n-m)\\&\hspace{3cm}-\sigma^2(m-n)^2\ketbra{m}{n},
\end{split}\end{align}
where we leave the coefficient $c_1$ unspecified because it can be eliminated by doing a rotation about $\hat{S}_z$ while leaving $\hat{O}$ intact (since they commute), and
\begin{equation}
    \sigma^2=\frac{\chi^2\alpha^2 e^{-\kappa t}}{\kappa^2}\Big[e^{\kappa t}-\Big(1+\kappa t+\frac{\kappa^2 t^2}{2}\Big)\Big].
\end{equation}For simplicity, we henceforth omit $c_1$. Under these approximations
\begin{equation}
    \hat{\rho}_t^{\text{spin}} \approx \mathcal{C}\sum_{mn}e^{-\frac{m^2+n^2}{N}-\sigma^2 (m-n)^2}\ketbra{m}{n},
\end{equation}
where $\sigma^2=\kappa \chi^2\alpha^2t^3/6$ and $\mathcal{C}$ is a normalization constant. If $\sigma\ll 1$ then we can approximate the summations by integrals. Furthermore, after a rescaling $m=\sqrt{N}x$, the state of the spin subsystem is given by
\begin{align}\begin{split}
    \hat{\rho}^{\text{spin}}_t&=\mathcal{C}'\int dx\,dy\, e^{-x^2-y^2-\sigma^2N(x-y)^2}\ketbra{x}{y},\\
    &=\mathcal{C}'e^{-\hat{x}^2}\Big(\int dx\,dy\, e^{-\sigma^2N(x-y)^2}\ketbra{x}{y}\Big)e^{-\hat{x}^2} ,\\
    &=\mathcal{C}''e^{-\hat{x}^2}e^{-\hat{p}^2/(4\sigma^2N)}e^{-\hat{x}^2} , \label{eqn:xyMagic}
\end{split}\end{align}
where $\mathcal{C}'$ and $\mathcal{C}'' $ are (distinct) normalization constants. Additionally, $x$ and $y$ are continuous variables defined with respect to the operators $\hat{x}$ and $\hat{p}$: $\hat{x}$ is a position-like operator satisfying $\hat{x}\ket{x'}=x'\ket{x'}$ and $\hat{p}$ is its conjugate momentum. 

The remaining exponential terms in the last line of Eq.~(\ref{eqn:xyMagic}) can be rewritten as
\begin{equation}
    \hat{\rho}_t^{\text{spin}}\propto e^{-(a^2\hat{x}^2+b^2\hat{p}^2)/2},
\end{equation}
where
\begin{align}\label{eqn:AppIntCoeff}
    \begin{split}
        \cosh(a b)&=1+\frac{1}{\sigma^2N},\\
        \frac{b^2}{a^2}&=\frac{1}{4(2\sigma^2N+1)} .
    \end{split}
\end{align}
It is now apparent that $\hat{\rho}_t^{\text{spin}}$ is diagonal in a basis of eigenstates of a harmonic oscillator with frequency $\omega=ab$:
\begin{equation}
    \hat{\rho}_t^{\text{spin}}=(1-e^{-\omega})e^{-\omega\hat{c}^{\dagger}\hat{c}},
\end{equation}
with
\begin{equation}
    \hat{c}=\sqrt{\frac{a}{2b}}\hat{x}+i\sqrt{\frac{b}{2a}}\hat{p} ,
\end{equation}
the associated bosonic annihilation operator. 
The eigenvalues are therefore
\begin{equation}
    \lambda_r=(1-e^{-\omega})e^{-\omega r},
\end{equation}
while the expectation values $\bra{r}\hat{O}\ket{s}$ (where $\ket{r}$ is now indexed by the excitation level of the harmonic oscillator) become, within the approximation $\chi\sqrt{N}t\ll 1$ (which will later shown to be valid as $\kappa$ limits the relevant timescales of interest), 
\begin{align}\begin{split}
   & \bra{r}\big[-i(e^{-i\chi\hat{S}_zt}-e^{i\chi\hat{S}_zt})\big]\ket{s}\\[5pt]
    &\approx \bra{r}\big[-i(e^{-i\chi\sqrt{N}\hat{x}t}-e^{i\chi\sqrt{N}\hat{x}t})\big]\ket{s}\\[5pt]
    &\approx-2\chi\sqrt{N}t\bra{r}\hat{x}\ket{s}\\[5pt]
    &= -\frac{2\chi\sqrt{Nb}t}{\sqrt{2a}}(\sqrt{r}\delta_{r,s+1}+\sqrt{s}\delta_{r+1,s}),
\end{split}\end{align}
where $\delta_{r,s}$ is a Kronecker delta. After squaring we get
\begin{equation}
    |\bra{r}\hat{O}\ket{s}|^2=\frac{2\chi^2Nbt^2}{a}(r\delta_{r,s+1}+s\delta_{r+1,s}) .
\end{equation}
Note that the cross terms have vanished. 

Since the expression for the Fisher information in Eq.~(\ref{eqn:AppFisherSpin}) is symmetric between $r$ and $s$, we just consider one of the Kronecker deltas and double the result. Then, Eq.~(\ref{eqn:AppFisherSpin}) simplifies into
\begin{align}\begin{split}
    \mathcal{F}_{Q}&=4+2\alpha^2e^{-\kappa t}\sum_{s=0}^{\infty}\frac{(\lambda_{s+1}-\lambda_s)^2}{\lambda_{s+1}+\lambda_s}\frac{4\chi^2 N bt^2(s+1)}{a}\\[8pt]
    &=4+\frac{8\chi^2Nb\alpha^2t^2e^{-\kappa t}(1-e^{-\omega })}{a}\\
    &\hspace{3cm}*\sum_{s=0}^{\infty}\frac{(e^{-\omega(s+1)}-e^{-\omega s})^2}{e^{-\omega(s+1)}+e^{-\omega s}}(s+1)\\[8pt]
    &=4+\frac{8\chi^2Nb\alpha^2 t^2e^{-\kappa t}(1-e^{-\omega })^3}{a(1+e^{-\omega })}\sum_{s=0}^{\infty}e^{-\omega s}(s+1)\\[8pt]
    &=4+\frac{8\chi^2Nb\alpha^2 t^2e^{-\kappa t} (1-e^{-w})}{a(1+e^{-w})}.
\end{split}\end{align}
We now plug in $\omega=ab$ and make repeated use of Eq.~(\ref{eqn:AppIntCoeff}) to obtain
\begin{align}\begin{split}
    \mathcal{F}_Q&=4+\frac{4\chi^2N\alpha^2 t^2e^{-\kappa t}}{1+2\sigma^2N},\\[8pt]
    &=4+\frac{4\chi^2N\alpha^2t^2e^{-\kappa t}}{1+\frac{2\chi^2\alpha^2 N}{\kappa^2}\Big[1-e^{-\kappa t}\big(1+\kappa t+\frac{\kappa^2 t^2}{2}\big)\Big]},\\[8pt]
    &\approx4+\frac{4\chi^2N\alpha^2t^2}{1+\kappa\chi^2Nt^3/3}, 
\end{split}\end{align}
where in the last line we have approximated $\kappa t\ll 1$. These are the results used for Eq.~(\ref{eqn:FisherTime}) of the main text.

\section{Sensitivity in the presence of cavity decay}\label{app:SensitivityKappa}
The operational sensitivity $(\delta \beta)^2$ achievable via the full time-reversal protocol and measurement of collective spin observables can also be calculated accounting for photon loss. Different to the prior computation of Fisher information, here the calculation is most simply performed in the Heisenberg picture. In particular, an operator $\hat{O}$ evolves according to:
\begin{equation}
    \hat{O}(\tau_1,\tau_2)=e^{\mathcal{L}^{\dagger}_{\chi}\tau_1}\mathcal{L}^{\dagger}_{\mathcal{D}(\beta)}e^{\mathcal{L}^{\dagger}_{-\chi} \tau_2}\hat{O}(0), \label{eqn:OevoSens}
\end{equation}
where $\mathcal{L}^{\dagger}$ are Hilbert-Schmidt adjoints, defined by
\begin{align}\begin{split}
    \mathcal{L}^{\dagger}_{\chi}\hat{O}&=i\chi\Big[\hat{a}^{\dagger}\hat{a}\hat{S}_z,\hat{O}\Big]+\kappa\Big(\hat{a}^{\dagger}\hat{O}\hat{a}-\frac{\{\hat{a}^{\dagger}\hat{a},\hat{O}\}}{2}\Big),\\[5pt]
    \mathcal{L}^{\dagger}_{\mathcal{D}(\beta)} \hat{O}&=\mathcal{D}^{\dagger}(\beta)\hat{O}\mathcal{D}(\beta).
\end{split}\end{align}
Relevant expectation values are then calculated with respect to $\hat{\psi}_0 = \vert \psi_0 \rangle \langle  \psi_0 \vert$ [Eq.~(\ref{eqn:IniState})]. 

Let us now consider the generic evolution of an operator of the following form: 
\begin{equation}\label{eqn:AppDOpForm}
    \hat{\Lambda}\equiv e^{\mathcal{L}^{\dagger}_{-\chi}\tau}\big[(\hat{S}^+)^m\normord{f}\big],
\end{equation}
where $f$ is any function of bosonic variables and $\normord{\,}$ denotes normal ordering. We consider this form because, as we will see, both the initial and final steps of the protocol fall under this category. Then $\hat{\Lambda}$ satisfies the following equation
\begin{equation}\label{eq:AppDOpMasterEquation}
    \partial_{\tau}\hat{\Lambda}=-i\chi\Big[\hat{a}^{\dagger}\hat{a}\hat{S}_z,\hat{\Lambda}\Big]+\kappa\Big(\hat{a}^{\dagger}\hat{\Lambda}\hat{a}-\frac{\{\hat{a}^{\dagger}\hat{a},\hat{\Lambda}\}}{2}\Big).
\end{equation}
For the sake of clarity, we remark that only $\hat{\Lambda}$ has any $\tau$ dependence. All the other operators in Eq.~\ref{eq:AppDOpMasterEquation} ($\hat{a}$, $\hat{a}^{\dagger}$, $\hat{S}_z$) are $\tau$ independent. We simplify Eq.~(\ref{eq:AppDOpMasterEquation}) by pulling out some factors
\begin{equation}
    \hat{\Xi}= e^{i\chi \hat{a}^{\dagger}\hat{a}\hat{S}_z \tau}e^{\frac{\kappa}{2}\hat{a}^{\dagger}\hat{a}\tau}\,\hat{\Lambda}\,e^{-i\chi \hat{a}^{\dagger}\hat{a}\hat{S}_z \tau}e^{\frac{\kappa}{2}\hat{a}^{\dagger}\hat{a}\tau}
\end{equation}
to obtain
\begin{equation}
    \partial_{\tau}\hat{\Xi}=\kappa e^{\kappa \tau} e^{i\chi\hat{S}_z \tau}\hat{a}^{\dagger}\,\hat{\Xi}\,\hat{a}\,e^{-i\chi\hat{S}_z \tau}
\end{equation}
with initial condition $\hat{\Xi}(0)=(\hat{S}^+)^m\normord{f}$. We try an ansatz for solution of the form
\begin{equation}
    \hat{\Xi}=(\hat{S}^+)^m\normord{\xi},
\end{equation}
where $\xi$ is a purely bosonic operator. This leads us into the following equation for $\normord{\xi}$
\begin{align}
    \begin{split}
        \partial_{\tau}(\normord{\xi})&=\kappa e^{\kappa\tau}e^{i\chi m\tau}\hat{a}^{\dagger}\normord{\xi}\,\hat{a}\\[5pt]
        &=\kappa e^{\kappa\tau}e^{i\chi m\tau}\normord{\hat{a}^{\dagger}\xi\,\hat{a}}\\[5pt]
        &=\kappa e^{\kappa\tau}e^{i\chi m\tau}\normord{\hat{a}^{\dagger}\hat{a}\,\xi},
    \end{split}
\end{align}
from which we find the solution
\begin{equation}
    \normord{\xi}=\,\normord{\exp\bigg[\frac{\kappa(e^{\kappa\tau+i\chi m \tau}-1)}{\kappa+i\chi m}\hat{a}^{\dagger}\hat{a}\bigg]f}\,.
\end{equation}
The final result for $\Xi$ is
\begin{align}\begin{split}\label{eqn:AppXiSol}
    \hat{\Xi}&=(\hat{S}^+)^m\normord{\exp\bigg[\frac{\kappa(e^{\kappa\tau+i\chi m \tau}-1)}{\kappa+i\chi m}\hat{a}^{\dagger}\hat{a}\bigg]f}.
\end{split}\end{align}

We now use this result to evaluate the evolution Eq.~(\ref{eqn:OevoSens}). The action of the first superoperator (corresponding to the \emph{last} step of the protocol) is
\begin{equation}
    (\hat{S}^+)^m(\tau_2)\equiv e^{\mathcal{L}^{\dagger}_{-\chi}\tau_2}\big[(\hat{S}^+)^m\big]\equiv\hat{\Lambda}_1,
\end{equation}
where $\hat{\Lambda}_1$ is of the form of Eq.~(\ref{eqn:AppDOpForm}) with $f=1$. Then
\begin{equation}
     \hat{\Xi}_1=(\hat{S}^+)^m\normord{\exp\bigg[\frac{\kappa(e^{\kappa\tau_2+i\chi m \tau_2}-1)}{\kappa+i\chi m}\hat{a}^{\dagger}\hat{a}\bigg]},
\end{equation}
and
\begin{align}
    \begin{split}
        \hat{\Lambda}_1&=\hat{\Xi}_1 \,e^{-(\kappa+i\chi m)\hat{a}^{\dagger}\hat{a}\tau_2}\\
        &=(\hat{S}^+)^m\,\normord{\exp\bigg[\frac{\kappa(e^{\kappa\tau_2+i\chi m \tau_2}-1)}{\kappa+i\chi m}\hat{a}^{\dagger}\hat{a}\bigg]}\,e^{-(\kappa+i\chi m)\hat{a}^{\dagger}\hat{a}\tau_2}\\[8pt]
    &=(\hat{S}^+)^m
    \bigg[\frac{\kappa(e^{\kappa\tau_2+i\chi m \tau_2}-1)}{\kappa+i\chi m}+1\bigg]^{\hat{a}^{\dagger}\hat{a}}\,e^{-(\kappa+i\chi m)\hat{a}^{\dagger}\hat{a}\tau_2}\\[8pt]
    &=(\hat{S}^+)^m\bigg[\frac{i\chi n(e^{-(\kappa+i\chi m)\tau_2}-1)}{\kappa+i\chi m}+1\bigg]^{\hat{a}^{\dagger}\hat{a}}\\[8pt]
    &=(\hat{S}^+)^m\,\normord{\exp\bigg[\frac{i\chi m(e^{-\kappa\tau_2-i\chi m\tau_2}-1)}{\kappa+i\chi m}\hat{a}^{\dagger}\hat{a}\bigg]}\\[8pt]
    &\equiv(\hat{S}^+)^m\normord{e^{\eta_{\tau_2,m}\hat{a}^{\dagger}\hat{a}}},
    \end{split}
\end{align}
where we have defined 
\begin{equation}
    \eta_{\tau,m}=\frac{i\chi m}{\kappa+i\chi m}(e^{-\kappa\tau-i\chi m\tau}-1),
\end{equation}
and have made repeated use of the identity $\normord{e^{(z-1)\hat{a}^{\dagger}\hat{a}}}=z^{\hat{a}^{\dagger}\hat{a}}$. 

Next, the displacement is effectuated by use of the identity $\mathcal{D}^{\dagger}(\beta) \hat{a} \mathcal{D}\beta) = \hat{a} + \beta$. Thus
\begin{equation}
    \mathcal{L}^{\dagger}_{\mathcal{D}(\beta)}[(\hat{S}^+(\tau_2))^m]=(\hat{S}^+)^m\normord{e^{\eta_{\tau,n}(\hat{a}^{\dagger}+\beta)(\hat{a}+\beta)}}
\end{equation}
Note that it is still of the form of Eq.~(\ref{eqn:AppDOpForm}). This allows us to again use the result of Eq.~(\ref{eqn:AppXiSol}) with
\begin{equation}
    f=\exp\Big[\eta_{\tau,n}(\hat{a}^{\dagger}+\beta)(\hat{a}+\beta)\Big]
\end{equation}
for the third evolution superoperator ($e^{\mathcal{L}_{\chi}^{\dagger}\tau_1}$) but with different $\hat{\Xi}$ and $\hat{\Lambda}$, which we denote $\hat{\Xi}_2$ and $\hat{\Lambda}_2$. Note that $\hat{\Lambda}_2$ is precisely the $(\hat{S}^+)^m(\tau_1,\tau_2)$ we are interested in. Taking into account that $-\chi\rightarrow\chi$ and $\tau\rightarrow \tau_1$, we can write the result directly:
\begin{align}
    \begin{split}
       \Xi_2&=(\hat{S}^+)^m\normord{\exp\bigg[\eta_{\tau_2,m}(\hat{a}^{\dagger}+\beta)(\hat{a}+\beta)\\[5pt]
       &\hspace{3cm}+\frac{\kappa(e^{\kappa \tau_1-i\chi m \tau_1}-1)}{\kappa-i\chi m}\hat{a}^{\dagger}\hat{a}\bigg]}\\[8pt]
       &=(\hat{S}^+)^me^{\beta^2\eta_{\tau_2,m}}e^{\eta_{\tau_2,m}\beta\hat{a}^{\dagger}}\normord{\exp\bigg[\eta_{\tau_2,m}\hat{a}^{\dagger}\hat{a}\\[5pt]
       &\hspace{0.2cm}+\frac{\kappa(e^{\kappa \tau_1-i\chi m \tau_1}-1)}{\kappa-i\chi m}\hat{a}^{\dagger}\hat{a}\bigg]}\,e^{\eta_{\tau_2,m}\beta\hat{a}}
    \end{split}
\end{align}
With $\hat{\Xi}_2$, $ (\hat{S}^+)^m(\tau_1,\tau_2)$ be computed using
\begin{equation}
    (\hat{S}^+)^m(\tau_1,\tau_2)= e^{i\chi \hat{a}^{\dagger}\hat{a}\hat{S}_z \tau_1}e^{-\frac{\kappa}{2}\hat{a}^{\dagger}\hat{a}\tau_1}\,\hat{\Xi}_2\,e^{-i\chi \hat{a}^{\dagger}\hat{a}\hat{S}_z \tau_1}e^{-\frac{\kappa}{2}\hat{a}^{\dagger}\hat{a}\tau_1},
\end{equation}
which results, after some simplifications, in
\begin{widetext}
\begin{align}\label{eqn:AppSfinal}
    \begin{split}
    (\hat{S}^+)^m(\tau_1,\tau_2)&=(\hat{S}^+)^m\exp\Big[\eta_{\tau_2,m}\beta\hat{a}^{\dagger}e^{i\chi(\hat{S}_z+m)\tau_1-\kappa \tau_1/2}\Big] e^{-(\kappa-i\chi m)\hat{a}^{\dagger}\hat{a}\tau_1} \normord{\exp\bigg[\eta_{\tau_2,m}\hat{a}^{\dagger}\hat{a}+\frac{\kappa(e^{\kappa \tau_1-i\chi m \tau_1}-1)}{\kappa-i\chi m}\hat{a}^{\dagger}\hat{a}\bigg]}\\
      &\hspace{0.5cm}*\exp\Big[\eta_{\tau_2,m}\beta\hat{a}e^{-i\chi\hat{S}_z\tau_1-\kappa \tau_1/2}\Big]e^{\beta^2\eta_{\tau_2,m}} ,\\[15pt]
        &=(\hat{S}^+)^m\exp\Big[\eta_{\tau_2,m}\beta\hat{a}^{\dagger}e^{i\chi(\hat{S}_z+m)\tau_1-\kappa \tau_1/2}\Big] \normord{\exp\bigg[(\eta_{\tau_2,m}e^{-(\kappa-i\chi m)\tau_1}+\eta_{\tau_1,m}^*)\hat{a}^{\dagger}\hat{a}\bigg]}\\
        &\hspace{0.5cm}*\exp\Big[\eta_{\tau_2,m}\beta\hat{a}e^{-i\chi\hat{S}_z\tau_1-\kappa \tau_1/2}\Big]e^{\beta^2\eta_{\tau_2,m}} ,\\[15pt]
        &=(\hat{S}^+)^m\normord{\exp\bigg[\big(\eta_{\tau_2,m}e^{-\kappa \tau_1+i\chi m\tau_1}+\eta_{\tau_1,m}^*\big)\hat{a}^{\dagger}\hat{a}+\eta_{\tau_2,m}\beta\hat{a}^{\dagger}e^{i\chi(\hat{S}_z+m)\tau_1-\kappa \tau_1/2}\\
        &\hspace{9cm}+\eta_{\tau_2,m}\beta\hat{a}e^{-i\chi\hat{S}_z\tau_1-\kappa \tau_1/2}+\beta^2\eta_{\tau_2,m}\bigg]} ,\\[15pt]
        &=(\hat{S}^+)^m\normord{\exp\bigg[\big(\eta_{\tau_2,m}e^{-\kappa \tau_1+i\chi m\tau_1}+\eta_{\tau_1,m}^*\big)\hat{a}^{\dagger}\hat{a}\\
        &\hspace{5cm}+\eta_{\tau_2,m}\beta e^{-\frac{\kappa \tau_1}{2}+\frac{i\chi m \tau_1}{2}}\big(\hat{a}^{\dagger}e^{i\chi(\hat{S}_z+\frac{m}{2})\tau_1}+\hat{a}e^{-i\chi(\hat{S}_z+\frac{m}{2})\tau_1}\big)+\beta^2\eta_{\tau_2,m}\bigg]}.
    \end{split}
\end{align}
\end{widetext}
Evaluating the expectation value of $\hat{S}^M(\tau_1,\tau_2)$ with respect to a photon coherent state then amounts to the replacements $\hat{a}\to\alpha$ and $\hat{a}^{\dagger}\to\alpha^*$ in the last line of Eq.~(\ref{eqn:AppSfinal}) since the expression is normal ordered. The relevant expectation values are then:

\begin{widetext}
\begin{align}\begin{split}\label{eqn:KappaExpectationValues}
    \braket{\hat{S}^+(\tau_1,\tau_2)}&=\frac{N}{2}\exp\Big[\alpha^2(\eta_{\tau_2,1}e^{-\kappa \tau_1+i\chi \tau_1}+\eta_{\tau_1,1}^*)+\eta_{\tau_2,1}\beta^2\Big]\\
    &\hspace{5cm}*\sum_{n=-\infty}^{\infty}i^nJ_n\Big(-2i\eta_{\tau_2,1}\alpha\beta e^{-\frac{\kappa}{2}\tau_1+i\frac{\chi}{2}\tau_1}\Big)\Big[\cos\Big(\frac{n\chi \tau_1}{2}\Big)\Big]^{N-1},\\[5pt]
    \braket{\hat{S}^{+2}(\tau_1,\tau_2)}&=\frac{N(N-1)}{4}\exp\Big[\alpha^2(\eta_{\tau_2,2}e^{-\kappa \tau_1+i2\chi \tau_1}+\eta_{\tau_1,2}^*)+\eta_{\tau_2,2}\beta^2\Big]\\
    &\hspace{5cm}*\sum_{n=-\infty}^{\infty}i^nJ_n\Big(-2i\eta_{\tau_2,2}\alpha\beta e^{-\frac{\kappa}{2}\tau_1+i\chi \tau_1}\Big)\Big[\cos\Big(\frac{n\chi  \tau_1}{2}\Big)\Big]^{N-2},\\[5pt]
    \braket{(\hat{S}^+\hat{S}^-)(\tau_1,\tau_2)}_{f}&=\frac{N(N+1)}{4}.
\end{split}\end{align}
\end{widetext}

\section{Sensitivity in the presence of spontaneous emission}\label{app:SensitivityGamma}
In this appendix we present the calculation of the expectation values required for the sensitivity in the presence of spontaneous emission. We present here the case of the resonant protocol. Even though the effect of spontaneous emission is different in the resonant and dispersive cases, the methods presented here can be adapted to tackle the case of the dispersive protocol.

Similar to the case of photon loss, in both cases we will compute the evolution within the Heisenberg picture, focusing on the collective observables. As discussed in the main text, since spontaneous emission is not the limiting factor of the sensitivity, we consider equal durations for the forward and backward evolutions:
\begin{equation}
    [(\hat{S}^+]^m(2\tau) =e^{\mathcal{M}_{\chi}^{\dagger}\tau}\mathcal{L}_{\mathcal{D}(\beta)}^{\dagger}e^{\mathcal{M}_{-\chi}^{\dagger}\tau}\big[(\hat{S}^+)^m\big] . \label{eqn:AppFEvo}
\end{equation}
Here, 
\begin{align}\begin{split}
    \mathcal{M}^{\dagger}_{\chi}\hat{O}&=i\chi[\hat{S}_z\hat{a}^{\dagger}\hat{a},\hat{O}]\\[5pt]&\hspace{1cm}+\gamma\sum_i\Big(2\hat{s}_z^i\hat{O} \hat{s}_z^i+\hat{s}_x^i\hat{O} \hat{s}_x^i+\hat{s}_y^i\hat{O} \hat{s}_y^i-\hat{O}\Big) ,
\end{split}\end{align}
and
\begin{equation}
    \mathcal{L}^{\dagger}_{\mathcal{D}(\beta)} \hat{O}=\mathcal{D}^{\dagger}(\beta)\hat{O}\mathcal{D}(\beta), 
\end{equation}
identically defined as per the previous section. 

We begin with
\begin{equation}
    (\hat{S}^+)^m(\tau)\equiv\hat{\Lambda}\equiv e^{\mathcal{M}_{-\chi}^{\dagger}\tau}\big[(\hat{S}^+)^m\big],
\end{equation}
which corresponds to the \emph{final} step of the protocol. By differentiating with respect to $\tau$, we find the $\hat{\Lambda}$ satisfies:
\begin{equation}\label{AppF:FirstMasterEquation}
    \partial_{\tau}\hat{\Lambda}=\mathcal{M}_{-\chi}^{\dagger} \hat{\Lambda},
\end{equation}
with initial condition $\hat{\Lambda}_0=[\hat{S}^+(0)]^m$.

The structure of $\mathcal{M}^{\dagger}_{-\chi}$ motivates the ansatz
\begin{equation}
    \hat{\Lambda}=f(\hat{n},\tau)(\hat{S}^+)^m,
\end{equation}
where $f$ is some operator valued function of $\hat{n}=\hat{a}^{\dagger}\hat{a}$ and is the only $\tau$ dependent part of the expression. Substitution of this ansatz into Eq.~(\ref{AppF:FirstMasterEquation}) results in a differential equation for $f$
\begin{equation}
    \partial_{\tau}f=-m\Big(\frac{3\gamma}{2}+i\hat{n}\Big)f .
\end{equation}
The solution is an exponential
\begin{equation}
    f(\hat{n},\tau)=e^{-\frac{3m\gamma\tau}{2}}e^{-i\hat{n}mt}
\end{equation} 
which is then substituted back into the ansatz to yield:
\begin{equation}
   (\hat{S}^+)^m(\tau)=e^{-\frac{3\gamma \tau}{2}}(\hat{S}^+)^m e^{-im\hat{n}\tau}.
\end{equation}

The next step involves a displacement, which is straightforward to implement once again using the identity $\mathcal{D}^{\dagger}(\beta)\hat{a}\mathcal{D}(\beta) = \hat{a} + \beta$ to yield:
\begin{align}\begin{split}
 (\hat{S}^+)^m(\tau)'&\equiv  \mathcal{L}^{\dagger}_{\mathcal{D}(\beta)}\big[(\hat{S}^+)^m(\tau)\big]\\[8pt]
 &=e^{-\frac{3\gamma \tau}{2}}(\hat{S}^+)^m e^{-im(\hat{a}^{\dagger}+\beta)(\hat{a}+\beta)\tau}. \label{eqn:AppEStep2}
\end{split}\end{align}

Before beginning the final step, it is convenient to do some preliminary manipulation of the last exponential in Eq.~(\ref{eqn:AppEStep2}),
\begin{align}\begin{split}
    e^{-im(\hat{a}^{\dagger}+\beta)(\hat{a}+\beta)\tau}&=e^{\beta^2(e^{-im\chi \tau}-1)}\bigg[\sum_{b\geq 0}e^{-\frac{i\chi m b \tau}{2}}(\hat{a}^{\dagger})^bO_b(\hat{n})\\[5pt]
    &+\sum_{b<0}e^{-\frac{i\chi m b \tau}{2}}O_{|b|}^{\dagger}(\hat{n})\,\hat{a}^{|b|}\bigg]e^{-i\chi m\hat{n} \tau}, \label{eqn:AppFPrelimStep3}
\end{split}\end{align}
where $O_{b}(\hat{n})$ are operator valued functions of $\hat{n}$ and $\tau$ that are to be determined. If we multiply by $e^{i\chi m\hat{n}\tau}$ on the right and take the expectation value of both sides with respect to a generic coherent state $\ket{\zeta}$ we find that the left and right hand sides of Eq.~(\ref{eqn:AppFPrelimStep3}) transform correspondingly to the new equality
\begin{widetext}
\begin{equation}
    \exp\Big[-2i\beta\sin(m\chi\tau/2)\big(\zeta^*e^{-im\chi\tau/2}+\zeta e^{im\chi\tau/2}\big)\Big]=\sum_{b\geq 0}e^{-\frac{i\chi m b \tau}{2}}(\zeta^*)^b\braket{O_b}+\sum_{b<0}e^{-\frac{i\chi m b \tau}{2}}\braket{O_{|b|}^{\dagger}}\,\zeta^{|b|} .
\end{equation}
Here, $\braket{O_b}=\bra{\zeta}O_b\ket{\zeta}$ is a function of $|\zeta|$ only. Expressing $\zeta$ in terms of its amplitude and phase $\zeta=|\zeta| e^{i\phi}$ we get
\begin{equation}
    \exp\Big[-4i|\zeta|\beta\sin(m\chi\tau/2)\cos(\phi+m\chi\tau/2)\Big]=\sum_{b\geq 0}e^{-\frac{i\chi m b \tau}{2}}|\zeta|^be^{-ib\phi}\braket{O_b}+\sum_{b<0}e^{-\frac{i\chi m b \tau}{2}}\braket{O_{|b|}^{\dagger}}|\zeta|^{|b|}e^{-ib\phi} . 
\end{equation}
\end{widetext}

Using the Jacobi-Anger expansion to re-express the LHS of this last equation and equating coefficients of $e^{ib\phi}$ we find that the following holds
\begin{equation}
    \bra{\zeta}O_b\ket{\zeta}=\frac{i^b}{|\gamma|^b}J_b\big[-4i|\zeta|\beta\sin(m\chi\tau/2)\big],
\end{equation}
where $J_b$ is the $b$th-order Bessel function of the first kind. The reason this expansion is convenient is that the evolution of the operators
\begin{equation}
    \hat{a}^b(\hat{S}^+)^m
\end{equation}
take a simple form in the final step of the evolution that we apply. The evolved operator at the end of the displacement step is, hence,
\begin{align}\begin{split}\label{eqn:AppEDisplacement}
 (\hat{S}^+)^m(\tau)'
 &=e^{-\frac{3\gamma \tau}{2}}(\hat{S}^+)^m e^{\beta^2(e^{-im\chi \tau}-1)}\\[5pt]
 &*\bigg[\sum_{b\geq 0}e^{-\frac{i\chi m b \tau}{2}}(\hat{a}^{\dagger})^bO_b(\hat{n})\\[5pt]
    &+\sum_{b<0}e^{-\frac{i\chi m b \tau}{2}}O_{|b|}^{\dagger}(\hat{n})\,\hat{a}^{|b|}\bigg]e^{-i\chi m\hat{n} \tau}
\end{split}\end{align}

Returning now to the evaluation of Eq.~(\ref{eqn:AppFEvo}), we finally must act with the superoperator $e^{\mathcal{M}^{\dagger}_{\chi}t}$ on each term of the sum in Eq.~(\ref{eqn:AppEDisplacement}), so we need to investigate the quantities
\begin{equation}
    \xi_{b,m}=e^{\mathcal{M}^{\dagger}_{\chi}\tau}\Big[\hat{a}^b(\hat{S}^+)^m\Big],
\end{equation}
which satisfy an equation similar to Eq.~(\ref{AppF:FirstMasterEquation}):
\begin{equation}\label{AppF:SecondMasterEqua}
    \partial_\tau\xi_{b,m}=\mathcal{M}^{\dagger}_{\chi}\xi_{b,m}
\end{equation}
This is all that we need to consider, since $\mathcal{M}^{\dagger}_{\chi}$ commutes with right and left multiplication by any function of $\hat{n}$. 

Once again, the structure of Eq.~(\ref{AppF:SecondMasterEqua}) motivates the ansatz
\begin{equation}\label{eqn:AppEAnsatz3}
    \xi_{b,m}=q_{b,m}(\hat{S}_z,\tau)\,\hat{a}^b(\hat{S}^+)^m e^{im\chi\hat{n}\tau}e^{-\frac{3\gamma \tau}{2}} ,
\end{equation}
where $q$ is a function to be solved for. 
Plugging the ansatz into Eq.~(\ref{AppF:SecondMasterEqua}) results in a differential-difference equation for $q_{b,m}(z,t)$:
\begin{align}\begin{split}\label{AppF:DiffEqn}
\partial_t q_{b,m}(z,t)&=-i\chi \,b \,z \,q_{b,m}(z,t)\\[5pt]&+\frac{\gamma N}{4}\Big[q_{b,m}(z+1,t)+q_{b,m}(z-1,t)-2q_{b,m}(z,t)\Big]\\[5pt]
&+\frac{\gamma z}{2}\Big[q_{b,m}(z-1,t)-q_{b,m}(z+1,t)\Big]\\[5pt]
& +\frac{\gamma m}{2}\Big[q_{b,m}(z,t)-q_{b,m}(z-1,t)\Big]
\end{split}\end{align}
subject to the initial condition $q_{b,m}(z,0)=1$, which is obtained by plugging $\tau=0$ in Eq.~(\ref{eqn:AppEAnsatz3}). To solve Eq.~(\ref{AppF:DiffEqn}), we plug in the ansatz solution $q_{b,m}(t,z)=\exp[v(t)+w(t)z]$, which leads into a set of ordinary nonlinear differential equations for $v$ and $w$:
\begin{align}\begin{split}
    \dot{v}&=\frac{\gamma N}{4}(e^{w}+e^{-w}-2)+\frac{\gamma m}{2}(1-e^{-w})\\[5pt]
    \dot{w}&=-i\chi b+\frac{\gamma}{2}(e^{-w}-e^w)
\end{split}\end{align}
These equations can be solved exactly, but the solution is not particularly illuminating. Instead, we solve it perturbatively in the ratio $\chi/\gamma$, which is very small for both protocols described in the main text, and assuming $N\gg 1$. Then,
\begin{align}\begin{split}\label{eqn:AppFV&W}
    v(t)&=-\frac{i\chi m b}{2\gamma}(e^{-\gamma t}+\gamma t-1)\\[5pt]
    &+\frac{N\chi^2 b^2}{8\gamma^2}(e^{-4\gamma t}-4e^{-\gamma t}-2\gamma t+3)\\[8pt]
    w(t)&=\frac{i\chi b}{\gamma}(e^{-\gamma t}-1),
\end{split}\end{align}
to order $(\chi/\gamma)^3$. We now express the evolved observable in terms of the $q_{b,m}$:
\begin{widetext}
\begin{align}\begin{split}
   (\hat{S}^+)^m(2\tau)f&=e^{-3m\gamma t}e^{\beta^2(e^{-im\chi t}-1)}\Bigg[\sum_{b\geq 0}e^{-\frac{im\chi b t}{2}}(\hat{a}^{\dagger})^b\hat{O}_bq_{-b,m}(\hat{S}^+)^m+\sum_{b<0}e^{-\frac{im\chi b t}{2}}\hat{O}^{\dagger}_{|b|}\hat{a}^bq_{-b,m}(\hat{S}^+)^m\Bigg],\\[8pt]
   (\hat{S}^+)^m(2\tau)f&=e^{-3m\gamma t}e^{\beta^2(e^{-im\chi t}-1)}\Bigg[\sum_{b\geq 0}e^{-\frac{im\chi b t}{2}}(\hat{a}^{\dagger})^b\hat{O}_bq_{-b,m}(\hat{S}^+)^m+\sum_{b<0}e^{-\frac{im\chi b t}{2}}\hat{O}^{\dagger}_{|b|}\hat{a}^be^{v+w\hat{S}_z}(\hat{S}^+)^m\Bigg].
\end{split}\end{align}
With this, we can take the expectation value with respect to $\hat{\rho}_0=\ketbra{N/2_x}{N/2_x}\otimes\ketbra{\alpha}{\alpha}$, which results in
\begin{align}\begin{split}
    \big\langle(\hat{S}^+)^m(2\tau)\big\rangle&=e^{-3m\gamma t}e^{\beta^2(e^{-im\chi t}-1)}\Bigg\{\sum_{b}i^bJ_b\big[-4i\alpha\beta\sin(m\chi t/2)\big]e^{-\frac{im\chi bt}{2}}\big[\bra{N/2_x}e^{v+w\hat{S}_z}(\hat{S}^+)^m\ket{N/2_x}\big]\Bigg\},\\[8pt]
     \big\langle(\hat{S}^+)^m(2\tau)\big\rangle&\approx \big\langle(\hat{S}^+)^m(0)\big\rangle e^{-3m\gamma t}e^{\beta^2(e^{-im\chi t}-1)}\Bigg\{\sum_{b}i^bJ_b\big[-4i\alpha\beta\sin(m\chi t/2)\big]e^{-\frac{im\chi bt}{2}}e^ve^{-N|w|^2/8}\Bigg\}.
\end{split}\end{align}
\end{widetext}
In Eq.~(\ref{eqn:AppFV&W}), we can omit the imaginary term on $v$ since it is small and has no $N$ enhancement factor, as compared to its second term. Furthermore, if we are probing times such that $\gamma t\ll 1$, we can keep only first order terms in $\gamma$ both in $v$ and $w$.
From this we obtain the final result for the expectation value:
\begin{widetext}
\begin{equation}\label{eqn:AppEExpectationValues}
\big\langle(\hat{S}^+)^m(2\tau)\big\rangle= \big\langle(\hat{S}^+)^m(0)\big\rangle e^{-3m\gamma t}e^{\beta^2(e^{-im\chi t}-1)}\Bigg\{\sum_{b}\exp\Big[\frac{\gamma\chi^2b^2N t^3}{24}\Big]i^bJ_b\big[-4i\alpha\beta\sin(m\chi t/2)\big]e^{-N\chi^2m^2 t^2/8}\Bigg\}.
\end{equation}
\end{widetext}

In comparison to the ideal case, Eq.~(\ref{AppA:IdealExpectationValues}), we note that spontaneous emission introduces two effects: the first is single-particle type decay of the expectation values, with the exponent $e^{-3\gamma m t}$, whilst the latter is the result of entanglement dynamics, given by the $e^{\gamma N\chi^2b^2t^3/24}$ factors. In the case of the resonant protocol, with $\chi=g/\alpha$, the latter can be made negligible by choosing a sufficiently large $\alpha$.

Lastly, to finally compute the necessary means and variances of $\hat{S}_{x,y}$, the expectation 
\begin{equation}
    \langle (\hat{S}^+\hat{S}^-)(2\tau) \rangle
\end{equation}
is required. This can be computed in a straightforward way by identifying that $\hat{S}^+\hat{S}^-$ initially commutes with the Hamiltonian $\hat{H}=\chi\hat{S}_z\hat{a}^{\dagger}\hat{a}$ and remains to do so throughout the evolution. Thus, its dynamics is given entirely by the spontaneous emission term, from which we can immediately write down the solution
\begin{equation}
     \langle (\hat{S}^+\hat{S}^-)(2\tau) \rangle=\frac{N}{2}+\frac{N(N-1)}{4}e^{-6\gamma t}
\end{equation}

The relevant expectation values are then:

\begin{widetext}
\begin{align}\begin{split}\label{eqn:AppResonantEV}
    \big\langle\hat{S}^{+}(2\tau)\big\rangle&=\frac{N}{2}e^{-3\gamma \tau}e^{\beta^2(e^{-i\chi \tau}-1)}\sum_{b=-\infty}^{\infty}e^{\frac{\gamma \chi^2b^2 N \tau^3}{24}}i^bJ_b\Big[-4\alpha\beta\sin\Big(\frac{\chi \tau}{2}\Big)\Big]e^{-Nb^2\chi^2\tau^2/8} ,\\[5pt]
     \big\langle\hat{S}^{+2}(2\tau)\big\rangle&=\frac{N(N-1)}{4}e^{-6\gamma \tau}e^{\beta^2(e^{-2i\chi \tau}-1)}\sum_{b=-\infty}^{\infty}e^{\frac{\gamma \chi^2b^2 N \tau^3}{24}}i^bJ_b\Big[-4\alpha\beta\sin(\chi \tau)\Big]e^{-Nb^2\chi^2\tau^2/8} ,\\[5pt]
    \big\langle(\hat{S}^+\hat{S}^-)(2\tau)\big\rangle&=\frac{N}{2}+\frac{N(N-1)}{4}e^{-6\gamma \tau}, 
\end{split}\end{align}
\end{widetext}

Keeping only the single particle decay terms, we find that the sensitivity to displacements (at $\beta=0$) when measuring $\hat{S}_\phi=\hat{S}_x\cos\phi+\hat{S}_y\sin(\phi)$ and $\phi\neq 0$ is
\begin{equation}
    (\delta\beta)^2=\frac{e^{N\chi^2t^2/8}}{4N\alpha\chi^2 \tau^2}\frac{(e^{6\gamma\tau}-\cos\phi^2)}{\sin\phi^2}.
\end{equation}
For short times $\chi\sqrt{N}\tau\ll 1$, we recover Eq.~(\ref{eqn:SensSpontEmit_Resonant}) in the main text.

A similar derivation shows that the relevant expectation values in the dispersive protocol are the following:

\begin{widetext}
\begin{align}\begin{split}\label{eqn:AppEDispersiveEV}
    \braket{\hat{S}^{+}(2\tau)}&=\frac{N}{2}e^{-2\gamma \tau}e^{\beta^2(e^{-i\chi \tau}-1)}\sum_{b=-\infty}^{\infty}e^{\frac{\gamma \chi^2(5b^2+8b+6) N \tau^3}{12}}i^bJ_b\Big[-4\alpha\beta\sin\Big(\frac{\chi \tau}{2}\Big)\Big]e^{-Nb^2\chi^2\tau^2/8} ,\\[5pt]
     \braket{\hat{S}^{+2}(2\tau)}&=\frac{N(N-1)}{4}e^{-4\gamma \tau}e^{\beta^2(e^{-2i\chi \tau}-1)}\sum_{b=-\infty}^{\infty}e^{\frac{\gamma \chi^2(5b^2+16b+24)N \tau^3}{12}}i^bJ_b\Big[-4\alpha\beta\sin(\chi \tau)\Big]e^{-Nb^2\chi^2t^2/8} ,\\[5pt]
     \braket{(\hat{S}^+\hat{S}^-)(2\tau)}&=\frac{N}{2}+\frac{N(N-1)}{4}e^{-4\gamma \tau} .
\end{split}\end{align}
\end{widetext}

\section{Simplified evaluation of quantum Fisher information}\label{app:SimplifiedFisherInformation}
In this appendix, we show how to simplify the calculation of the quantum Fisher information of a mixed state when this state has only a few eigenvalues that are nonzero. This is relevant for the manipulations that lead to Eq.~(\ref{eqn:FisherSimplified}), to Eq.~(\ref{eqn:FisherSpin}) and for the numerical evaluation of Eq.~(\ref{eqn:FisherSpin}) in the main text. 

For generality, we consider
\begin{equation}
    \hat{\rho}=\sum_{a}\lambda_{a}\ket{a}\bra{a},
\end{equation}
where $\hat{\rho}$ is some density matrix, $\ket{a}$ are the eigenstates of $\hat{\rho}$ and $\lambda_a$ the corresponding eigenvalues. Then, the Fisher information with respect to a generator $\hat{O}$ is
\begin{equation}
    \mathcal{F}_{Q}=2\sum_{a,b}\frac{(\lambda_a-\lambda_b)^2}{\lambda_a+\lambda_b}\big|\langle a|\hat{O}|b\rangle\big|^2 .
\end{equation}

Let us now assume that the eigenvalues can be separated into two sets, $I$ and $I^c$ (complement of $I$) such that for any $\lambda_r\in I$ and $\lambda_s\in I^c$, $\lambda_r\gg\lambda_s$. Given that the eigenvalues of $\rho$ are always less than 1, this effectively means that $\lambda_s\in I^c$ are much closer to zero than $\lambda_r\in I$. Then we can split the sum into 4 parts
\begin{equation}
    \sum_{a,b}= \sum_{a,b\,\in I}+\sum_{a\in I^c,b\,\in I}+\sum_{a\,\in I,b\in I^c}+\sum_{a\in I^c,b\in I^c}
\end{equation}
We can neglect the last term since it is of the first order in the small eigenvalues. The second and third sums are equal and $\lambda_a\in I$ dominate over $\lambda_a\in I^c$ so they can be simplified to
\begin{align}\begin{split}
    \sum_{a\in I^c,b\,\in I}\lambda_b\big|\langle a|\hat{O}|b\rangle\big|^2&=  \sum_{a\in I^c I,b\,\in I}\lambda_b\langle a|\hat{O}|b\rangle\langle b|\hat{O}^{\dagger}|a\rangle\\[5pt]
    &= \sum_{a\in I^c,b\,\in I}\langle a|\hat{O}\hat{\rho}|b\rangle\langle b|\hat{O}^{\dagger}|a\rangle\\[5pt]
    &=\sum_{a\in I^c\in I}\langle a|\hat{O}\hat{\rho}\hat{P}_I\hat{O}^{\dagger}|a\rangle\\[5pt]
    &=\mathrm{Tr}\bigg(\sum_{a\not\in I}|a\rangle\langle a|\hat{O}\hat{\rho}\hat{P}_I\hat{O}^{\dagger}\bigg)\\[5pt]
    &=\mathrm{Tr}\Big[(1-\hat{P}_I)\hat{O}\hat{\rho}\hat{P}_I\hat{O}^{\dagger}\Big],
\end{split}\end{align}
where
\begin{equation}
    \hat{P}_I=\sum_{b\in I}|b\rangle\langle b|
\end{equation}
is the projector into subspace $I$. The Fisher information then becomes
\begin{align}\begin{split}
    \mathcal{F}_Q&=2\sum_{a,b\in I}\frac{(\lambda_a-\lambda_b)^2}{\lambda_a+\lambda_b}\big|\langle a|\hat{O}|b\rangle\big|^2\\[5pt]
    &\hspace{1cm}+4\mathrm{Tr}\Big(\hat{P}_I\hat{O}^{\dagger}\hat{O}\hat{\rho}\Big)-4\mathrm{Tr}\Big(\hat{P}_I\hat{O}^{\dagger}\hat{P}_I\hat{O}\hat{\rho}\Big),
\end{split}\end{align}
which is the result Eq.~(\ref{eqn:FisherSimplified}) quoted in the main text.

\end{document}